\documentclass[onecolumn,superscriptaddress,nofootinbib,floatfix,notitlepage,prd]{revtex4-1}

\usepackage{graphicx}
\usepackage{amssymb,amsmath}
\usepackage[usenames]{color}
\usepackage{wasysym}
\usepackage{soul}
\usepackage{times}
\usepackage{mathrsfs}
\usepackage{hyperref}
\usepackage{booktabs}
\usepackage{chngcntr}
\usepackage{tikz}
\usepackage{mathtools}
\usepackage{cleveref}
\usepackage{verbatim}

\hypersetup{
  colorlinks=true,          linkcolor=blue,           citecolor=cyan,         }
\usepackage{float}

\usepackage{tensor}

\definecolor{orange}{rgb}{1,0.5,0}

\def\lesssim{\mathrel{\hbox{\rlap{\hbox{\lower4pt\hbox{$\sim$}}}\hbox{$<$}}}}
\def\gtrsim{\mathrel{\hbox{\rlap{\hbox{\lower4pt\hbox{$\sim$}}}\hbox{$>$}}}}
\def\alt{\mathrel{\hbox{\rlap{\hbox{\lower4pt\hbox{$\sim$}}}\hbox{$<$}}}}
\def\agt{\mathrel{\hbox{\rlap{\hbox{\lower4pt\hbox{$\sim$}}}\hbox{$>$}}}}

\def\gta{\ifmmode {\mathbin{\lower 3pt\hbox       {$\,\rlap{\raise 5pt\hbox{$\char'076$}}\mathchar"7218\,$}}}
    \else {${\mathbin{\lower 3pt\hbox
    {$\rlap{\raise 5pt\hbox{$\char'076$}}\mathchar"7218\,$}}}
    $}\fi}
\def\lta{\ifmmode {\,\mathbin{\lower 3pt\hbox       {$\,\rlap{\raise 5pt\hbox{$\char'074$}}\mathchar"7218\,$}}}
    \else {${\mathbin{\lower 3pt\hbox
    {$\rlap{\raise 5pt\hbox{$\char'074$}}\mathchar"7218\,$}}}
    $}\fi}

\newcommand{\beq}{\begin{equation}}
\newcommand{\eeq}{\end{equation}}
\newcommand{\bea}{\begin{eqnarray}}
\newcommand{\eea}{\end{eqnarray}}
\newcommand{\ba}{\begin{align}}
\newcommand{\ea}{\end{align}}

\DeclareMathOperator{\sgn}{sgn}
\DeclarePairedDelimiter\abs{\lvert}{\rvert}
\DeclarePairedDelimiter\norm{\lVert}{\rVert}
\makeatletter
\let\oldabs\abs
\def\abs{\@ifstar{\oldabs}{\oldabs*}}
\let\oldnorm\norm
\def\norm{\@ifstar{\oldnorm}{\oldnorm*}}
\makeatother

\renewcommand{\BibitemShut}[1]{}

\newcommand{\Ekin}{\mathcal{E}_{kin}}
\newcommand{\Emag}{\mathcal{E}_{mag}}
\graphicspath{{figures/}}
\DeclareGraphicsExtensions{.pdf}

\newcommand{\Reyn}{\operatorname{\mathit{Re}}}

\allowdisplaybreaks
\begin{document}
\title{Artificial neural network subgrid models of 2-D compressible magnetohydrodynamic turbulence}

\author{Shawn G. Rosofsky}
\affiliation{NCSA, University of Illinois at Urbana-Champaign, Urbana, Illinois 61801, USA}
\affiliation{Department of Physics, University of Illinois at Urbana-Champaign, Urbana, Illinois 61801, USA}
\author{E. A. Huerta}
\affiliation{NCSA, University of Illinois at Urbana-Champaign, Urbana, Illinois 61801, USA}
\affiliation{Department of Astronomy, University of Illinois at Urbana-Champaign, Urbana, Illinois 61801, USA}

\begin{abstract}
We explore the suitability of deep learning to capture the physics of subgrid-scale ideal magnetohydrodynamics turbulence of 2-D simulations of the magnetized Kelvin-Helmholtz instability. We produce simulations at different resolutions to systematically quantify the performance of neural network models to reproduce the physics of these complex simulations. We compare the performance of our neural networks with gradient models, which are extensively used in the extensively in the magnetohydrodynamic literature. Our findings indicate that neural networks significantly outperform gradient models in accurately computing the subgrid-scale tensors that encode the effects of magnetohydrodynamics turbulence. To the best of our knowledge, this is the first exploratory study on the use of deep learning to learn and reproduce the physics of magnetohydrodynamics turbulence. 
\end{abstract}

\maketitle

\section{Introduction}
\label{sec:intro}

In astrophysical simulations of magnetohydrodynamics (MHD) such as magnetized binary neutron star (BNS) mergers, we confront turbulent phenomena in the limit of infinite Reynolds number $\Reyn$ \cite{Schmidt2015}.  While these turbulent effects are often ignored, ultra-high resolution simulations of magnetized BNS mergers have demonstrated that MHD turbulence can amplify the magnetic field by several orders of magnitude, and occur only at resolutions that are too computationally expensive to run in bulk~\cite{Kiuchi2015a,Baiotti2017}.  This amplification is due to the magnetized Kelvin-Helmholtz Instability (KHI), which occurs when two fluids flow past each other in opposite directions. In order to help resolve turbulence originating from the KHI, we examine the methods employed in more traditional hydrodynamical turbulence.

To resolve turbulent effects, the computational fluid dynamics (CFD) community uses several classes of simulations that provide varying degrees of accuracy.  Direct numerical simulations (DNS) provide the most accurate results by capturing all the effects at all scales relevant to the problem being studies.  To resolve the turbulent effects these simulations, DNS require extremely high resolutions that scale as the cube of $\Reyn$.  This resolution require renders DNS feasible only for a small number of simulations.  Moreover, our problems of interest has extremely high $\Reyn$, resulting in DNS becoming too computationally expensive for our work.

The other prominent techniques, Reynolds-averaged Navier-Stokes (RANS), and large eddy simulations (LES), employ subgrid-scale (SGS) models to reproduce the most important effects of DNS such as the energy transfer rate at much lower resolutions. RANS is the most widely used, but is best suited for steady state phenomena.  For an instability such as the KHI, LES serve as the preferred approach. The goal behind LES is to evolve the equations with sufficient resolution to resolve the largest eddies and rely on the SGS model to compute the contribution of the smaller eddies. 

Recent work has sought to develop SGS models of MHD turbulence using traditional LES models \cite{Muller2002,Muller2002a,Miesch2015,Grete2015,Grete2016,Grete2017a,Grete2017b,Kessar2016,Vlaykov2016,Vigano2019a,Carrasco2019a,Grete2017}.  However, MHD turbulence presents some unique challenges not observed in standard hydrodynamical turbulence.  Although the boundary conditions for problems of interest are typically much simpler, the equations are more complex.  These complexities include a dynamo mechanism for the conversion between kinetic and magnetic energy as well as anisotropies arising from the magnetic field \cite{Beresnyak2015,Beresnyak2019,Grete2017}.  Moreover, there exists a much weaker understanding of MHD turbulence compared to the hydrodynamical variety. 

To resolve these complexities without exerting significant efforts studying the intricacies of MHD turbulence, we explore the use of artificial neural networks (ANN) to act as SGS models.  Significant work has been done in examining and evaluating ANN models of hydrodynamical turbulence for both RANS and LES in recent years \cite{Ling2016,Maulik2017,Fang2018,Wang2018,Xie2019,Xie2019b,Xie2020a,Pawar2020,Brunton2020}.  These studies indicate that ANNs may outperform traditional approaches used to model turbulence. 

In this article we develop a proof-of-concept neural network model to quantify the performance of deep learning algorithms to reproduce the true dynamics of turbulent magnetic field amplification at manageable resolutions of MHD simulations of the KHI in the LES formalism. We use as a driver for this study 2-D MHD simulations, and compare the performance of our neural networks to traditional models, such as the  \textit{a priori} study introduced in~\cite{Vigano2019a}. For reference, \textit{a priori} study involves evaluating the performance of the models in how closely they reproduce the SGS effects compared to the filtered DNS data. In contrast, \textit{a posteriori} study would implement these models in an actual simulation to observe how the SGS models compare to the higher resolution DNS simulations.  We leave the more computationally expensive 3-D case as well as the implementation of these neural network models, and subsequent \textit{a posteriori} comparison of the models, to future work.  Herein, we will perform a more in-depth analysis of the conditions each model performs best in our \textit{a priori} study to gain as much insight as possible before moving to the more complicated tests.

This article is organized as follows: \Cref{sec:LES} provides an overview of the LES formalism and its application to the MHD equations.  In \Cref{sec:models} we describe the SGS models used in this work, including our proposed ANN model and the traditional gradient model.  \Cref{sec:simulation} describes the simulations used to train and evaluate our SGS models. We describe the methods in which those simulations were employed to train the ANN model in \Cref{sec:training}.  In \Cref{sec:methodology} we define the metrics used to evaluate the SGS models.  We provide the results of our \textit{a priori} study of the ANN SGS turbulence model and compare its performance with that of the gradient model.  \Cref{sec:conclusions} summarizes our findings and outline future directions of work.

\section{LES Formalism}
\label{sec:LES}
In this section we introduce the mathematical formalisms that we will use throughout the article. We describe the LES formalism, and briefly describe the compressible MHD equations, which will be used as the science driver of our analysis.

\subsection{Filtering}
\label{sec:Filtering}

In the LES formalism, one views the grid resolution as a spatial filter applied to a continuous variable. In this approach, the size of the grid $\Delta$ corresponds to the size of the filter.  Typically, we start with very high resolution data taken from DNS or experimental results and apply a filter with a cutoff size $\Delta_f$, where $\Delta_f>\Delta$ is the lower resolution grid on which we want to perform our simulation on. We apply the kernel $G$ to a field $f$ as

\begin{align}
\overline{f}(\mathbf{x}, t)&=\int_{-\infty}^{\infty} G\left(\mathbf{x}-\mathbf{x}^{\prime}\right) f\left(\mathbf{x}^{\prime}, t\right) d\mathbf{x}^{\prime}. \label{eq:filter}
\end{align}

For implicit LES simulations which are employed in this work, the filtering operator of size $\Delta_f$ is applied to the high resolution simulation of grid size $\Delta$ when calibrating SGS models.  In turn, this filtering provides insight into the effect of moving to a lower to a lower grid resolution. The choice of filter depends on the numerical method employed.  For finite volume schemes like those used in this work, a box or top-hat filter is used to simulate the spatial averaging that occurs during such schemes.  This filter kernel is given in real space for $D$ spatial dimensions as 
\begin{align}
G\left(\left|\mathbf{x}-\mathbf{x}^{\prime}\right|\right)&=\prod_{i=1}^{D} G_{i}\left(\left|x_{i}-x_{i}^{\prime}\right|\right)\,, \label{eq:box_filter}
\end{align}
where
\begin{align}
G_{i}\left(\left|x_{i}-x_{i}^{\prime}\right|\right)&=\left\{\begin{array}{ll}{1 / \Delta_{f}} & {\text { if }\left|x_{i}-x_{i}^{\prime}\right| \leq \Delta_{f} / 2}\,. \\ {0} & {\text { otherwise }}\end{array}\right. \label{eq:box_filter_1D}
\end{align}

\noindent Filtering operators commute with linear terms.  However, nonlinearities in the MHD equations fail to commute with the filtering operator.  This results in a residual term known as the SGS tensor.  We will provide examples of these SGS tensors in the next section.

For compressible fluids, we use a specific type of filtering called Favre or density weighted to simplify our problem by eliminating the SGS tensor in the continuity equation.  For some quantity $f$ weighted by some density $\rho$, we define the Favre filtered quantity $\widetilde{f}$ as 
\begin{align}
\widetilde{f}&=\frac{\overline{\rho f}}{\overline{\rho}}.
\end{align}
This also gives us the identity $\overline{\rho f} = \overline{\rho}\widetilde{f}$.

\subsection{Compressible MHD Equations}
\label{sec:MHD_eq}
\subsubsection{Unfiltered MHD Equations}
\label{sec:MHD_unfilt}

For the evolution of our system, we used the conservative form of the ideal compressible Newtonian MHD equations.  Each equation continuity, momentum, induction, and energy evolution respectively represents the local evolution of a globally conserved quantity.  The equations are given by 
\begin{gather} 
\partial_{t} \rho+\partial_{i}\left[\rho v^{i}\right]=0\,, \label{eq:continuity} \\ 
\partial_{t}\left(\rho v^{j}\right)+\partial_{i}\left[\rho v^{i} v^{j}-B^{i} B^{j}+\delta^{i j}\left(p+\frac{B^{2}}{2}\right)\right]=0 \,,\label{eq:momentum} \\ 
\partial_{t} B^{j}+\partial_{i}\left[v^{i} B^{j}-v^{j} B^{i}\right]=0\,, \label{eq:induction} \\ 
\partial_{t} u+\partial_{i}\left[\left(u+p+B^{2}\right) v^{i}-\left(v_{j} B^{j}\right) B^{i}\right]=0\,, \label{eq:energy}
\end{gather}
where the total energy density $u$ is defined as
\begin{align}
u&=e+\frac{\rho v^2}{2}+\frac{B^2}{2}.
\end{align}
Here, the indices are spacial components assuming Einstein summation convention, $\delta^{ij}$ is the Kronecker delta, $\rho$ is the mass density, $p$ is the pressure, $e$ is the internal energy density, $v^i$ is the velocity, and $B^i$ is the magnetic field.  The units of this expression are such that the speed of light $c$ and the magnetic permeability $\mu_0$ are $c=\mu_0=1$. For this system, we used an ideal gas equation of state (EOS) to define $p$ as 
\begin{align}
p&=\left( \gamma-1 \right)e\,,
\end{align}
where $\gamma$ is the adiabatic index set to $\gamma=4/3$ for a relativistic gas in this work.  We note that we intentionally did not exploit any simplifications made using the fact that we have an ideal gas EOS to ensure that our ANN turbulence model can be used for any generic EOS.  This is done to ensure that the model can be easily employed by BNS simulations where the EOS is a variable parameter.

\subsubsection{Filtered MHD Equations}
\label{sec:MHD_filt}

To derive the filtered equations, we apply \Cref{eq:box_filter} to \Crefrange{eq:continuity}{eq:energy} \cite{Vigano2019a}.  We find these equations become
\begin{gather}
\partial_t\overline{\rho}+\partial_i\left[\overline{\rho}{\widetilde{v}}^i\right]=0\,, \label{eq:filt_continuity}\\
\partial_t\left(\overline{\rho}{\widetilde{v}}^j\right)+\partial_i\left[\overline{\rho}{\widetilde{v}}^i{\widetilde{v}}^j-{\overline{B}}^i{\overline{B}}^j+\delta^{ij}\left(\widetilde{p}+\frac{{\overline{B}}^2}{2}\right)\right]=-\partial_i\tau_{mom}^{ij}\,, \label{eq:filt_momentum} \\
\partial_t{\overline{B}}^j+\partial_i\left[{\widetilde{v}}^i{\overline{B}}^j-{\widetilde{v}}^j{\overline{B}}^i\right]=-{\partial_i\tau}_{ind}^{ij}\,, \label{eq:filt_induction} \\
\partial_t\overline{u}+\partial_i\left[\left(\overline{u}+\widetilde{p}+{\overline{B}}^2\right){\widetilde{v}}^i-\left({\widetilde{v}}_j{\overline{B}}^j\right){\overline{B}}^i\right]=-\partial_i\tau_{eng}^i+\Sigma_{eng}\,, \label{eq:filt_energy}
\end{gather}
where the merged SGS tensor terms are given by
\begin{align}
\tau_{mom}^{ij}&=\overline{\rho}\tau_{kin}^{ij}-\tau_{mag}^{ij}+\delta^{ij}\left(\frac{1}{2}\delta_{kl}\tau_{mag}^{kl}+\left(\overline{p}-\widetilde{p}\right)\right)\,, \label{eq:tau_mom} \\
\tau_{eng}^{i}&=\tau_{enth}^i+\tau_{mom}^{ij}{\widetilde{v}}_j+\tau_{ind}^{ij}{\overline{B}}_j\,, \label{eq:tau_eng}
\end{align}
and the scalar SGS tensor terms denoted by $\Sigma$ are given by
\begin{align}
\Sigma_{eng}&=\Sigma_{pres}+\Sigma_{mom}+\Sigma_{ind}\,,\\
\Sigma_{pres}&=\overline{v^i\partial_i p} -{\widetilde{v}}^i\partial_i\widetilde{p}\,,\\
\Sigma_{mom}&=\frac{1}{2}\left(\partial_i{\widetilde{v}}_j+\partial_j{\widetilde{v}}_i\right)\tau_{mom}^{ij}\,,\\
\Sigma_{ind}&=\frac{1}{2}\left(\partial_i{\overline{B}}_j-\partial_j{\overline{B}}_i\right)\tau_{ind}^{ij}\,.
\end{align}
In the above expressions, we have defined 
\begin{align}
\widetilde{e} &= \overline{u}-\frac{\overline{\rho} \widetilde{v}^{2}}{2}-\frac{\overline{B}^{2}}{2}\,, \\
\widetilde{p} &= (\gamma-1) \widetilde{e}\,,
\end{align}
and will define the enthalpy $h$ and its filtered version $\widetilde{h}$ as
\begin{align}
h&=\rho+e+p\,, \\
\widetilde{h}&=\overline{\rho}+\widetilde{e}+\widetilde{p}.
\end{align}

For modeling the SGS terms in \Crefrange{eq:filt_continuity}{eq:filt_energy}, we only care about $\tau_{kin}$ describing turbulent motion, $\tau	_{mag}$ describing the contribution of the turbulent magnetic field to the motion, $\tau_{ind}$ describing the turbulent amplification of the magnetic field, and $\tau_{enth}$ describing the effect of turbulence on the energy transfer.  We neglect the terms $\left(\overline{p}-\widetilde{p}\right)$ and $\Sigma_{pres}$ as we expect their contributions to be small and EOS dependent, which reduces the robustness of our models.  The rest of the terms in \Crefrange{eq:filt_continuity}{eq:filt_energy} are combinations of the aforementioned terms. The four SGS tensors we want to model are defined formally as 
\begin{align}
\tau_{kin}^{ij}&=\widetilde{v^iv^j}-{\widetilde{v}}^i{\widetilde{v}} \label{eq:tau_kin}^j\,, \\
\tau_{mag}^{ij}&=\overline{B^iB^j}-{\overline{B}}^i{\overline{B}}^j \label{eq:tau_mag}\,, \\
\tau_{ind}^{ij}&=\left(\overline{v^iB^j}-\overline{v^jB^i}\right)-\left({\widetilde{v}}^i{\overline{B}}^j-{\widetilde{v}}^j{\overline{B}}^i\right) \label{eq:tau_ind}\,, \\
\tau_{enth}^i&=\overline{hv^i}-\widetilde{h}{\widetilde{v}}^i\,. \label{eq:tau_enth}
\end{align}

\noindent The astute reader may notice that $\tau_{enth}^{i}$ is actually a vector, but we will refer to it as an SGS tensor throughout this work for the sake of conciseness.

\section{Modeling SGS Tensors}
\label{sec:models}

In this section we introduce the gradient model, which currently represents the state-of-the-art in the LES MHD literature, and our deep learning algorithm. In what follows, we will present direct comparisons between these two methodologies to highlight their key differences, and to furnish evidence that deep learning outperforms the gradient approach.

\subsection{Gradient Model}
\label{sec:grad}

The gradient model is extensively used in the LES MHD literature \cite{Grete2017,Vigano2019a}. The prevalence of this model in other LES MHD turbulence studies promotes it as a good baseline to test the performance of our neural network model. The gradient model is derived using the Taylor expansion of the SGS stress tensor under a particular filtering operator.  Here we use the leading order expansion of our box filtering operator which is also valid for a Gaussian filter~\cite{Grete2017} to obtain 

\begin{align}
\overline{f g} &\simeq \overline{f} \overline{g}+\frac{\Delta_f^2}{12} \partial^{i} \overline{f} \partial_{i} \overline{g}\,, \label{eq:filt_expansion} \\
\widetilde{f g} &\simeq \widetilde{f} \widetilde{g}+\frac{\Delta_f^2}{12} \partial^{i} \widetilde{f} \partial_{i} \widetilde{g} \,,\label{eq:favre_filt_expansion} \\
\overline{f g} &\simeq \widetilde{f} \overline{g}+\frac{\Delta_f^2}{12} \partial^{i} \widetilde{f} \left( \partial_{i} \overline{g} - \frac{\partial_i \overline{\rho}}{\overline{\rho}} \overline{g} \right)\,, \label{eq:mixed_filt_expansion}
\end{align}
for regular filtered terms, Favre filtered terms, and mixed filtered terms, respectively~\cite{Vigano2019a}. This results in the following expressions for the SGS tensors~\cite{Vigano2019a}

\begin{align}
\tau_{kin}^{ij}=&C_{kin}^{ij}\frac{\Delta_f^2}{12}\partial_k{\widetilde{v}}^i\partial^k{\widetilde{v}}^j\,, \label{eq:grad_tau_kin} \\
\tau_{mag}^{ij}=&C_{mag}^{ij}\frac{\Delta_f^2}{12}\partial_k{\overline{B}}^i\partial^k{\overline{B}}^j\,, \label{eq:grad_tau_mag} \\
\tau_{ind}^{ij}=&C_{ind}^{ij}\frac{\Delta_f^2}{12}\Biggl[\partial_k\widetilde{v}^i\left(\partial^k{\overline{B}}^j-\frac{\partial^k \overline{\rho}}{\overline{\rho}} {\overline{B}}^j\right) \notag \\  &-\partial_k{\widetilde{v}}^j\left(\partial^k{\overline{B}}^i-\frac{\partial^k \overline{\rho}}{\overline{\rho}}{\overline{B}}^i\right)\Biggr]\,, \label{eq:grad_tau_ind} \\
\tau_{enth}^i=&C_{enth}^i\frac{\Delta_f^2}{12}\frac{\gamma}{\gamma-1}\left[\partial_j\widetilde{p}-\widetilde{p}\frac{\partial_j \overline{\rho}}{\overline{\rho}}\right]\partial^j\widetilde{v}^i\,. \label{eq:grad_tau_enth}
\end{align}	

The coefficient $C^{ij}$ is determined by the best fit of the data to a time slice of filtered DNS data for each component of $\tau^{ij}_{grad}$ independently.  The fitting is determined by 

\begin{align}
C^{ij}&=\frac{\sum\limits_{\mathbf{x}_f} \left(\tau^{ij}_{DNS}\left(\mathbf{x}_f\right)\tau^{ij}_{grad}\left(\mathbf{x}_f\right)\right)}{\sum\limits_{\mathbf{x}_f} \tau^{ij}_{grad}\left(\mathbf{x}_f\right)}\,, \label{eq:Cij_coef}
\end{align}
where $\tau^{ij}_{grad}$ is the SGS tensor calculated by the gradient model in \Crefrange{eq:grad_tau_kin}{eq:grad_tau_enth}, $\tau^{ij}_{DNS}$ is the true SGS tensor computed directly from the DNS data, $\mathbf{x}_f$ represents the filtered grid, and Einstein summation notation is not used. When employing this model in an \textit{a posteriori} test, one would estimate $C^{ij}$ with a secondary filter \cite{Vollant2016,Xie2019}.  In the LES literature, this is known as a dynamical model.  However, we do not use a secondary filter for our \textit{a priori} study and instead filter the DNS data directly.  We acknowledge that this may overestimate the performance of the gradient model compared to an \textit{a posteriori} study.

\subsection{Neural network model}
\label{sec:ANN}

Artificial neural networks (ANN) are the building blocks of deep neural networks (DNN). The basic units of calculation in ANNs are called neurons, which are connected via weighted inputs that resemble synapses. These biologically inspired models have the proven capability of learning from data, which has accelerated the data-driven discovery revolution over the last decade~\cite{LeCun:Nature,miotto,IsmailFawaz2019,SCHMIDHUBER201585,2019NatRP6}. 

\begin{figure*}
\centering

\def\layersep{1.5cm}
\def\Ninput{4}
\def\Nneuron{5}
\def\Nhidden{2}
\def\Noutput{4}

\begin{tikzpicture}[
   shorten >=1pt,->,
   draw=black!50,
    node distance=\layersep,
    every pin edge/.style={<-,shorten <=1pt},
    neuron/.style={circle,fill=black!25,minimum size=17pt,inner sep=0pt},
    input neuron/.style={neuron, fill=green!50},
    output neuron/.style={neuron, fill=red!50},
    hidden neuron/.style={neuron, fill=blue!50},
    annot/.style={text width=4em, text centered}
]

    \foreach \name / \y in {1,...,\Ninput}
        \node[input neuron, pin=left:Input \#\y] (I-\name) at (0,-\y) {};

    \foreach \N in {1,...,\Nhidden} {
       \foreach \y in {1,...,\Nneuron} {
          \path[yshift=0.5cm]
              node[hidden neuron] (H\N-\y) at (\N*\layersep,-\y cm) {};
           }
    \node[annot,above of=H\N-1, node distance=1cm] (hl\N) {Hidden layer \N};
    }

	\foreach \name / \y in {1,...,\Noutput}{
	    	\node[output neuron,pin={[pin edge={->}]right:Output \#\y}] (O-\name) at ({(\Nhidden+1)*\layersep},-\y cm) {};
}

    \foreach \source in {1,...,\Ninput}
        \foreach \dest in {1,...,\Nneuron}
            \path (I-\source) edge (H1-\dest);

    \foreach [remember=\N as \lastN (initially 1)] \N in {2,...,\Nhidden}
       \foreach \source in {1,...,\Nneuron}
           \foreach \dest in {1,...,\Nneuron}
               \path (H\lastN-\source) edge (H\N-\dest);

    \foreach \source in {1,...,\Nneuron}
		\foreach \dest in {1,...,\Noutput}
	        \path (H\Nhidden-\source) edge (O-\dest);

    \node[annot,left of=hl1] {Input layer};
    \node[annot,right of=hl\Nhidden] {Output layer};

\end{tikzpicture}
 \caption{Schematic illustration of a neural network. A multilayer perceptron with two hidden layers is presented. Circles represent neurons, whereas arrows correspond to weights.} 
\label{fig:ann_diagram}
\end{figure*}
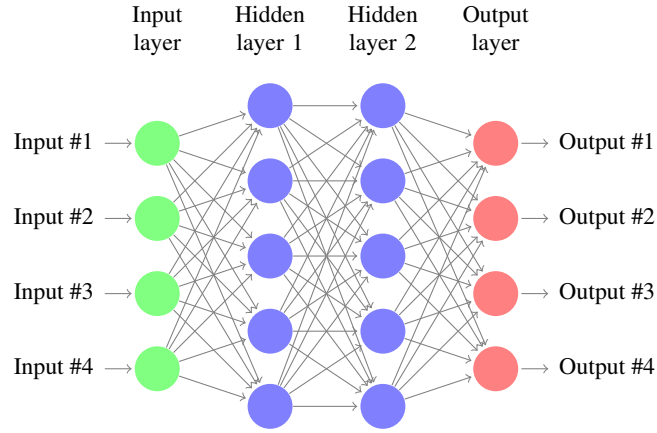

As shown in~\Cref{fig:ann_diagram}, a neural network creates a relationship between the inputs and outputs. This relation uses multiple layers of neurons connected through a series of linear or nonlinear functions.  The input layer takes the input data and applies these operations to calculate its outputs $X^{1}_i$ for each input $i$.  Then, each of the ANN's subsequent layers $l$ takes the outputs of the previous layer $X^{l-1}_j$ of layer $l-1$ and applied this same calculations to calculate the outputs $X^{l}_i$ of each of is neurons.  The calculation is performed as 

\begin{align}
X_{i}^{l}&=g\left(s_{i}^{l}+b_{i}^{l}\right)\,, \\
s_{i}^{l}&=\sum_{j} W_{i j}^{l} X_{j}^{l-1}\,,
\end{align}

\noindent where $g$ is a nonlinear function known as an activation function, and the parameters to be tuned during training are the weights, $W_{ij}^{l}$, and biases, $b^{l}_{i}$. The values of $W_{ij}^{l}$ and $b^{l}_{i}$ are continually adjusted during the training stage until training data with the same labels consistently yield similar results in the output layer $X^{L}_{i}$. In our case, the output of the neural network model corresponds to the SGS tensor components. For the activation function of the hidden layers, we selected the rectified linear unit (ReLU), which is common in machine learning for its fast training speed.  The ReLU is defined as $g(x)=\max(0,x)$.  For the output layer, we used a linear activation function, defined simply as $g(x)=x$.

Most ANN models of turbulence use a multilayer perception (MLP) network~\cite{Maulik2017,Wang2018,Xie2019} or some slight variation of an MLP~\cite{Ling2016,Fang2018}.  In this work, we also employ an MLP network to implement our model.  The network acts on individual grid cells.  The network configuration used in this work had an input layer with $N_I$ inputs, a hidden layer with 64 neurons followed by another hidden layer with 32 neurons, and finally output layer with $N_O$ outputs.

There is some variation in the literature in selecting the input features for ANN models of hydrodynamical turbulence~\cite{Ling2016,Maulik2017,Wang2018,Fang2018,Xie2019,Xie2019b,Pawar2020}.  The inputs for ANN model $\tau_{ANN}$ were all quantities defined for the SGS tensors in \Crefrange{eq:tau_kin}{eq:tau_enth}, the first and second derivatives of those quantities, and the value of all aforementioned terms in cells adjacent to the cell of interest.  All derivatives were computed using 4th order centered finite differencing.  For the mixed filtered quantities $\tau_{ind}$ and $\tau_{enth}$, we add the mass density $\rho$ to our collection of variables that we include in the inputs in the same manner described above.  The inputs to each ANN are explicitly given in \Cref{app:ann_inputs}.

In our case, the outputs are all unique components of the desired SGS tensor which vary depending on the tensor of interest.  Thus, we have $N_O=3$ for $\tau_{kin}$ and $\tau_{mag}$, $N_O=1$ for $\tau_{ind}$, and $N_O=2$ for $\tau_{enth}$  This differs from most of the literature where a different ANN is used to find each individual component of the SGS tensor~\cite{Maulik2017,Wang2018,Xie2019,Xie2019b}.  By computing all components of the SGS tensor, we hope to incorporate physical symmetries and constraints into future models of $\tau_{ANN}$ such as Galilean invariance, though we do not attempt to do so in this work. 

For reference, we have chosen mean-squared error (MSE) as the loss function to optimize the performance of our neural network model. We describe in detail the high resolution simulations of the magnetized KHI used to train and test our models in~\Cref{sec:simulation}.  The hyperparameters of our neural network model are presented in~\Cref{sec:training}.

\begin{figure}[h]
\centering
\includegraphics[width=0.33\linewidth]{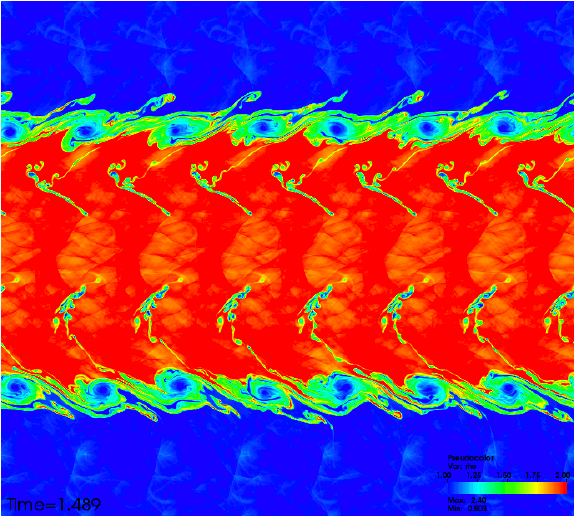}
\includegraphics[width=0.33\linewidth]{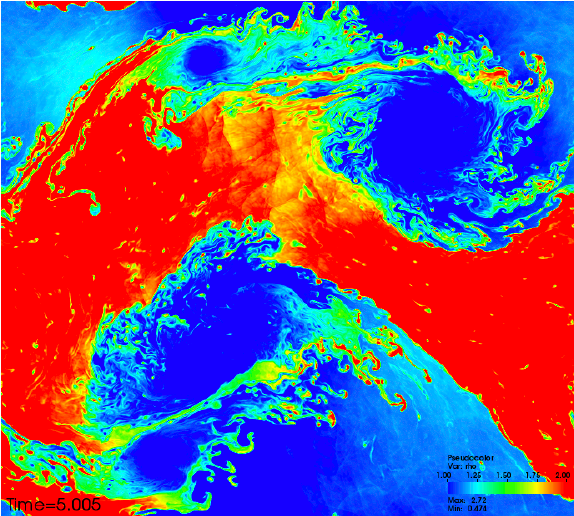}
\includegraphics[width=0.33\linewidth]{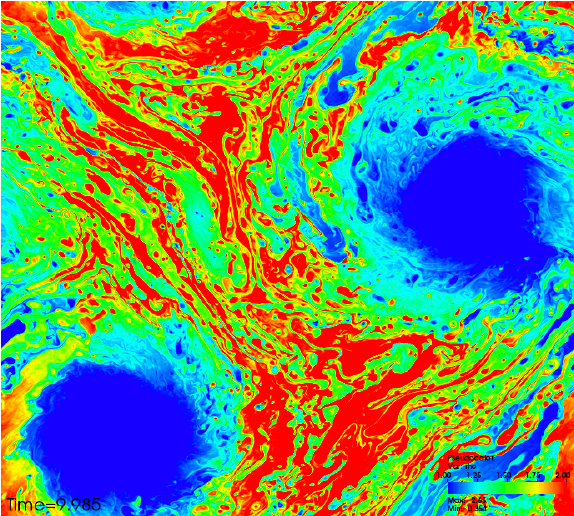}
\caption{Plots of the mass density distribution $\rho$ time slices at $t=1.5,5,10$ of our magnetized KHI simulations with a resolution of $N=2048^2$.  On the left, we have the $t=1.5$ time slice in which we can observe vortexes begin to form between the two fluids.  The number and size of these vortexes are controlled by an initial sinusoidal perturbation of the fluid velocity in the $y$-direction.  The center plot at $t=5$ occurs after many of the aforementioned vortexes have merged together at which point the flow has become unsteady.  This unsteady flow manifests itself in the high density fluid beginning to break apart.  The rightmost plot at $t=10$ depicts the fluids in a turbulent mixing process with two low density vortexes helping to drive this mixing.}
\label{fig:simulation}
\end{figure}

\section{Simulation}
\label{sec:simulation}

To train and evaluate the model, we ran 2-D magnetized KHI simulations.  As described above, the KHI instability occurs when two fluids are moving in opposite direction.  When magnetic fields are included, the instability accelerates and the magnetic fields are amplified throughout the process. The KHI was selected because BNS mergers, the targeted application of this work, experience a KHI-like process during the merger phase.

\begin{figure}[h]
\centering
\includegraphics[width=0.33\linewidth]{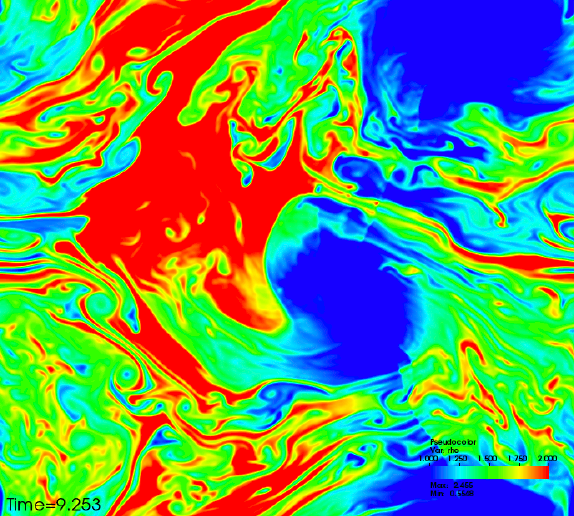}
\includegraphics[width=0.33\linewidth]{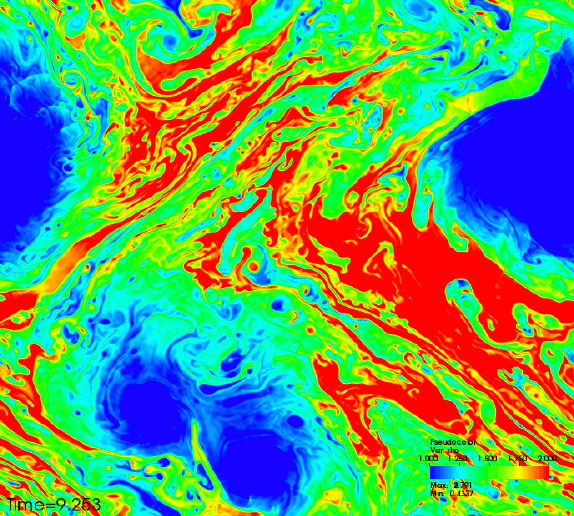}
\includegraphics[width=0.33\linewidth]{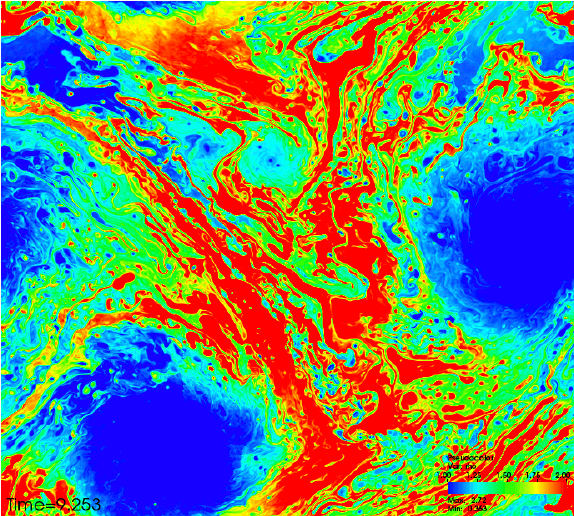}
\caption{Plots of the density distribution $\rho$ of the test dataset at time slices $t=9.25$ for resolutions $N=512^2,1048^2,2048^2$ in the left, middle, and right image respectively.  All three datasets appear to be undergoing a turbulent mixing process at $t=9.25$.  We observe that while the $N=1024^2$ and $N=2048^2$ runs appear to share many of the same general characteristics, the $N=512^2$ run fails to reproduce these same feature.  This failure implies that $N=512^2$ is not enough to capture the turbulent effects of the magnetized KHI without a SGS model.  We note that the testing time slice is of particular importance because it was used to evaluate the SGS models.}
\label{fig:simulation_test}
\end{figure}

The simulations were run using the open-source \textit{Simflowny} code~\cite{simflowny1,simflowny2}. For these simulations, the grid was a Cartesian square with $x,y \in \left[-L/2,L/2\right]$, with length $L=1$. These simulations were performed at three grid sizes with the number of points $N=512^2$, $N=1024^2$, and $N=2048^2$ for the low, medium, and high resolutions, respectively.  The boundary conditions were chosen to be periodic in all directions.  We evolved the equations for 10 units of time.  Using a RK4 time integration scheme, we evolved the MHD equations in \Crefrange{eq:continuity}{eq:energy} with timestep of $\Delta t = \frac{0.25}{\sqrt{N}}$.  We show density plots of these simulations in \Cref{fig:simulation,fig:simulation_test}.

To assist in triggering the instability, we add velocity perturbations to the system in both coordinate directions.  The specific setup for the initial conditions for the grid functions in this simulation is given by

\begin{gather}
\rho=\rho_{0}+\sgn(y) \left[\delta\rho \tanh\left( \frac{\abs{y}-y_l}{a_l} \right)\right]\,, \\
v_{x}= \sgn(y) \left[v_{x0} \tanh\left( \frac{\abs{y}-y_l}{a_l} \right)\right]+\delta v_{x} \sin(2\pi n_x y) \\
v_{y}=\sgn(y) \left\{\delta v_{y} \sin(2\pi n_y x) \exp\left[{-\left(\frac{\abs{y}-y_{l}}{\sigma}\right)^2}\right]\right\}\,, \\
B_{x}=B_{x0}\,, \\
B_{y}=B_{y0}\,, \\
p=p_0\,.
\end{gather}

\noindent In the above expressions, $\rho_{0}=1.5$ and $\delta \rho=-0.5$ are the average and difference of the low density region $\rho_1=1$ and high density region $\rho_2=2$ respectively.  $y_l=0.25$ is the $y$-coordinate where the transition from $\rho_1$ to $\rho_2$ occurs.  $a_l=0.01$ is the characteristic size of this transition region, providing a smooth transition that mitigates some of the numerical instabilities of the transition between the different density regions.  $v_{x0}=0.5$ is the initial velocity of the fluid in the $x$ direction.  $\delta v_{x}=0.01$ is a sinusoidal perturbation of $v_{x0}$ with $n_x=4$ periods going along the $y$ direction.  $\delta v_{y}=0.2$ is a sinusoidal perturbation of the $y$ component of the velocity with $n_y=7$ periods along the $x$ direction.  $\sigma=0.1$ is the characteristic Gaussian falloff of $\delta v_y$ away from $y_l$.  We note that for $|y|>0.45$, $\delta v_y$ is set to $0$.  $B_{x0}=0.001$, $B_{y0}=0$, and $p_0=1$ are the initial $x$-component of the magnetic field, initial $y$-component of the magnetic field, and initial pressure respectively.

Like \cite{Vigano2019a}, we desired to evolve with similar numerical methods to those used in numerical relativity simulations of BNS mergers.  We employed the Method of Lines (MoL) to discretize our system of equations.  We used a finite volume scheme with MP5 reconstruction and Local Lax Friedrichs (LLF) flux splitting for the evolution of our system, which provides numerical stability even in the presence of shocks.  This scheme views the ideal MHD equations in \Crefrange{eq:continuity}{eq:energy} as 
\begin{align}
\partial_{t} \boldsymbol{U}+\partial_{i} \boldsymbol{F}^i=\boldsymbol{S}\,,
\end{align}
where $\boldsymbol{U}=\left\{\rho,\rho v^j,B^j,u\right\}$ are our conserved quantities, $\boldsymbol{F}$ are the fluxes for those conserved fields, and $\boldsymbol{S}=\left\{0,0,0,0\right\}$ are the source terms.  The source term is set to zero in our case, but is nonzero in general if say an external force like gravity is applied to the fluid. $\boldsymbol{F}$ is allowed to depend on the conserved variables, but not on their derivatives.  The SGS tensors, which depend on derivatives of the conserved variables, would be placed in $\boldsymbol{S}$ rather than $\boldsymbol{F}$ when implementing one of the aforementioned SGS models in a simulation.

To preserve the divergence free condition on the magnetic field, we used a hyperbolic divergence cleaning \cite{Dedner2002}.  This divergence cleaning adds another evolution equation to our system for $\phi$ to ensure the magnetic field divergence decays to 0 and is defined as 
\begin{align}
\partial_{t}\phi + c_h^2 \partial_i B^i &= -\frac{c_h}{c_r}\phi
\end{align}
\noindent where $c_h=1$ and $c_r=0.18$.

\section{Training}
\label{sec:training}

The KHI simulation data was filtered using a box filter with filter sizes $f=2,4,8,16$ where $f$ is defined as $f=\frac{\Delta_f}{\Delta}$. For each of the filter sizes and resolutions, SGS tensors and inputs to the ANNs were calculated after $t=1$ every $\sim{0.1}$ time units until the simulation ended at $t=10$\footnote{Due to memory consumption issues, we used less data to train the $N=1024^2$ $f=2$ and $N=2048^2$ $f=2,4$ models. Specifically, the $N=1024^2$ $f=2$ and $N=2048^2$ $f=4$ models sampled training data every $\sim{0.5}$ time units.  The $N=2048^2$ $f=2$ model sampled training data every $\sim{1}$ time units.}.  
The test data evaluated these same quantities at $t \approx 9.25$.  This approach ensures that even models that are trained with low resolution simulations are exposed to data with sufficient size and variety.  We found that this approach prevents overfitting. In 3D, we expect to use fewer time slices as each time slice contains significantly more samples than in 2D. Another observation is that we experimented with data augmentation methods, as those described in~\cite{Maulik2017} which consist of augmenting the data by providing multiple copies of each time-slice, but choosing a different point after filtering~\cite{Maulik2017}. However, we found that this approach does not generalize well during testing. To address that problem, we chose multiple time-slices during training. 

\begin{center}
\begin{figure}[h]
\centering
\includegraphics[width=0.49\linewidth]{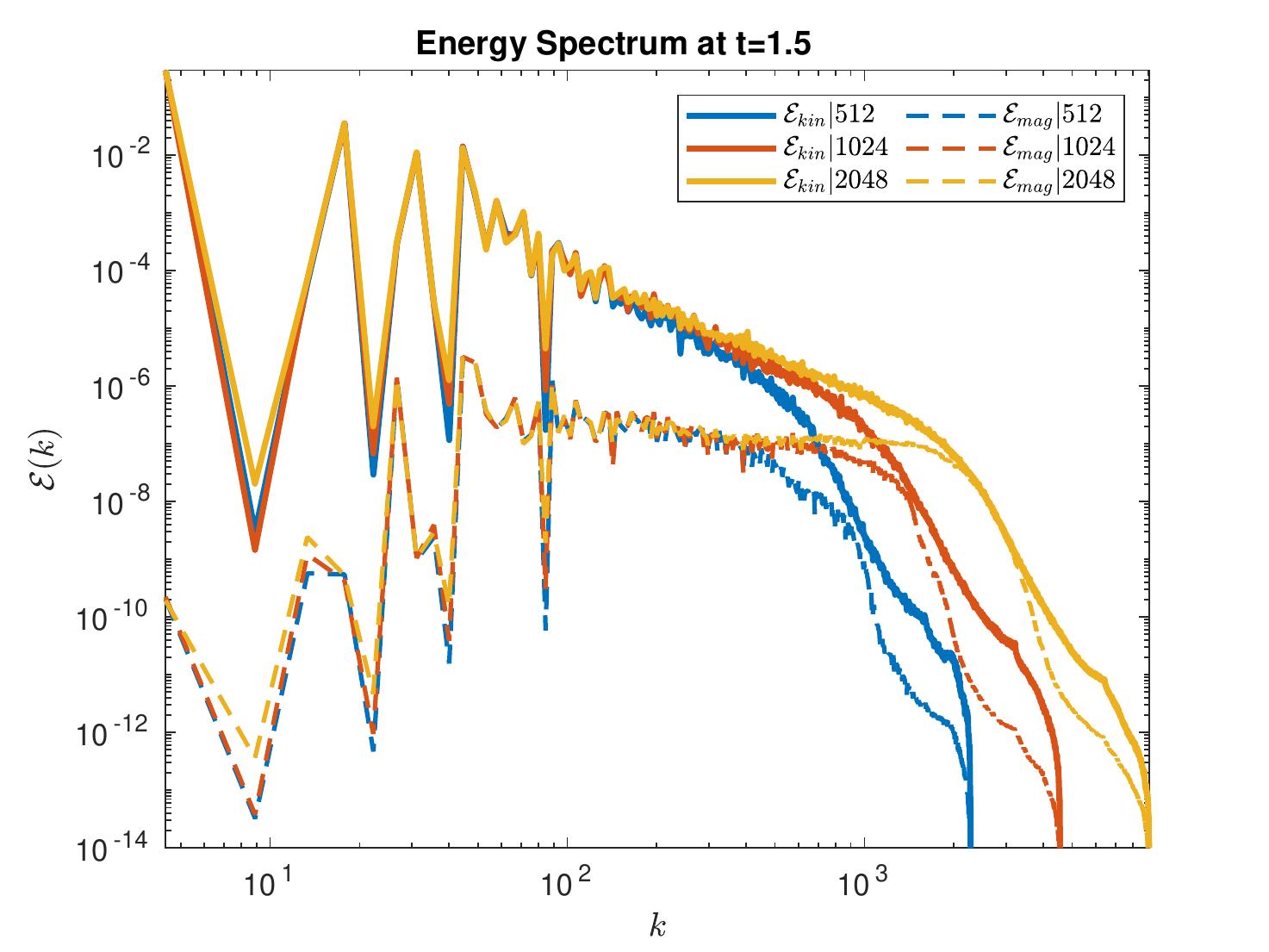}
\includegraphics[width=0.49\linewidth]{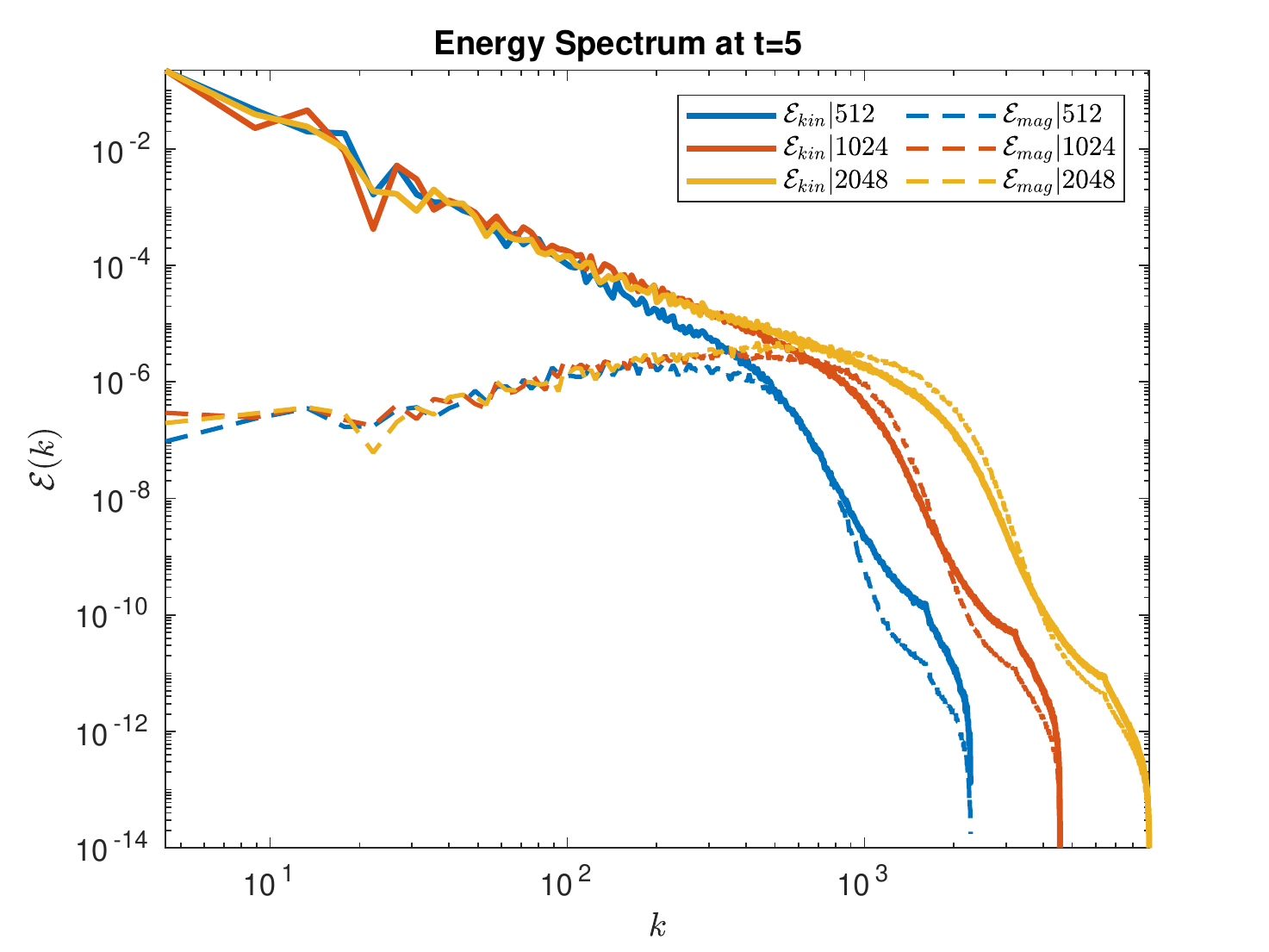}
\includegraphics[width=0.49\linewidth]{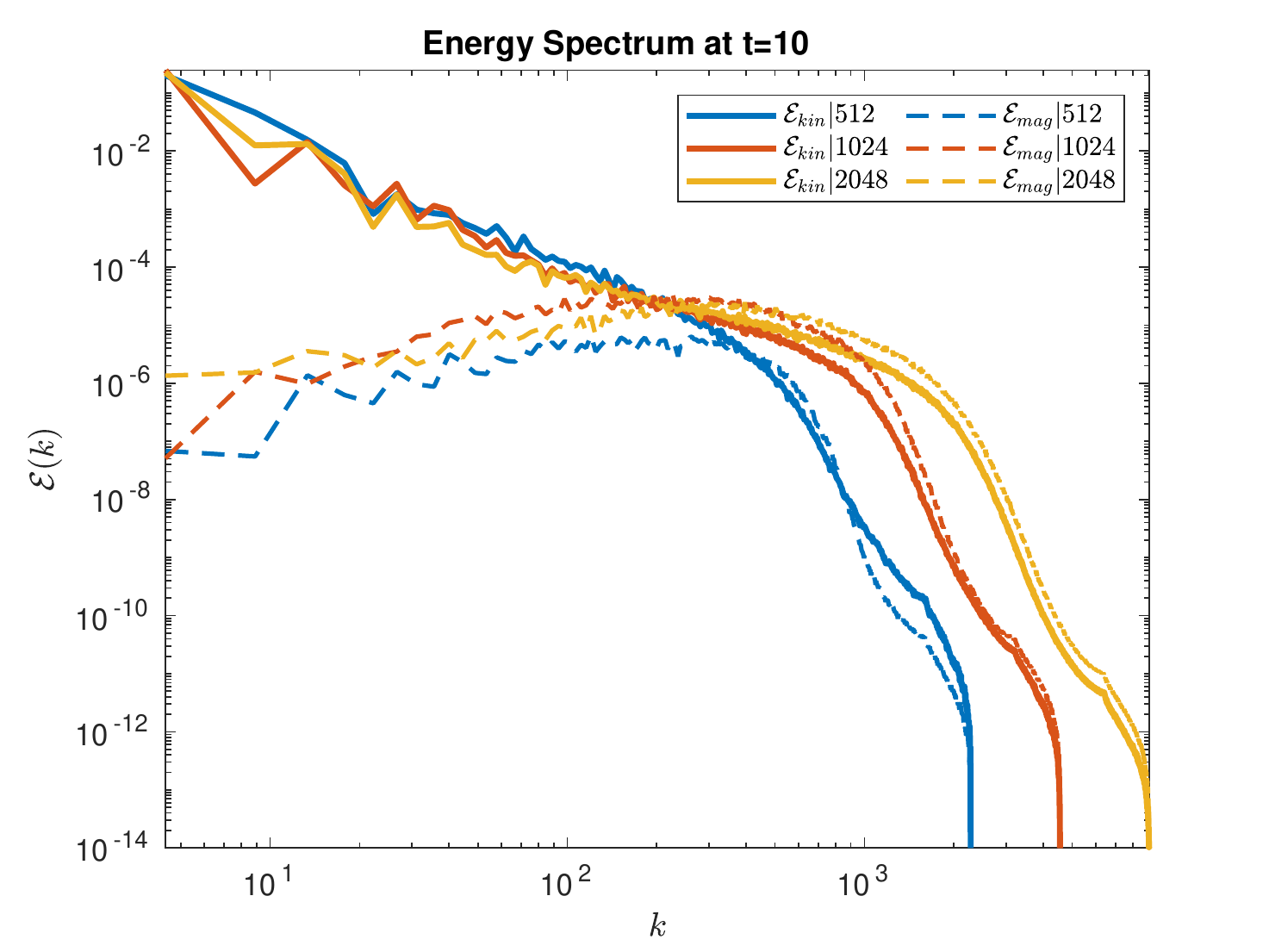}
\includegraphics[width=0.49\linewidth]{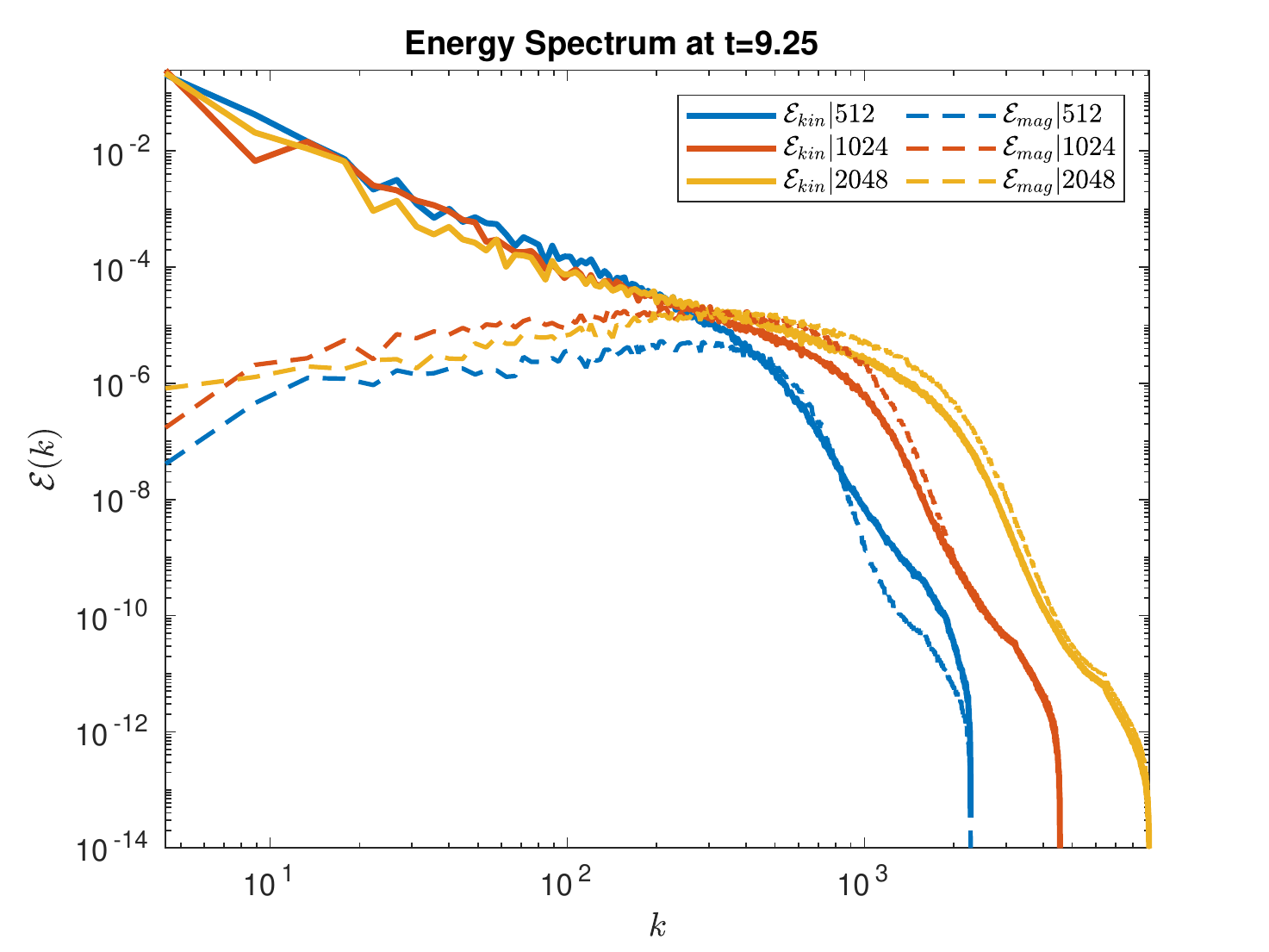}
\caption{Plots of the energy spectra $\mathcal{E}(k)$ at various time steps throughout the simulation for each resolution.  The kinetic energy spectra $\Ekin(k)$ is given by the solid lines, while the magnetic energy spectra $\Emag(k)$ is denoted by the dashed lines.  The resolutions $N=512^2,1024^2,2048^2$ are given by the blue, red, and green lines respectively.  The top left, top right, and bottom left images are taken at approximately $t=1.5,5,10$ respectively and correspond to the timesteps displayed of the density distribution plots in \Cref{fig:simulation}. The bottom right plot provides the spectra of the test dataset used for evaluating the models whose density distribution is featured in \Cref{fig:simulation_test}.}
\label{fig:spectra}
\end{figure} \end{center}

After calculating the SGS tensors and the necessary inputs to the ANN models, we exported the data to train the model in \texttt{TensorFlow}~\cite{tensorflow2015-whitepaper}. The data was normalized to have zero mean and unit standard deviation.  We used 10\% of the simulated data for validation purposes. For the training of the neural network model, we used an ADAM optimizer with early stopping \cite{Kingma2014}.  The maximum number of epochs was 100.  A batch size of 1000 was used during training.

\section{Methodology}
\label{sec:methodology}

In this section we describe quantities that we will use to test our neural network model, and metrics to assess its ability to correctly reproduce true features and properties of the testing data set.

\subsection{Spectra Calculation}
\label{sec:spectra_calc}

The energy spectrum $\mathcal{E}(k)$ represents the spatial scale at which the energy is distributed in a given process.  For low wave number $k$, we see the large scale features of the energy spectrum.  On the other hand, high $k$ values give the small scale features of the spectrum.  The ultimate goal of the large eddy simulation is to reproduce the energy spectrum of the DNS simulations as closely as possible. Appendix~\ref{app:specta_calc} describes how to compute these quantities.

In MHD turbulence, we are concerned about the energy spectra of the kinematic motion $\mathcal{E}_{kin}(k)$ and the magnetic field $\mathcal{E}_{mag}(k)$.  We note that these energy spectra have a different expected distribution.  The kinetic energy spectrum falls of as $\mathcal{E}_{kin}(k)\propto k^{-5/3}$ at high wave numbers.  However, the magnetic energy spectrum rises as $\mathcal{E}_{mag}(k)\propto k^{3/2}$ at large $k$ values \cite{Vigano2019a}.  Thus, we expect the small scale behavior will be especially significant in the overall magnetic energy contribution and must be modeled carefully.

Moreover, we are interested in the total energy obtained by integrating over all the spectra.  By examining how the total energy changes over time, we can extract useful information about characteristics of the simulation.  In particular we would like to measure how the the kinetic energy $E_{kin}$ and magnetic energy $E_{mag}$ change through the effect of the KHI.

\begin{center}
\begin{figure}[h]
\centering
\includegraphics[width=0.49\linewidth]{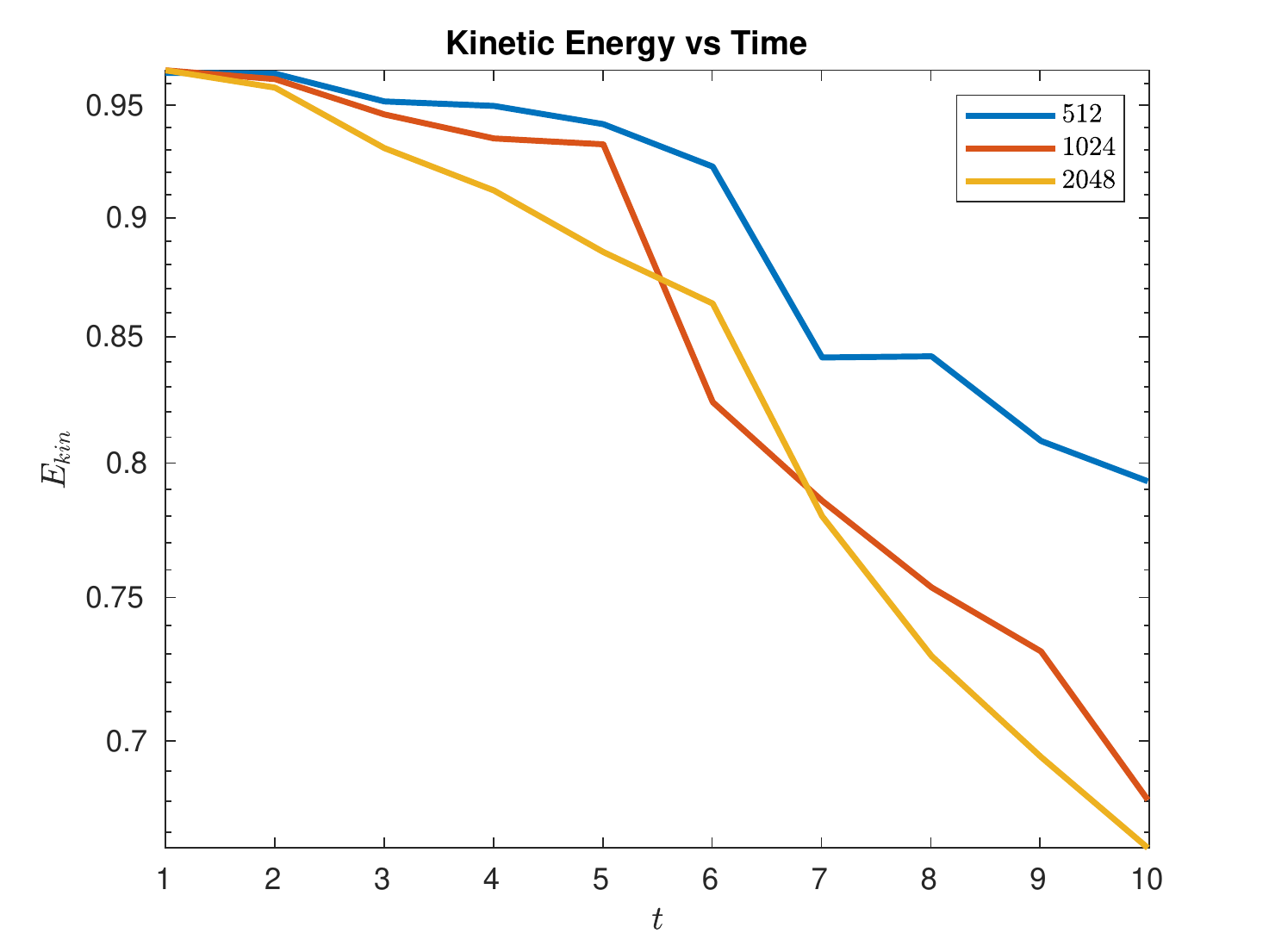}
\includegraphics[width=0.49\linewidth]{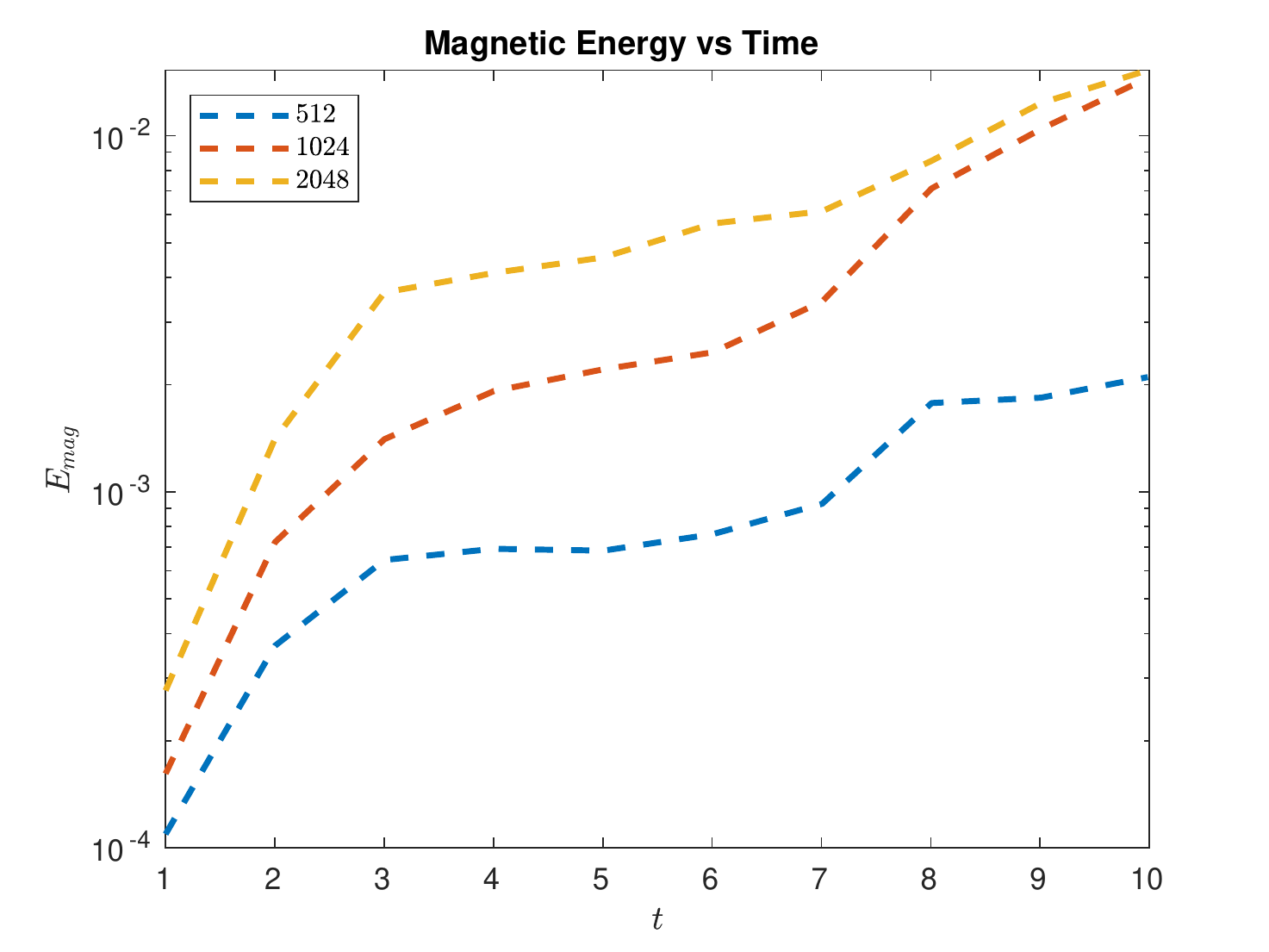}
\caption{Time evolution of the total kinetic energy, $E_{kin}$, (left panel) and total magnetic energy, $E_{mag}$, (right panel).  The kinetic energy decreases over time as it is being converted into magnetic energy.  This energy conversion can be seen in the right panel where $E_{mag}$ increases over time.  We notice that $E_{kin}$ decreases more rapidly for the high resolution runs while $E_{mag}$ increases more rapidly.  This would indicate that this conversion from $E_{kin}$ to $E_{mag}$ occurs most efficiently at small scales that high resolutions simulations can best resolve.}
\label{fig:energy}
\end{figure}

 \end{center}

\subsection{Model Performance Criteria}
\label{sec:performance_criteria}

To quantify the performance of our models, we will use several common turbulence statistics.  The first of these statistics is the correlation coefficient $C$ which shows how well the data and the model follow one another.  We define $C$ as
\begin{gather}
C=\frac{\left\langle\left(\tau_{DNS}-\left\langle\tau_{DNS}\right\rangle\right)\left(\tau_{model}-\left\langle\tau_{model}\right\rangle\right)\right\rangle}{\sqrt{\left\langle\left(\tau_{DNS}-\left\langle\tau_{DNS}\right\rangle\right)^{2}\right\rangle\left\langle\left(\tau_{model}-\left\langle\tau_{model}\right\rangle\right)^{2}\right\rangle}}\,,
\label{eq:corr}
\end{gather}
where $\tau_{DNS}$ is  the SGS tensor computed from filtering the high resolution data, $\tau_{model}$ is the SGS tensor computed from the SGS model we are testing, and $\left\langle x \right\rangle$ is the volumetric average of the quantity $x$.  $C$ can range from $-1$ to $1$ with values near to $-1$ being anti-correlated, values near to $0$ being uncorrelated, and values near to $1$ being well correlated.  Simply put, the closer $C$ is to $1$, the better the model.  We use $C$ as our primary measure of performance for our models. We will also look at the relative error between the model and the simulation denoted by $E$.  $E$ is defined as 
\begin{gather}
E=\frac{\sqrt{\left\langle\left(\tau_{DNS}-\tau_{model}\right)^{2}\right\rangle}}{\sqrt{\left\langle \tau_{DNS}^{2}\right\rangle}}\,,
\label{eq:err}
\end{gather}
with all quantities defined in the same manner \Cref{eq:corr}.  We note that the lower the value of $E$ is for a model, the better the model. The root-mean-square ($RMS$) of the model tells us the degree to which the model deviates from the average.  The $RMS$ of a quantity $x$ is given by 
\begin{gather}
RMS(x)=\sqrt{\left\langle(x-\left\langle x \right\rangle)^{2}\right\rangle}\,.
\label{eq:rms}
\end{gather}
Here, we will calculate the $RMS$ for $\tau_{model}$ and $\tau_{DNS}$.  The goal here is for the $RMS$ of $\tau_{model}$ is to be as close to $RMS$ of $\tau_{DNS}$ as possible.  In addition, we would like to use the absolute value of $RMS$ of $\tau_{DNS}$ to tell us more about the features of $\tau$ for the various models, resolutions, and filter sizes.

\begin{center}
\begin{figure}[h]
\centering
\includegraphics[width=0.33\linewidth]{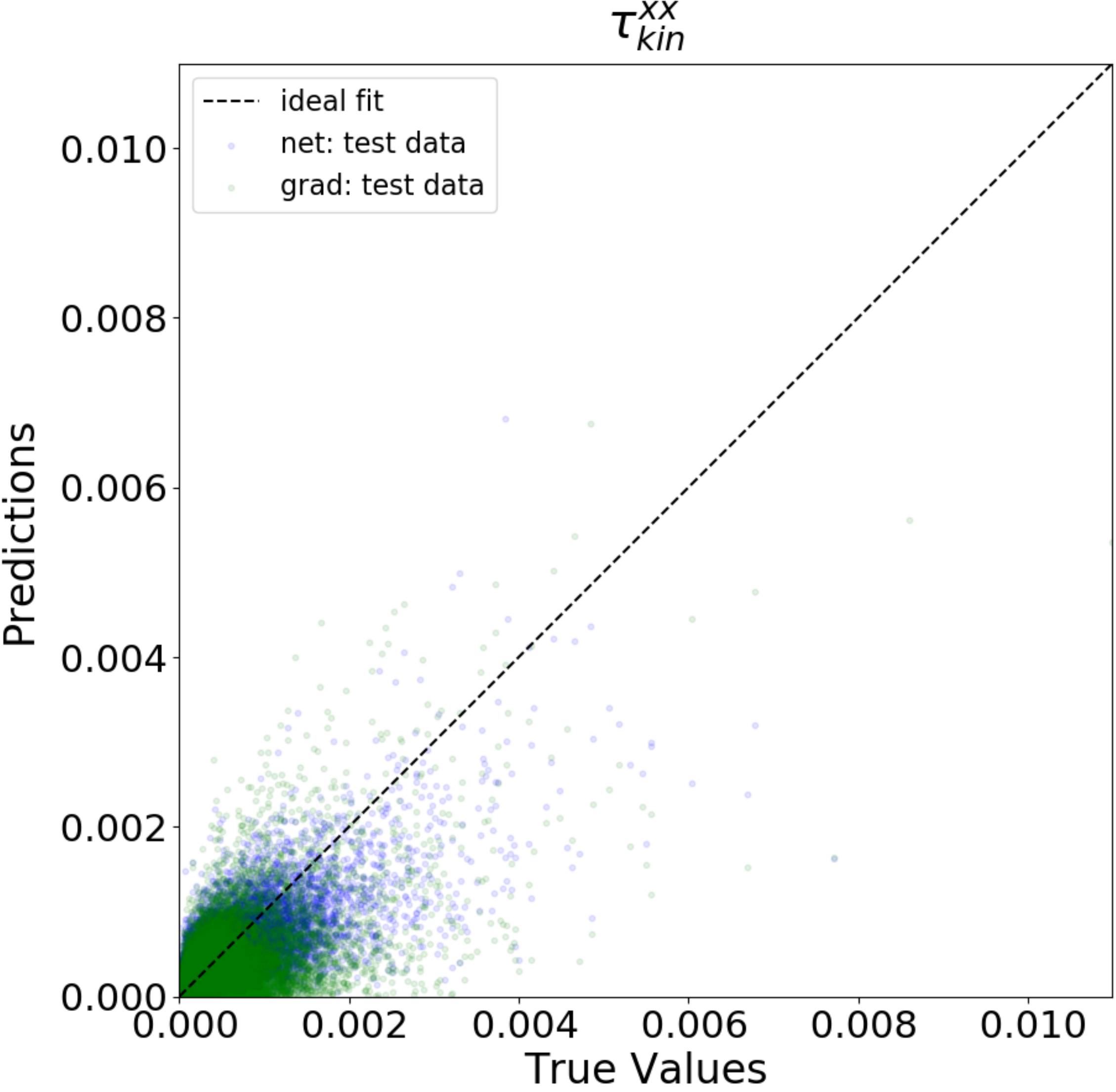}
\includegraphics[width=0.33\linewidth]{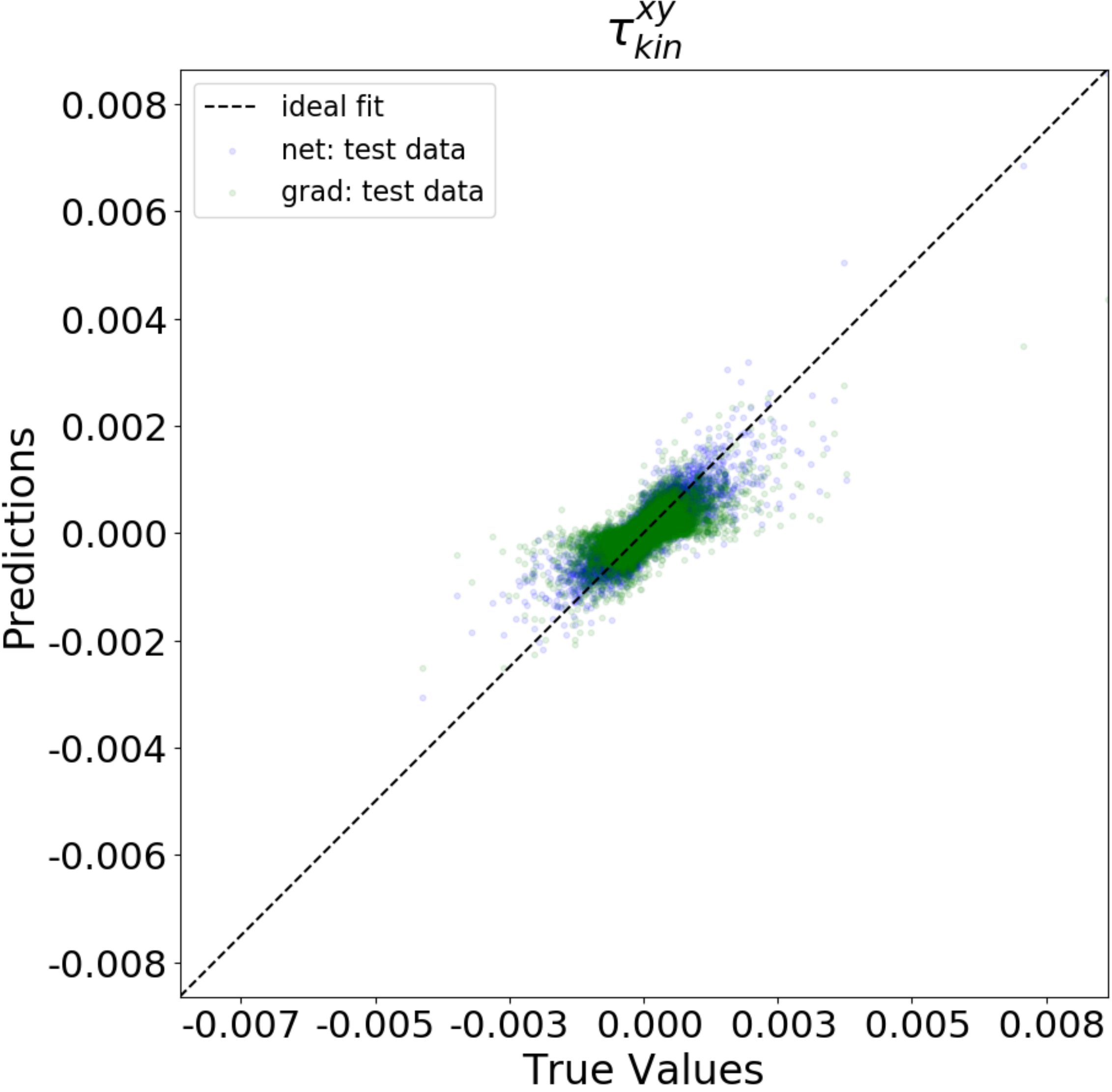}
\includegraphics[width=0.33\linewidth]{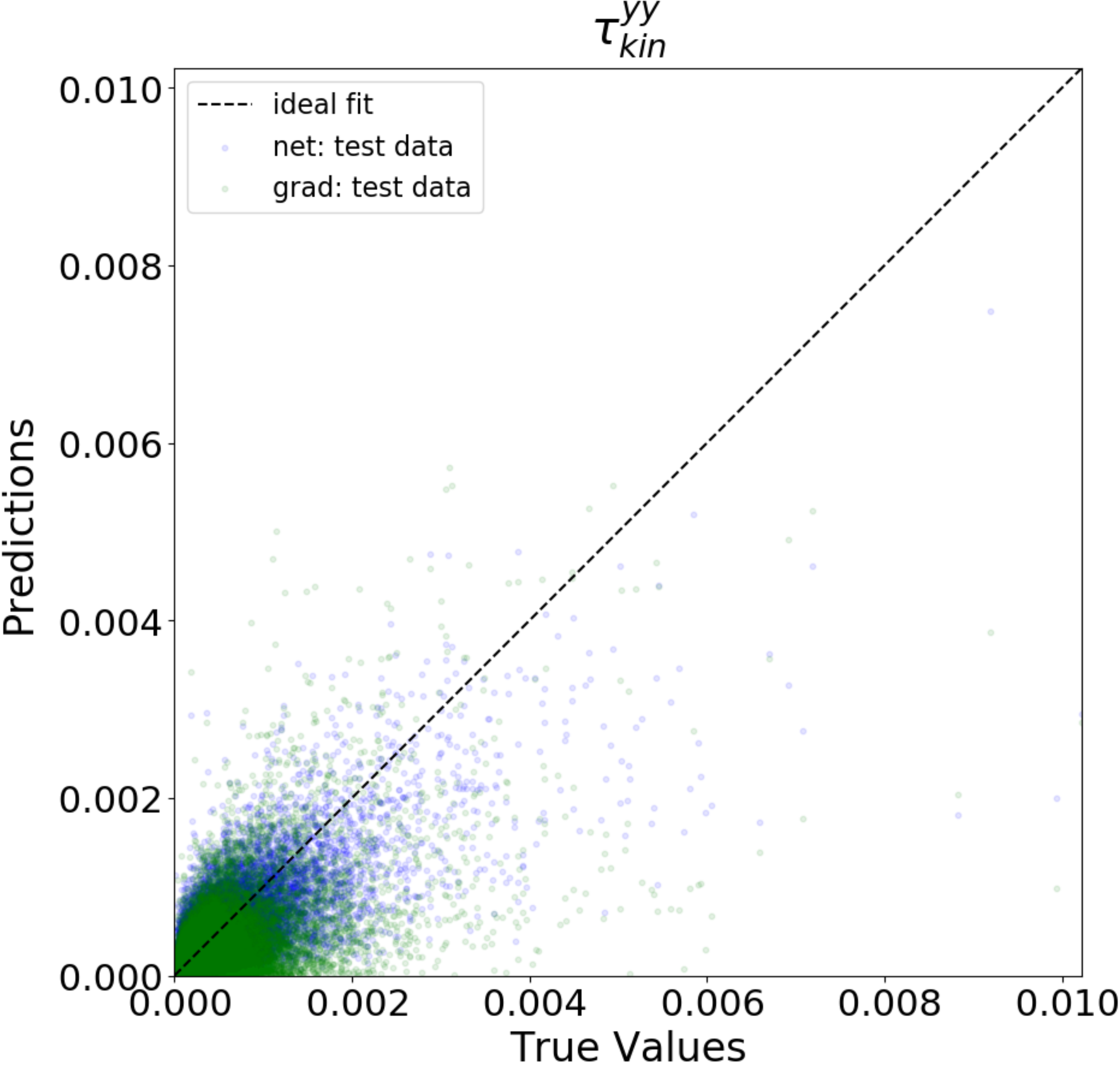}

\includegraphics[width=0.33\linewidth]{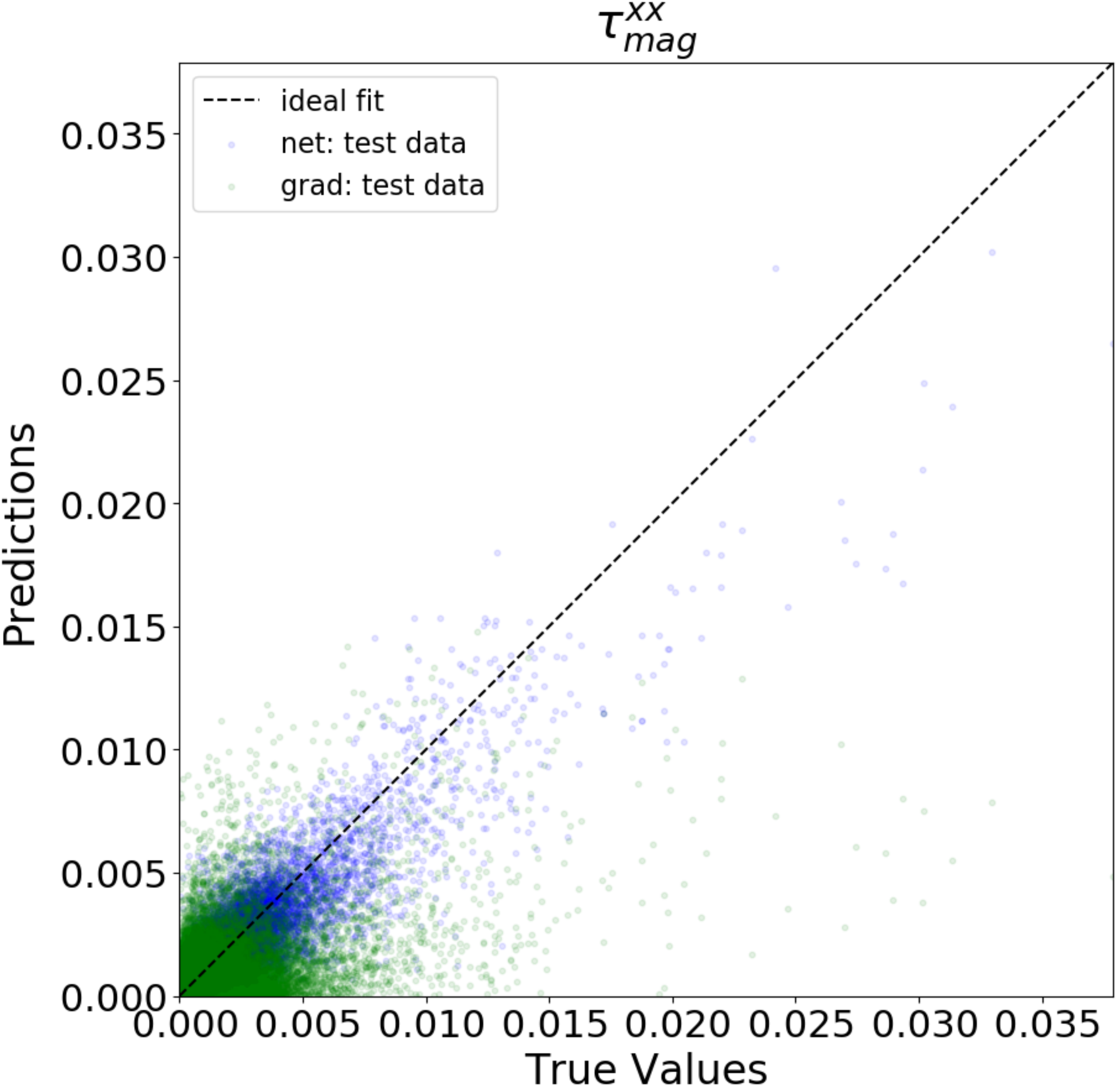}
\includegraphics[width=0.33\linewidth]{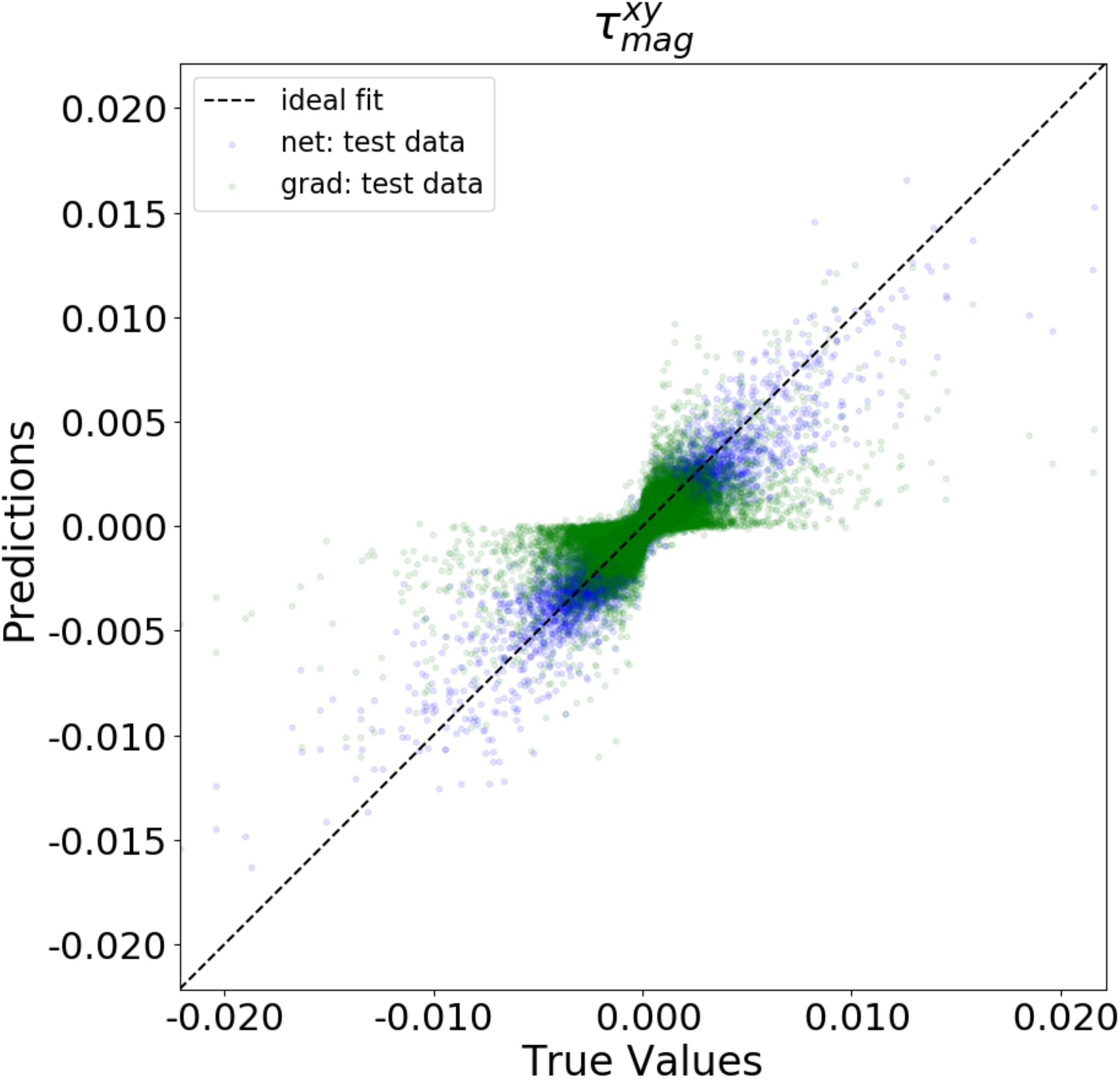}
\includegraphics[width=0.33\linewidth]{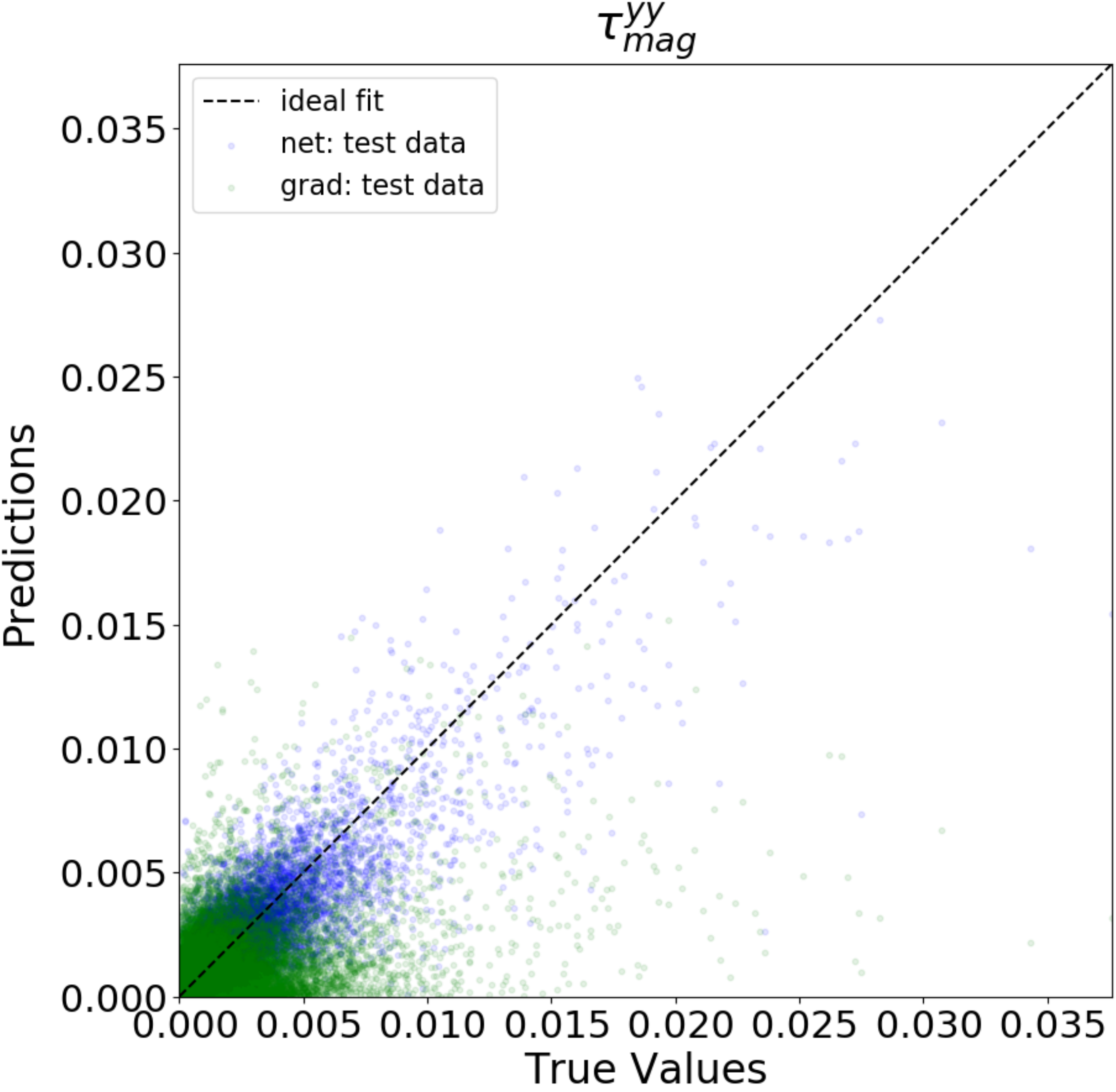}

\includegraphics[width=0.33\linewidth]{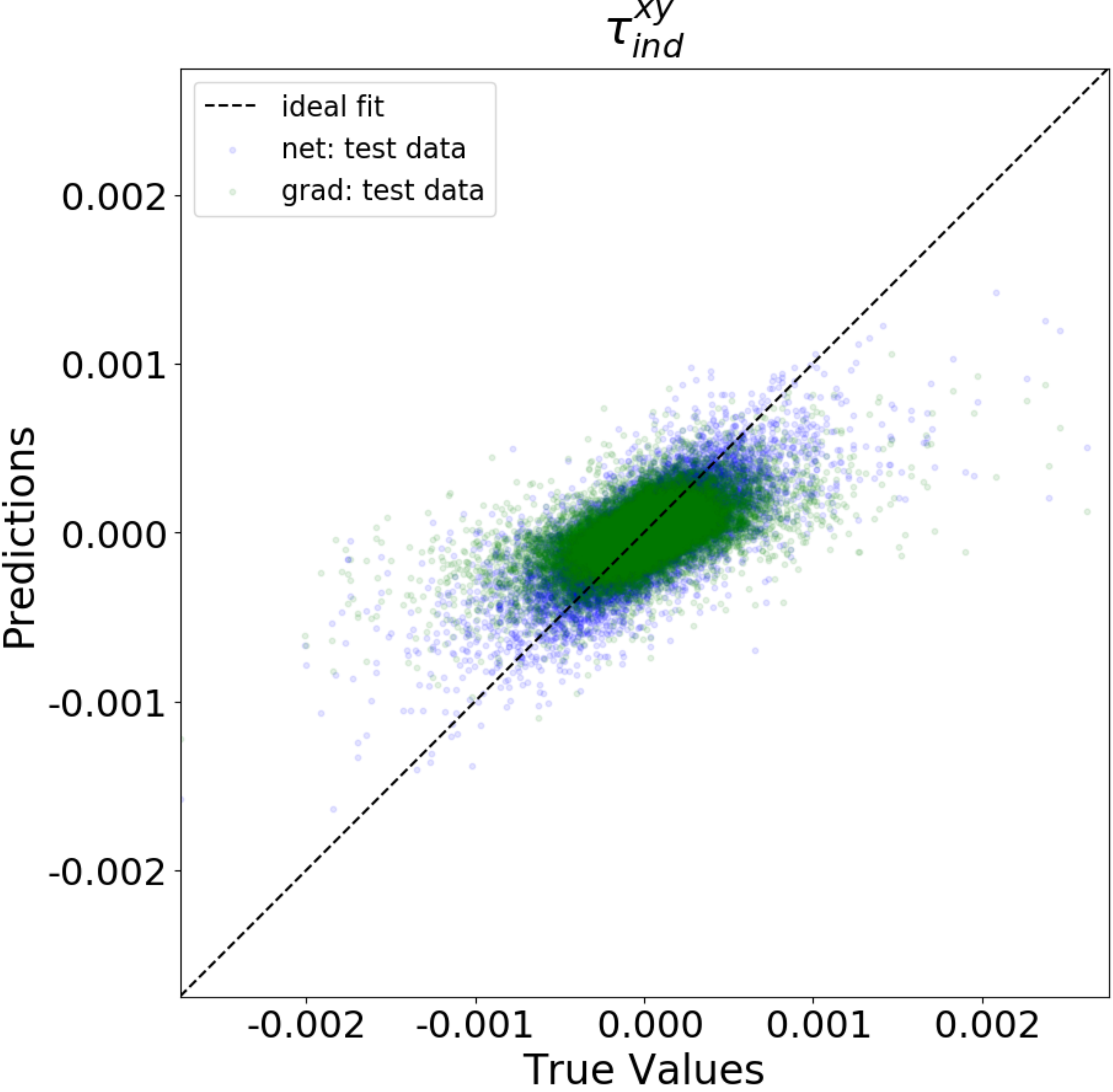}
\includegraphics[width=0.33\linewidth]{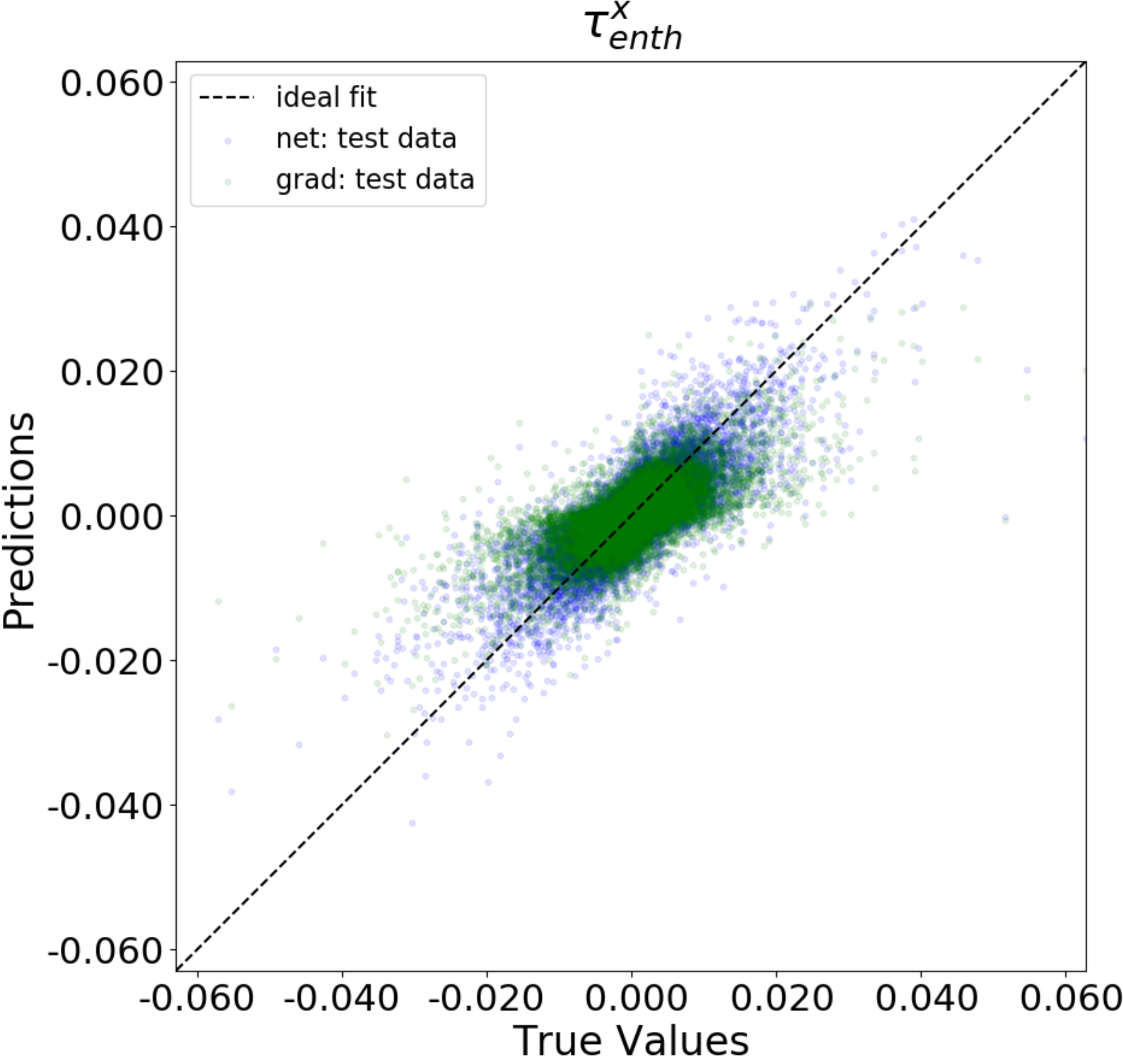}
\includegraphics[width=0.33\linewidth]{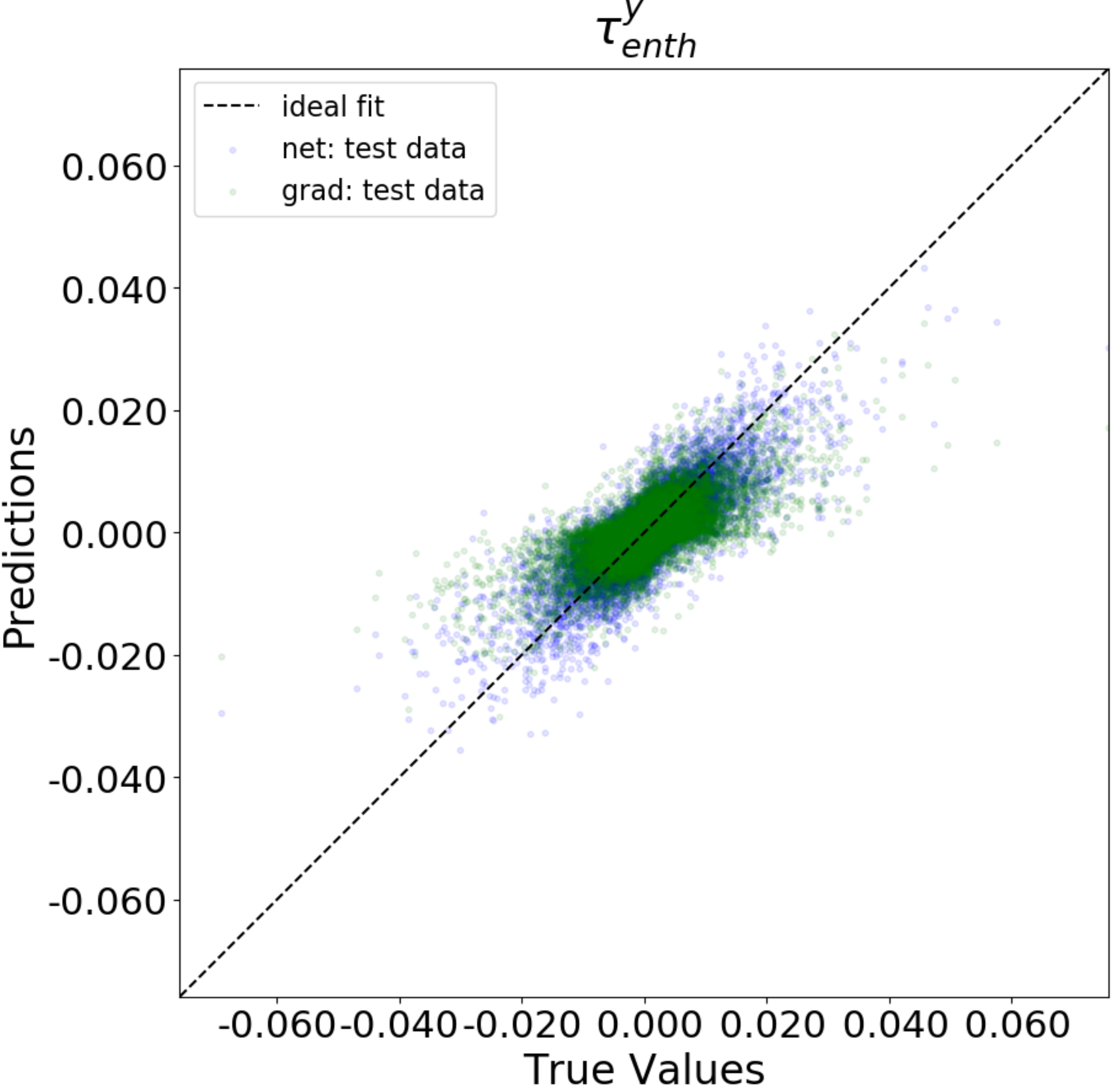}
\caption{Predictions vs target values of SGS tensors for the $N=2048^2$ resolution test dataset with filter size $f=8$ for each of our SGS models.  The small transparent green circles represent values of the gradient model and are overlaid on top of the small transparent blue circles which represent the values of the ANN model.  The black dashed line depicts a perfect one to one matching between the targets and the predictions.  The first row displays the $xx$, $xy$, and $yy$ components of $\tau_{kin}$ from left to right respectively.  The middle row shows the same information for the components of $\tau_{mag}$ SGS tensor.  The bottom row shows the $xy$ component of $\tau_{ind}$ on the left, the $x$ component of $\tau_{enth}$ in the middle, and the $y$ component of $\tau_{enth}$ on the right.  We observe that the blue circles of the ANN model appear much closer to the dashed black line for high SGS tensor values than the green circles of the gradient model especially in the $\tau_{mag}$ and $\tau_{ind}$ SGS tensors.}
\label{fig:targets_vs_predictions}
\end{figure}
 \end{center}

\section{Results}
\label{sec:results}

In this section we present results of several tests we conducted to assess the reliability of our neural network model to accurately capture the physics of our testing data sets.

\subsection{Spectra}
\label{sec:spectra}

We will begin the discussion of the results by analyzing the spectra of the simulations.  The first three images in Fig.~\ref{fig:spectra} illustrates the spectra at the time slices of the simulation that were featured in the density plots of Fig.~\ref{fig:simulation} with all simulation resolutions included.  These selected time slices occur at approximately $t=1.5,5,10$.  The last plot of Fig.~\ref{fig:spectra} depicts the spectrum of the test dataset whose density distribution can be seen in Fig.~\ref{fig:simulation_test}.

Fig.~\ref{fig:energy} includes both $\Ekin$ and $\Emag$. We observe that the $\Ekin$ of the plots is fairly similar at low $k$ values.  The obvious exception to this is the $t=1.5$ plot where the low $k$ spectrum appears to still be settling down for both energy types, though this effect does not appear to be resolution dependent. We also notice that the $2048^2$ resolution simulation has reduced values of $\Ekin$ at low $k$ compared to the other simulations at later times, likely due to the kinetic energy being converted into magnetic energy more efficiently at high resolutions.  At high $k$ values, we observe a faster $\Ekin$ falloff at low resolution.  This drop off is likely due to the effect of the finite grid resolution on the small scale features.

The magnetic field spectra at low $k$ is significantly smaller than its kinetic energy counterpart.  As $k$ increases, the magnetic field spectra increases, it may eventually surpass the kinetic energy spectra before decaying.  It appears that much of this decay is an effect of the finite grid resolution.  At later times, all $\Emag$ spectra increase considerably.  The high resolution simulations have noticeably greater $\Emag$ than those at lower resolutions.  This effect is likely caused by the conversion of kinetic energy to magnetic energy being more efficient at high resolutions.

The last plot of Fig.~\ref{fig:spectra} presents the spectra of our testing dataset.  We observe that this plot shares simulate characteristics to the $t=10$ spectra plot.  However, we note that the high $k$ region of the $N=1024^2$ simulation's $\Emag$ spectra is weaker relative to the $N=2048$ simulation's $\Emag$ spectra than in the $t=10$ spectra plot.
\begin{figure}
\centering
\includegraphics[width=0.33\linewidth]{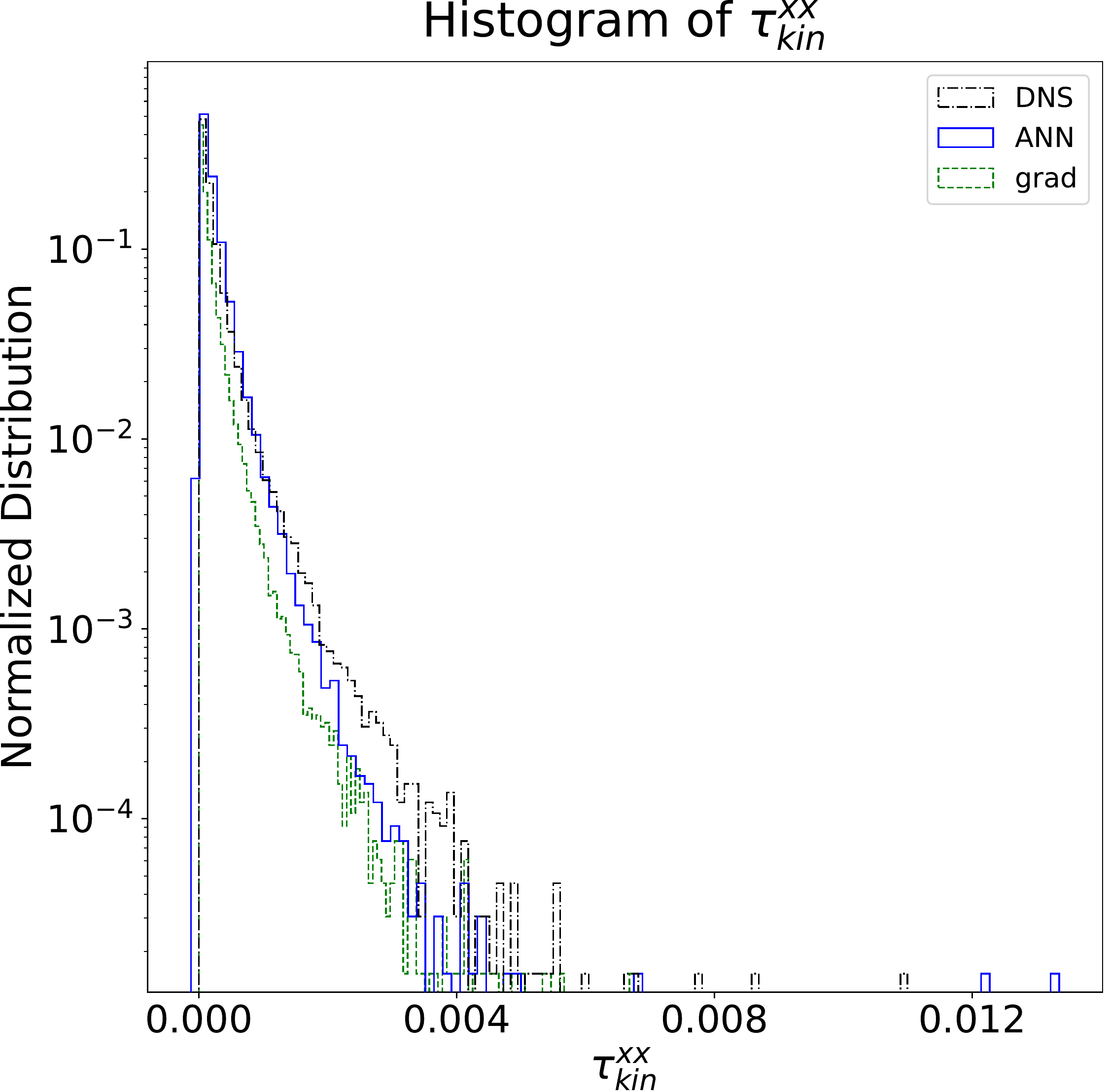}
\includegraphics[width=0.33\linewidth]{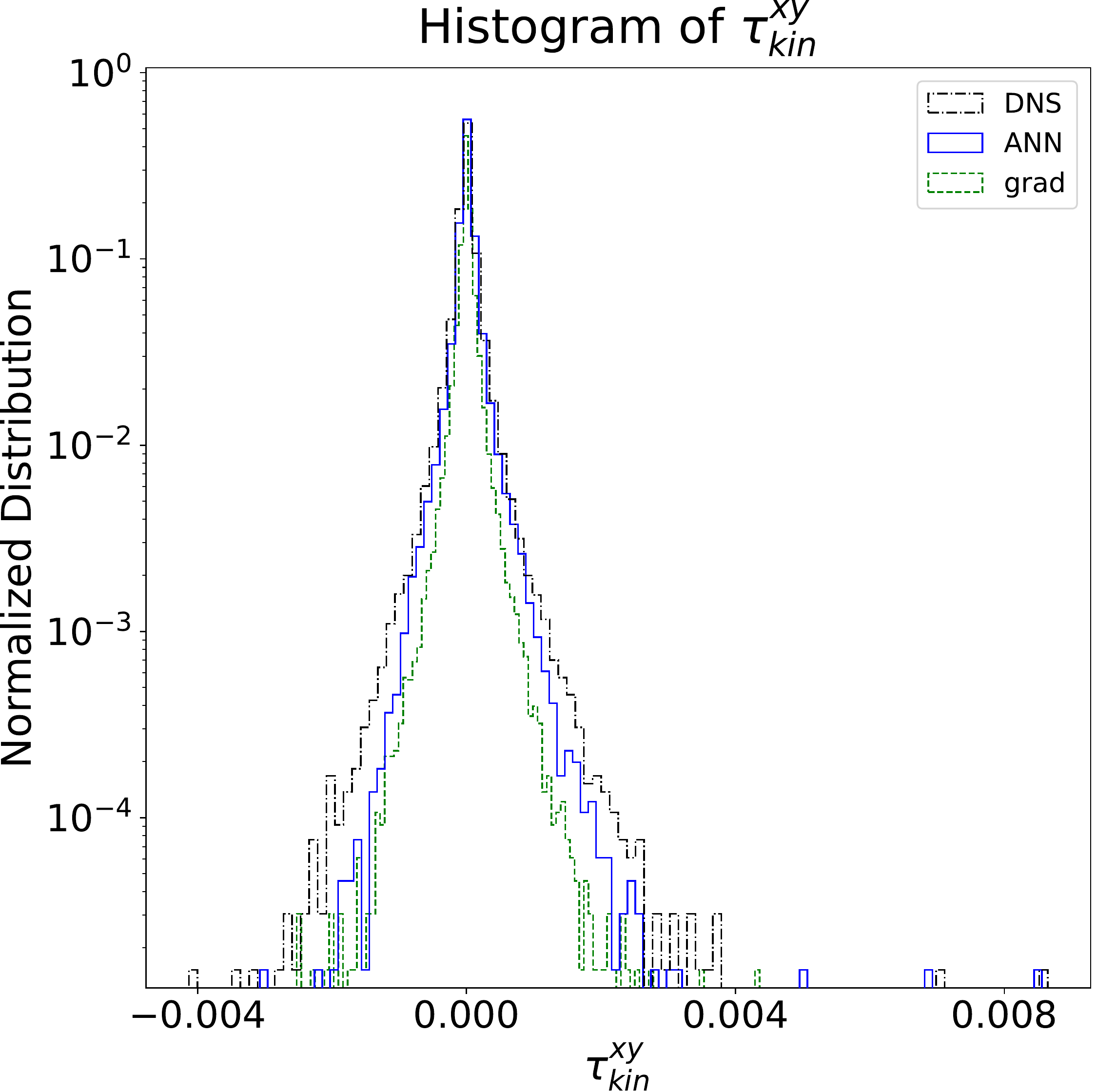}
\includegraphics[width=0.33\linewidth]{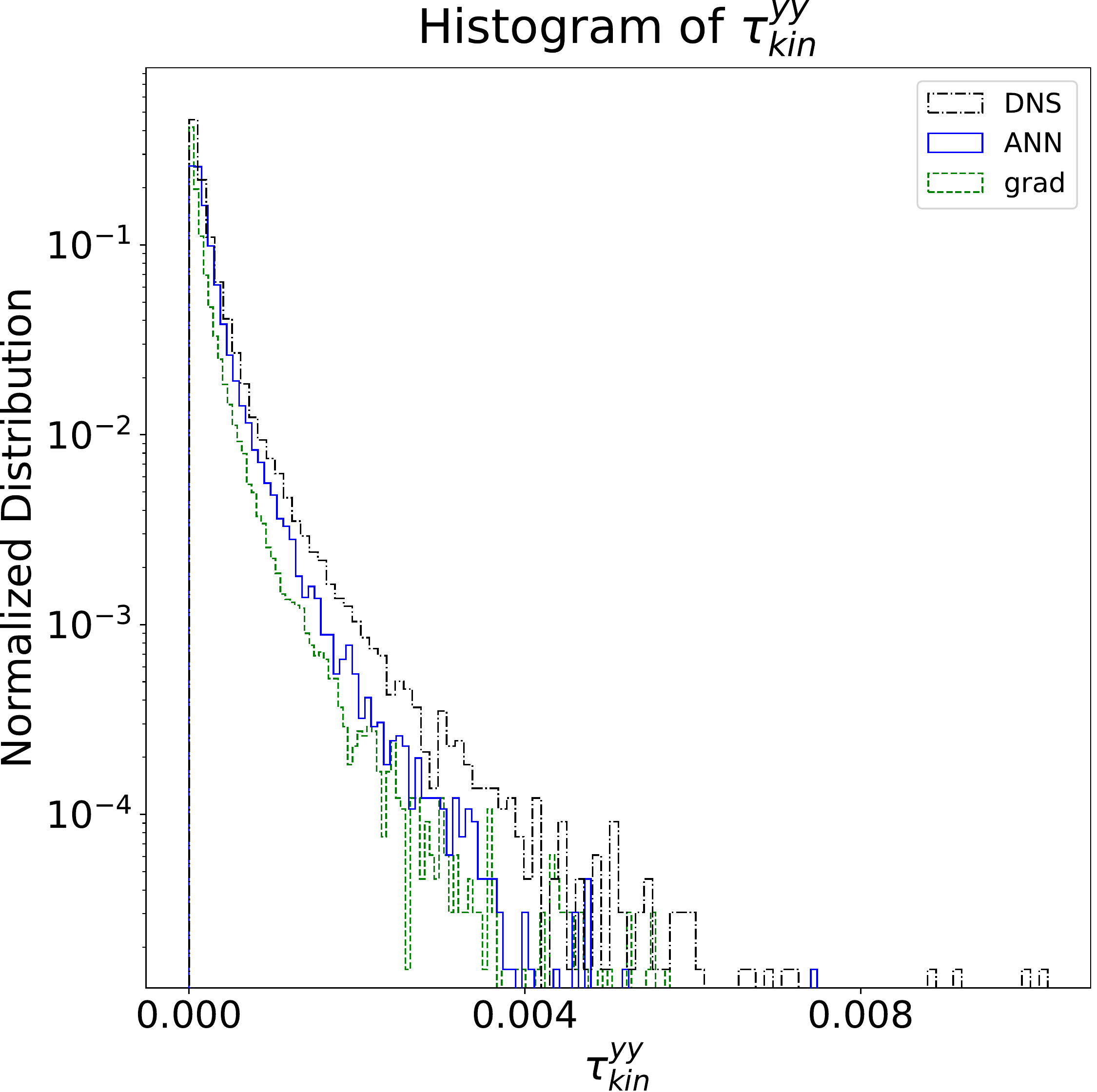}

\includegraphics[width=0.33\linewidth]{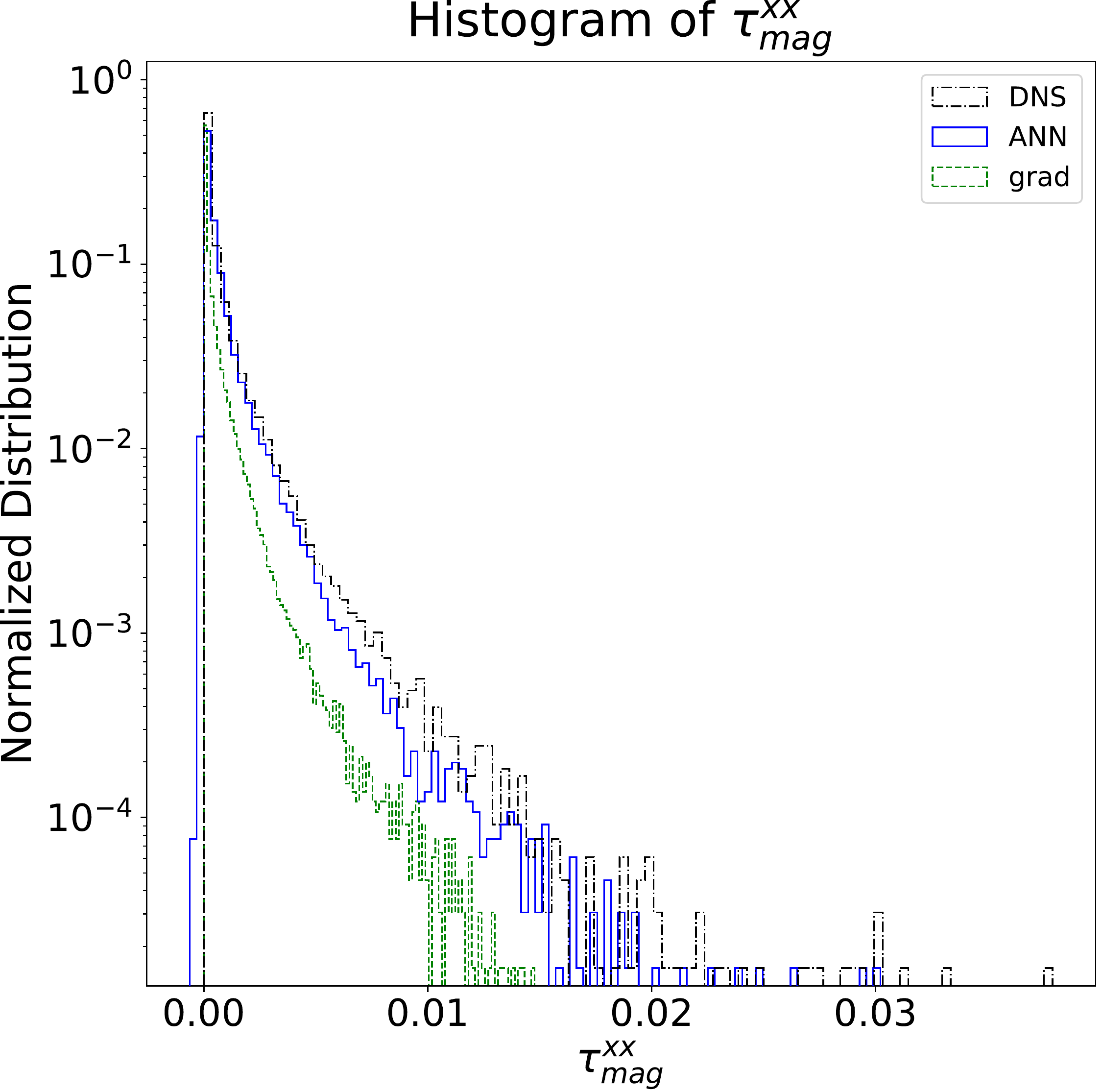}
\includegraphics[width=0.33\linewidth]{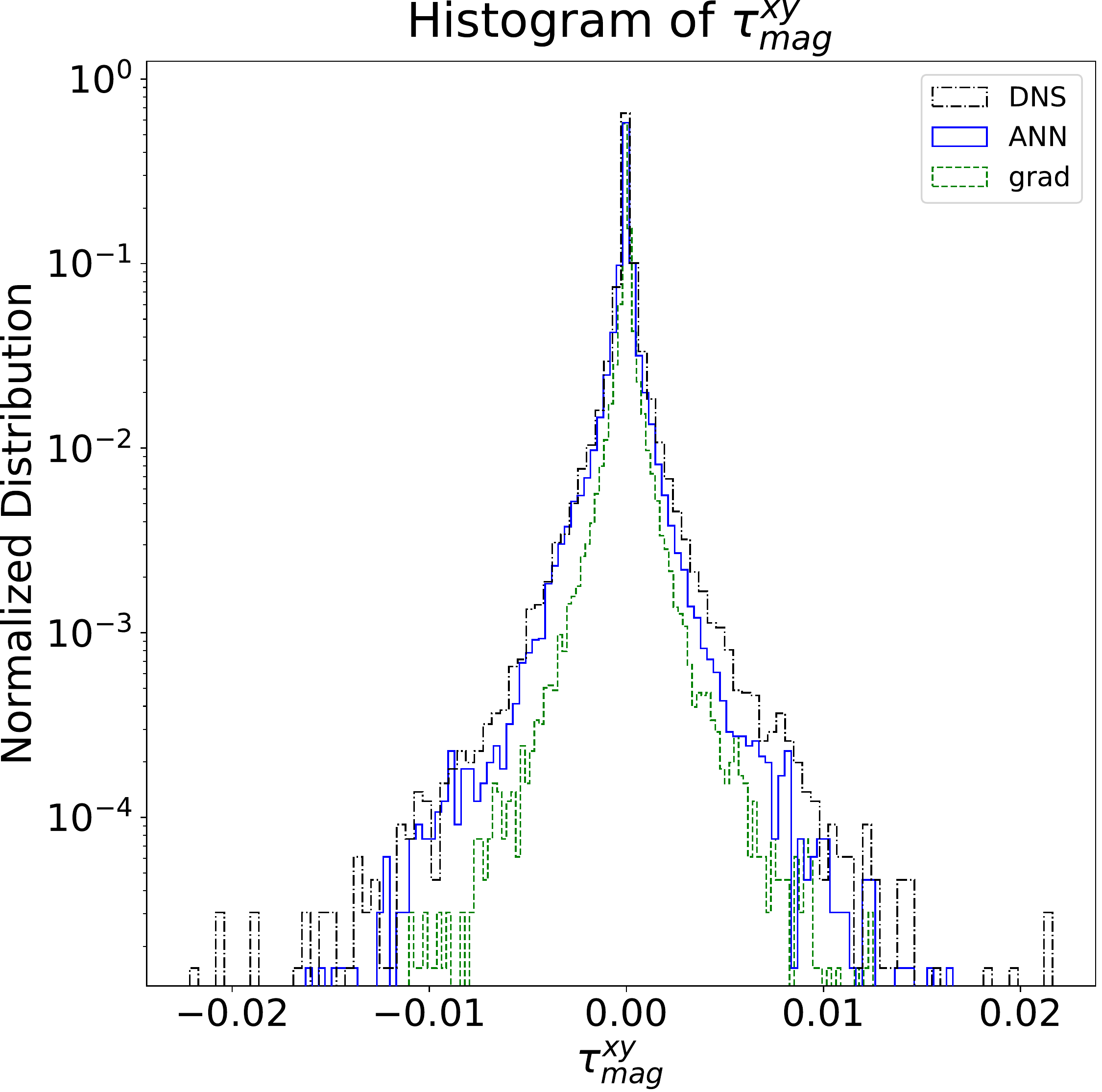}
\includegraphics[width=0.33\linewidth]{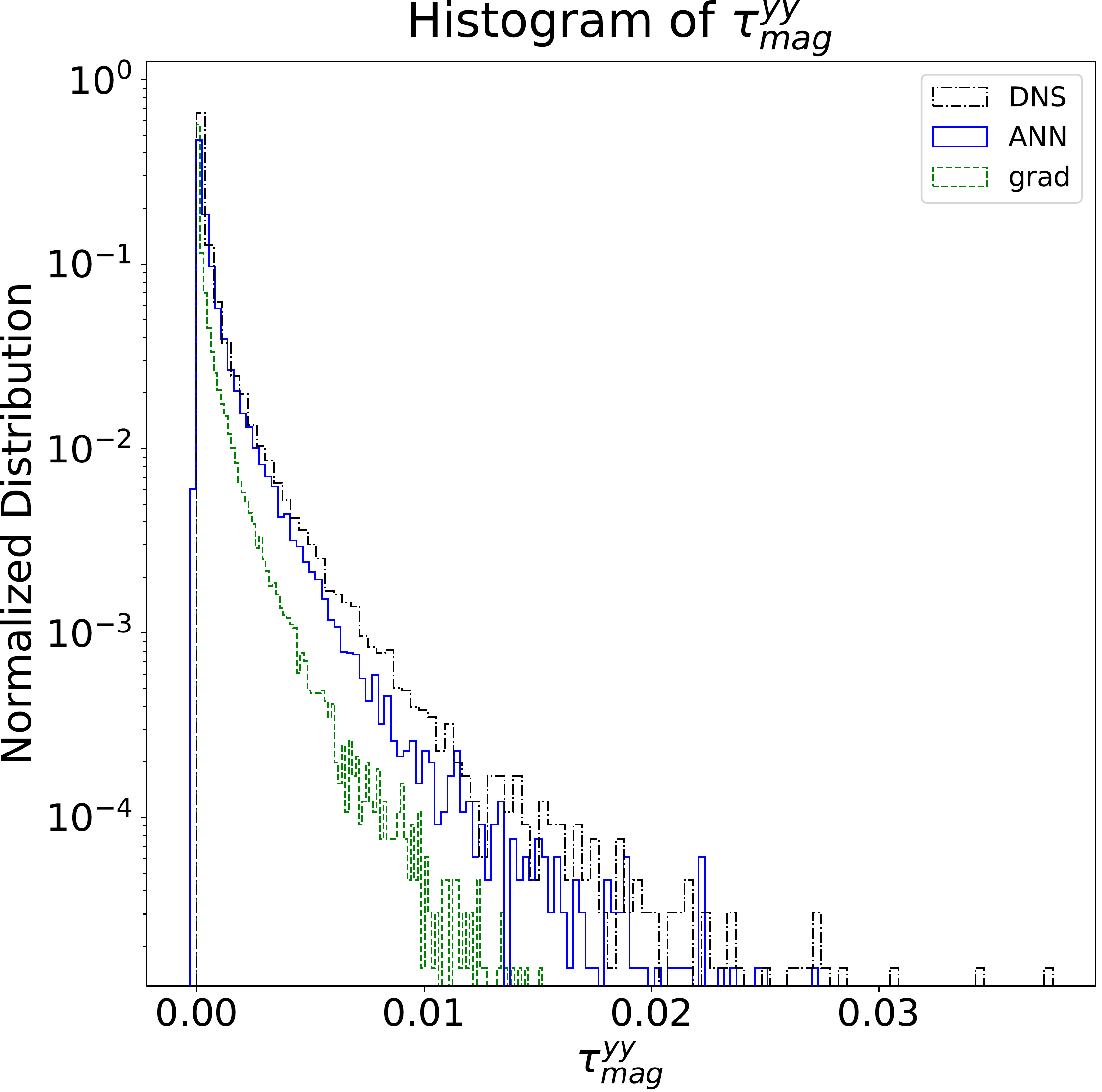}

\includegraphics[width=0.33\linewidth]{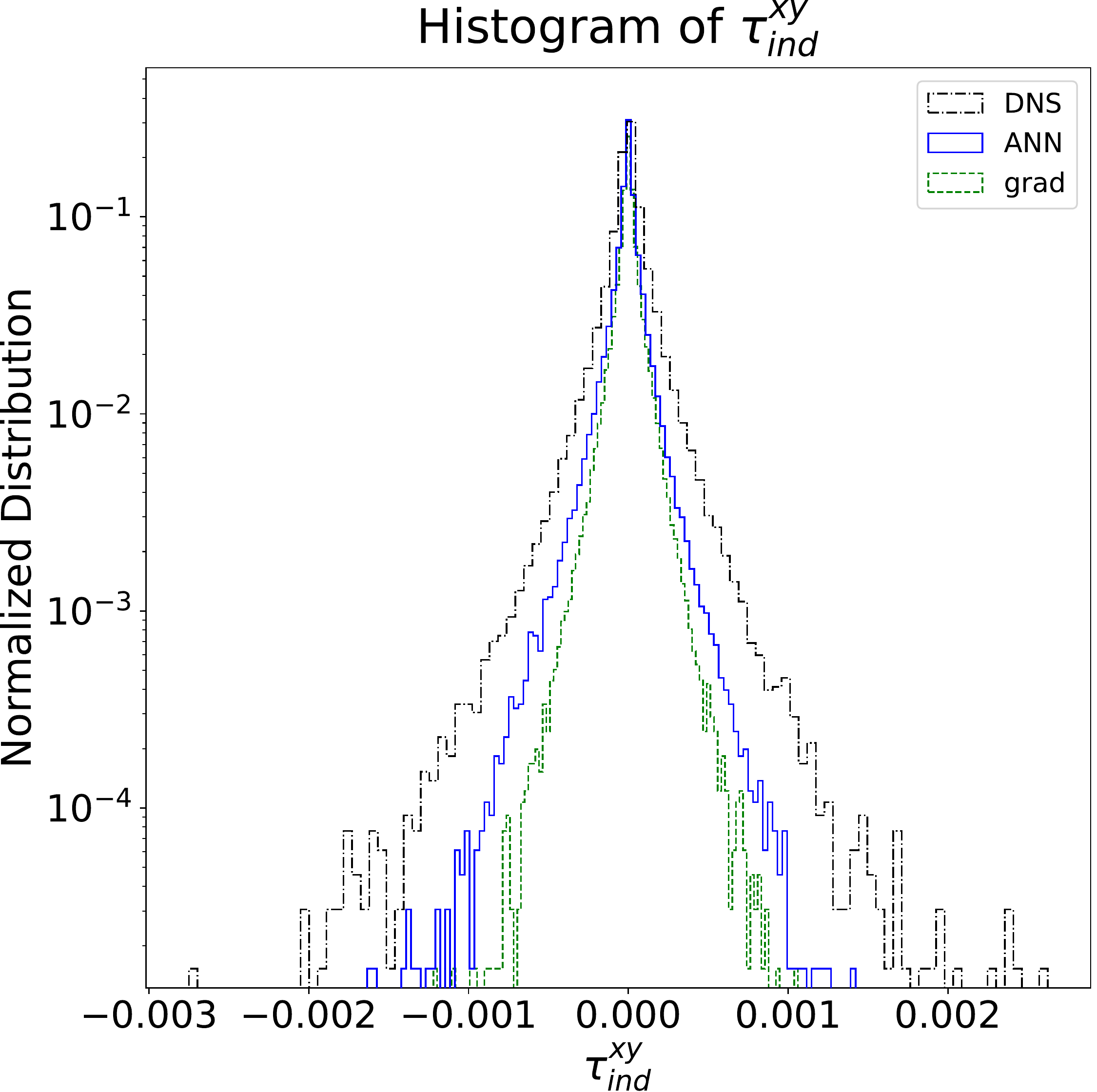}
\includegraphics[width=0.33\linewidth]{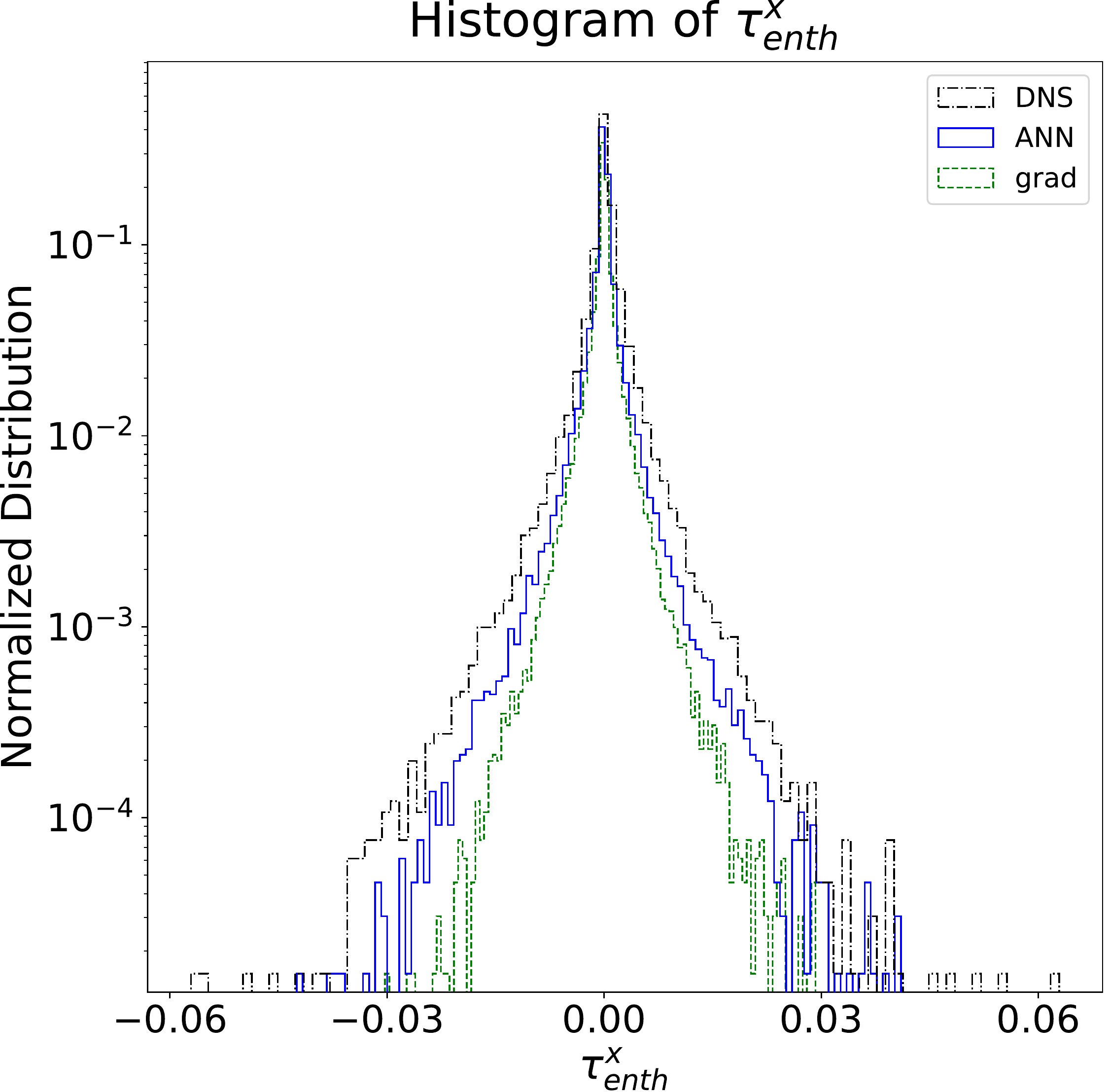}
\includegraphics[width=0.33\linewidth]{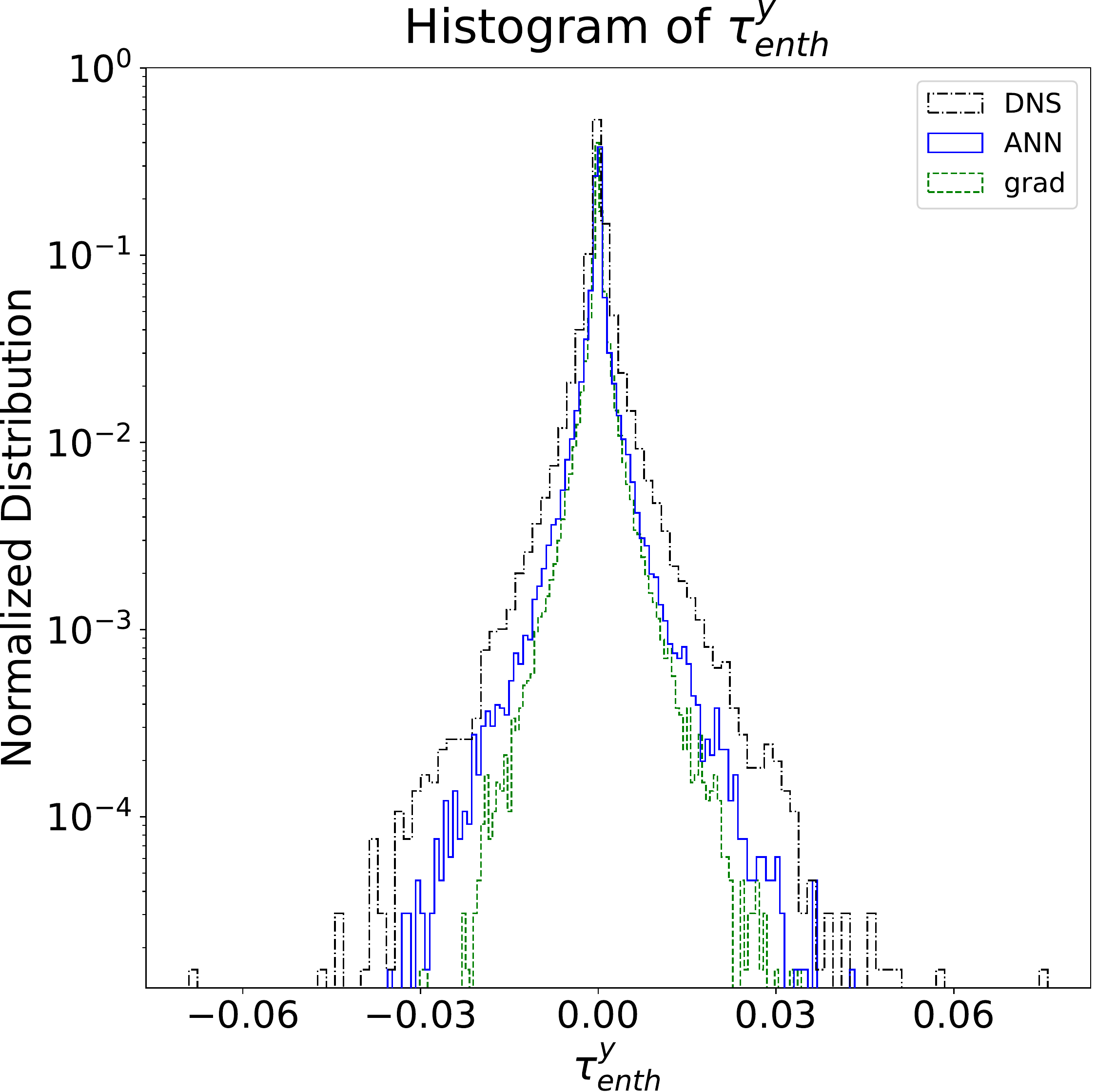}
\caption{Histogram of the normalized probability distribution of the values of SGS tensors for the $N=2048^2$ resolution test dataset with filter size $f=8$ for each of our SGS models.  The black dot-dashed line represents the distribution of the DNS data, the solid blue line represents the distribution of the ANN model predictions, and the green dashed line represents the distribution of the gradient model predictions.  The first row displays the $xx$, $xy$, and $yy$ components of $\tau_{kin}$ from left to right respectively.  The middle row shows the same information for the components of $\tau_{mag}$ SGS tensor.  The bottom row shows the $xy$ component of $\tau_{ind}$ on the left, the $x$ component of $\tau_{enth}$ in the middle, and the $y$ component of $\tau_{enth}$ on the right.  We observe that the ANN model predictions more closely resemble the distribution of the DNS values than those of the gradient model.}
\label{fig:histogram}
\end{figure}

Fig. \ref{fig:energy} shows the integrated energy spectrum or total energy vs time starting at $t=1$ for both the kinetic energy $E_{kin}$ and the magnetic energy $E_{mag}$ in the first and second plots respectively.  We notice that $E_{kin}$ starts the same for all simulations, but decreases over time.  The higher resolution simulations decreased in $E_{kin}$ faster than those at lower resolutions.  This may indicate that $E_{kin}$ is being converting into $E_{mag}$. On the other hand, $E_{mag}$ started fairly similar in magnitude for all resolution with deviations of order unity.  We then see an increase in the magnetic energy with the higher resolution simulations increasing much faster than their lower resolution counterpart.  At $t \sim 7$, the $N=1024^2$ simulation is observed to rise faster than the $N=2048^2$ simulation, which results in both simulations having nearly equal energy by the end of the simulation at $t=10$.  $E_{mag}$ still appears to be increasing at $t=10$, which may indicate that the process of magnetic amplification may still be ongoing.
\begin{figure*}
\centering
\includegraphics[height=0.23\textheight]{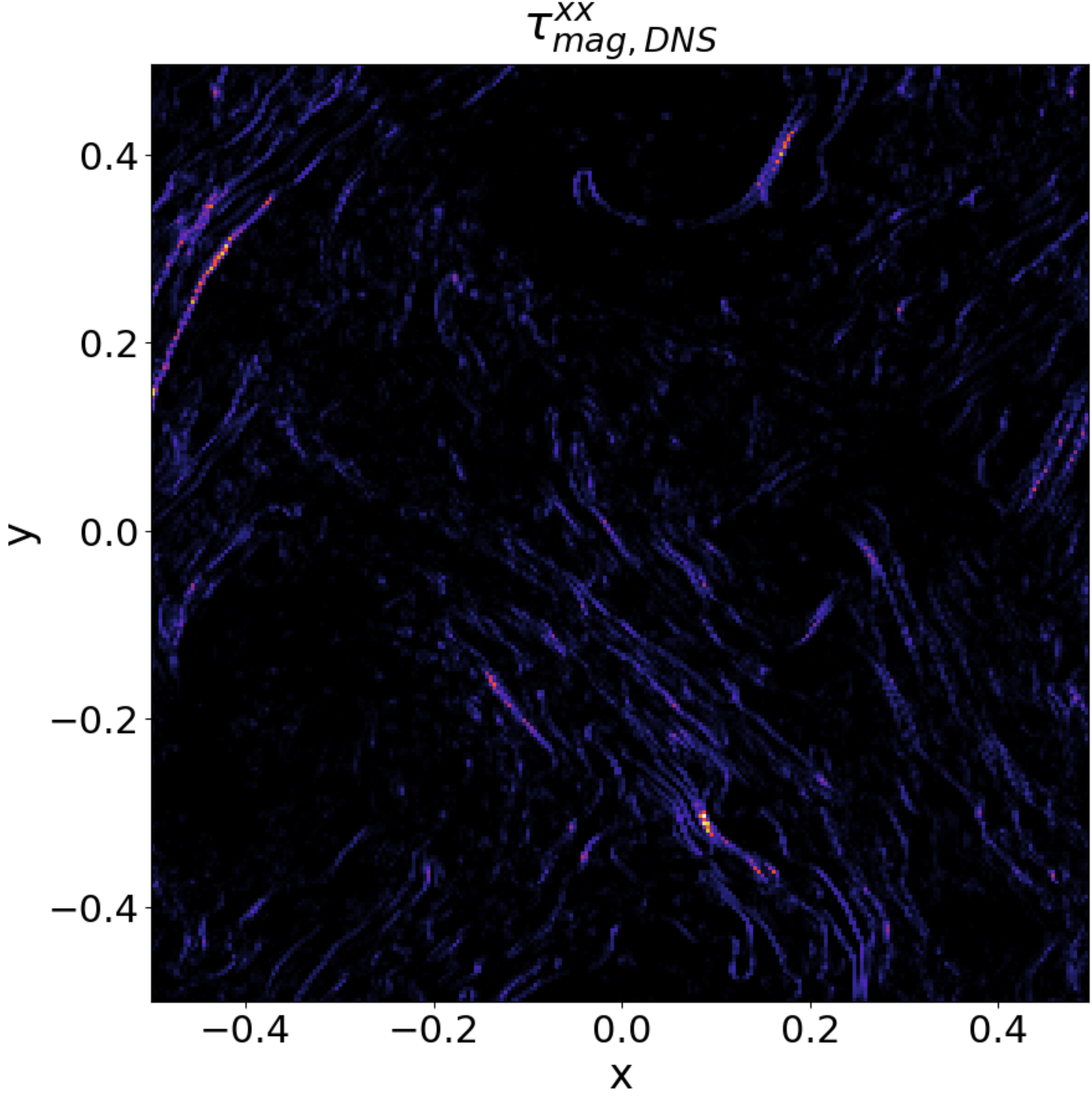}
\includegraphics[height=0.23\textheight]{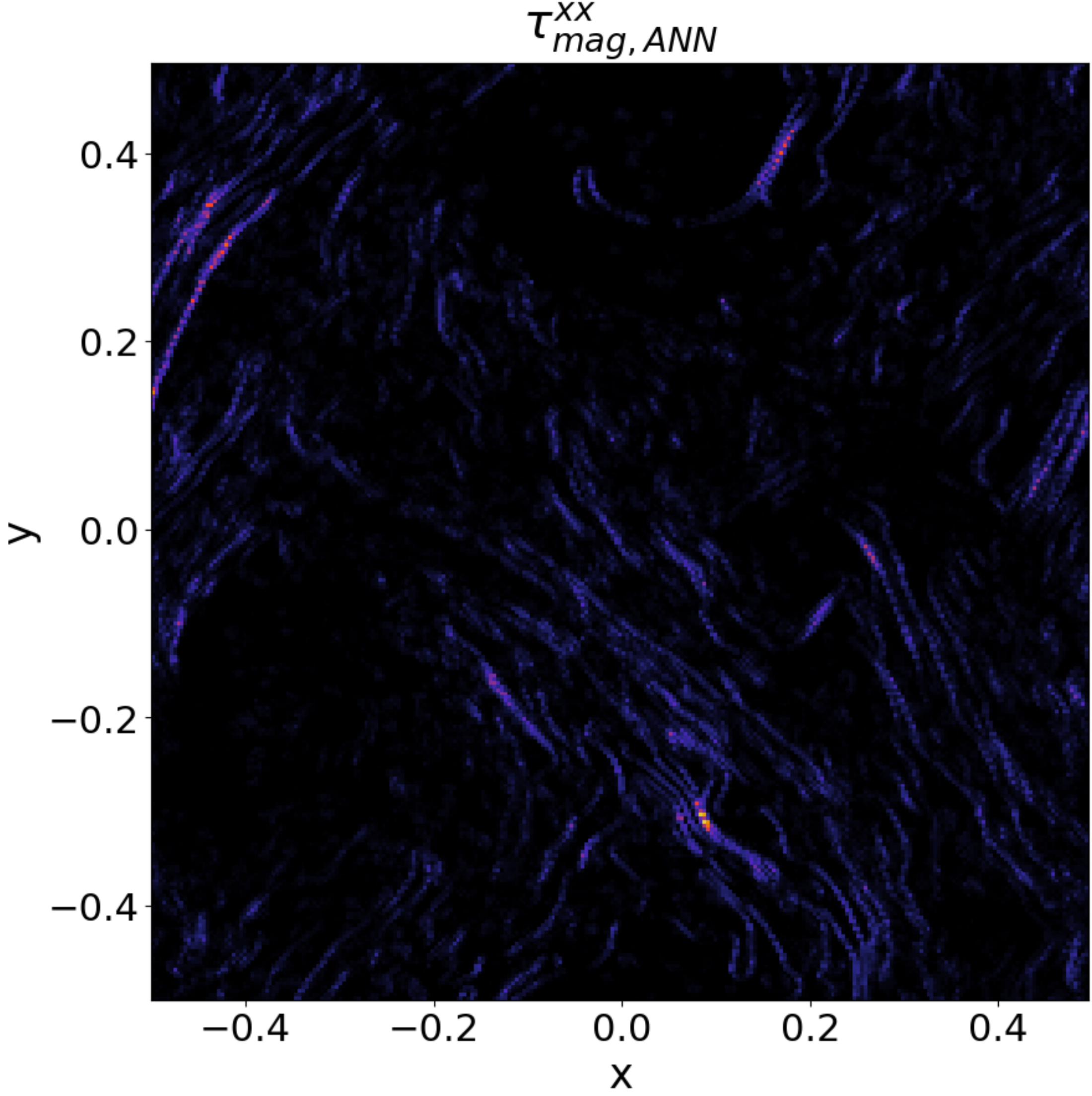}
\includegraphics[height=0.23\textheight]{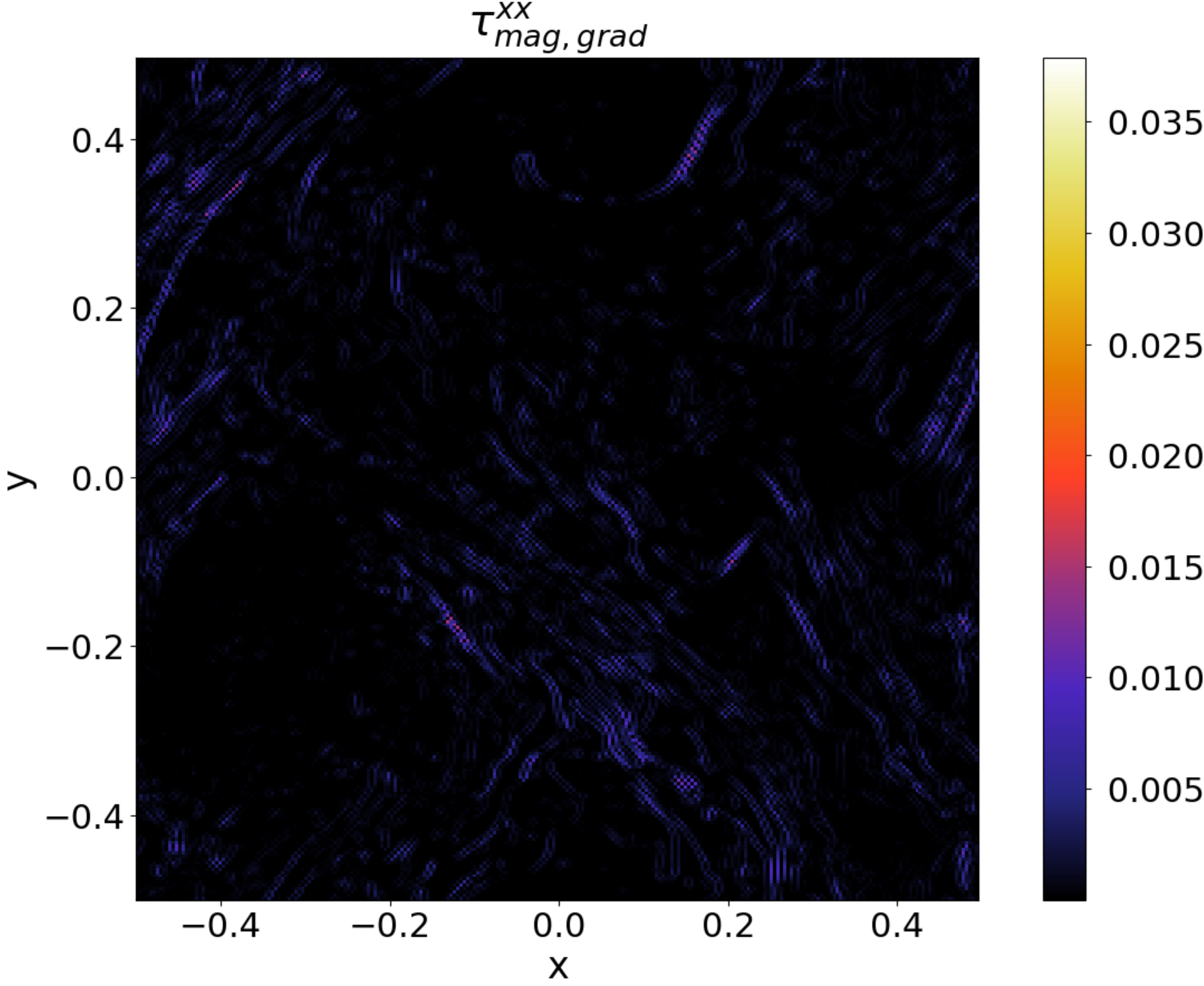}

\includegraphics[height=0.23\textheight]{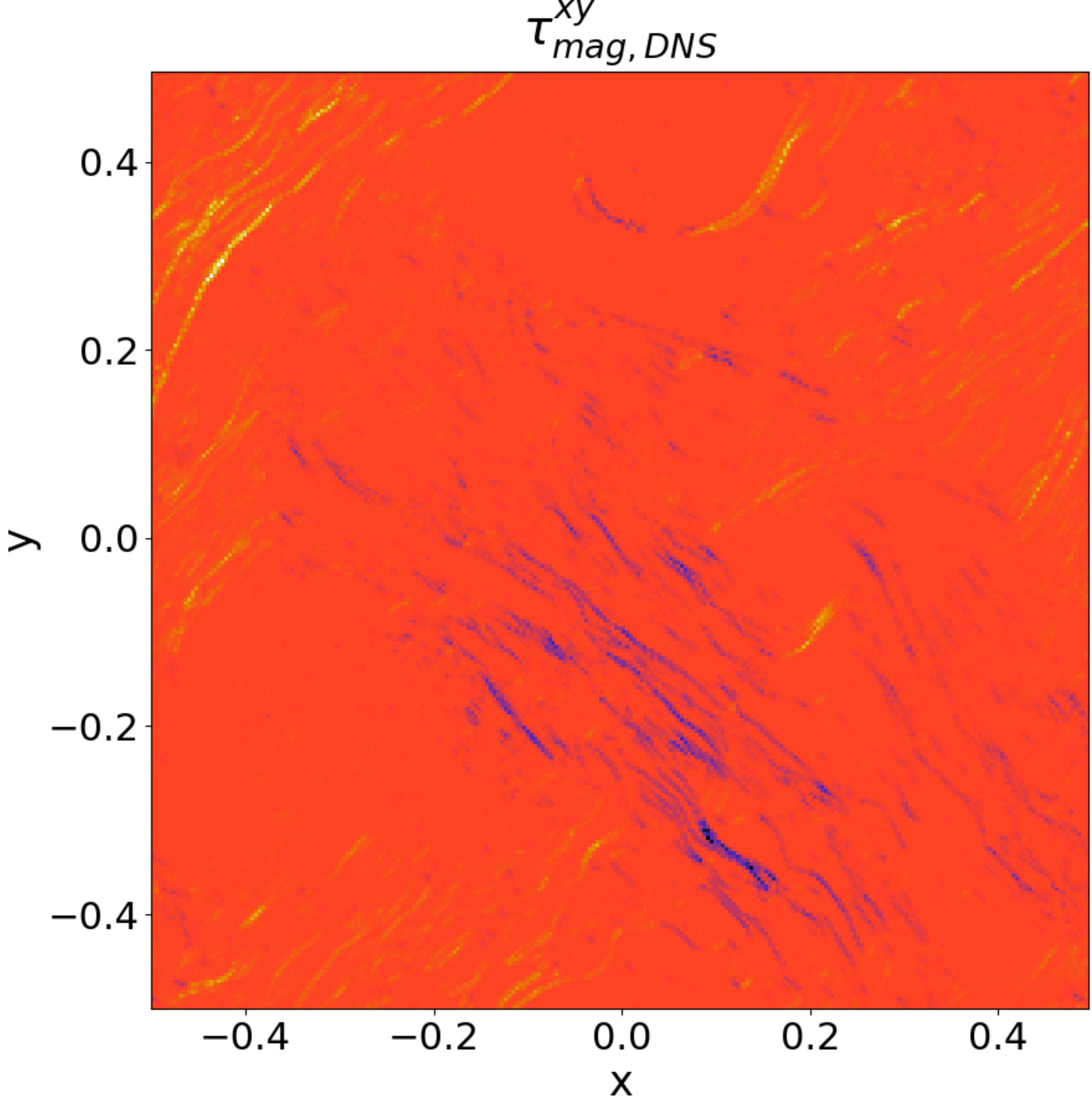}
\includegraphics[height=0.23\textheight]{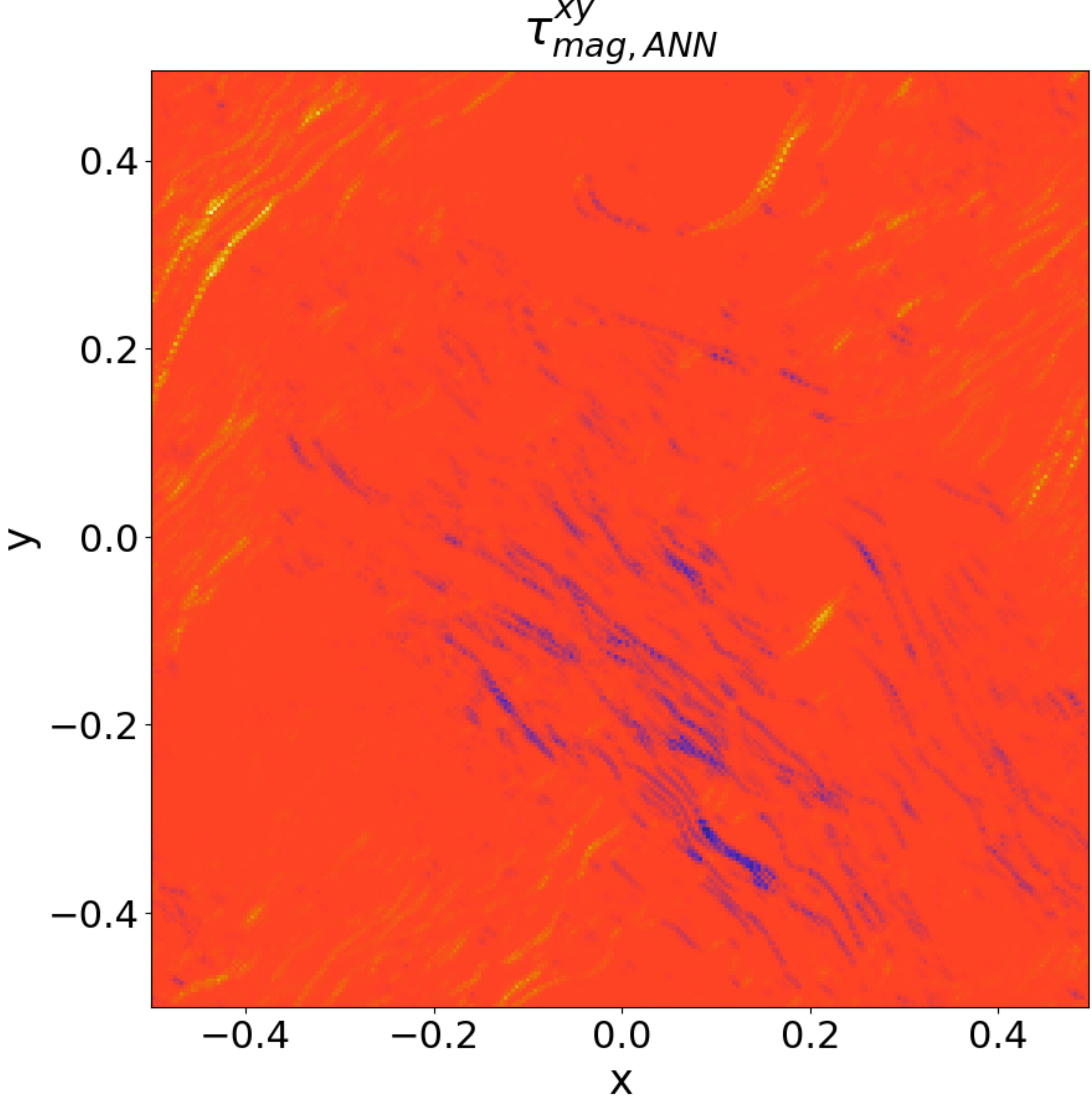}
\includegraphics[height=0.23\textheight]{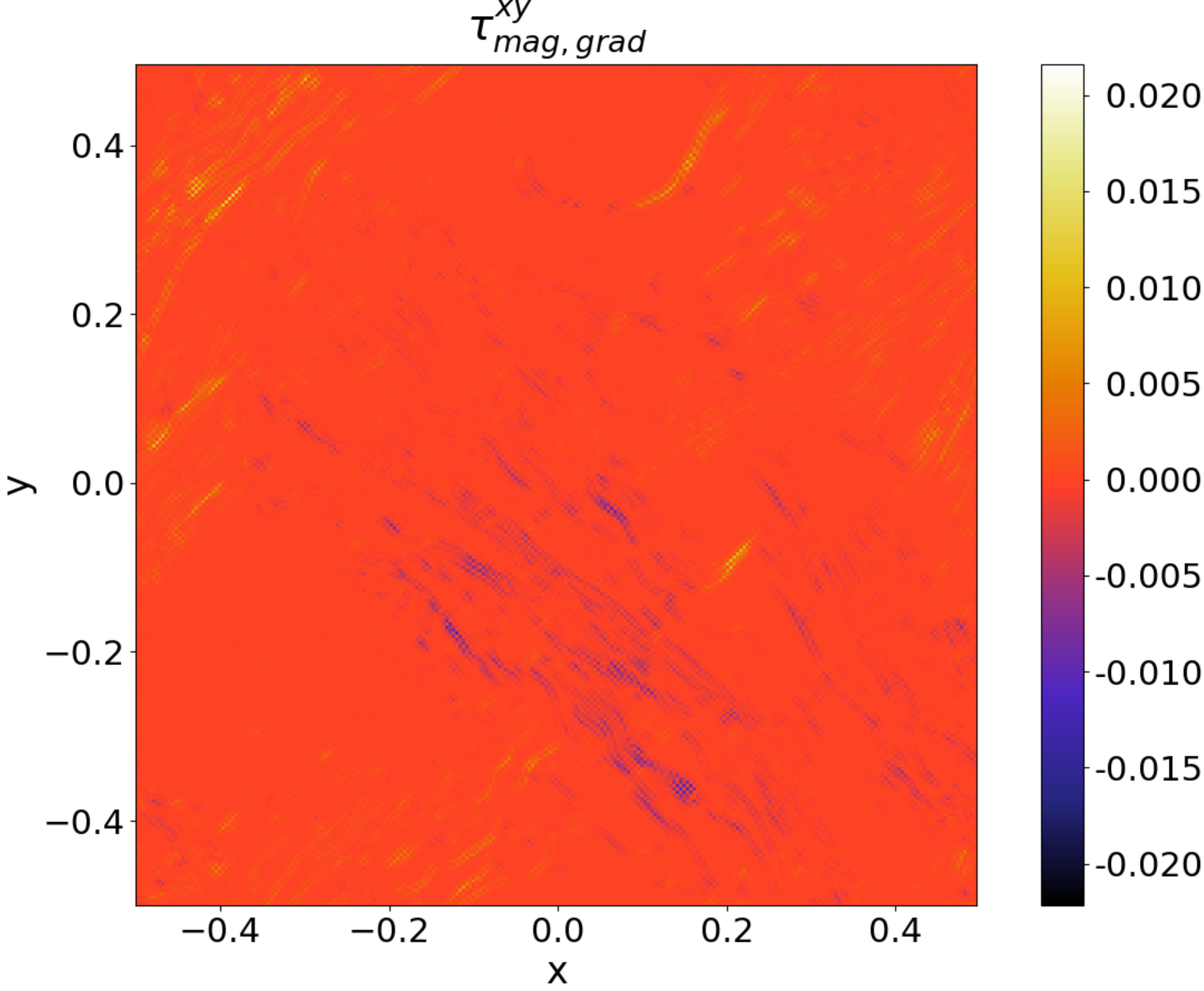}

\includegraphics[height=0.23\textheight]{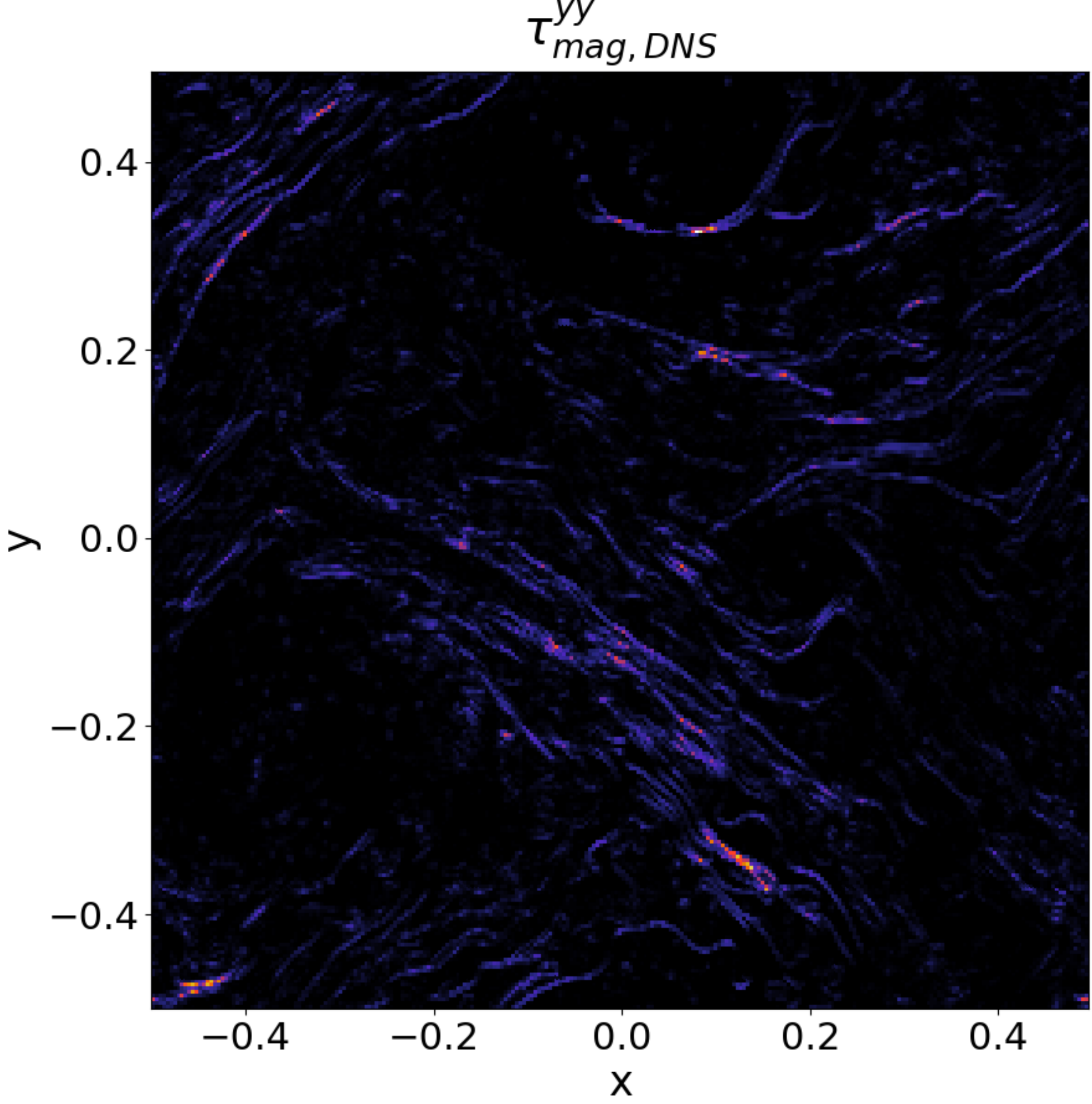}
\includegraphics[height=0.23\textheight]{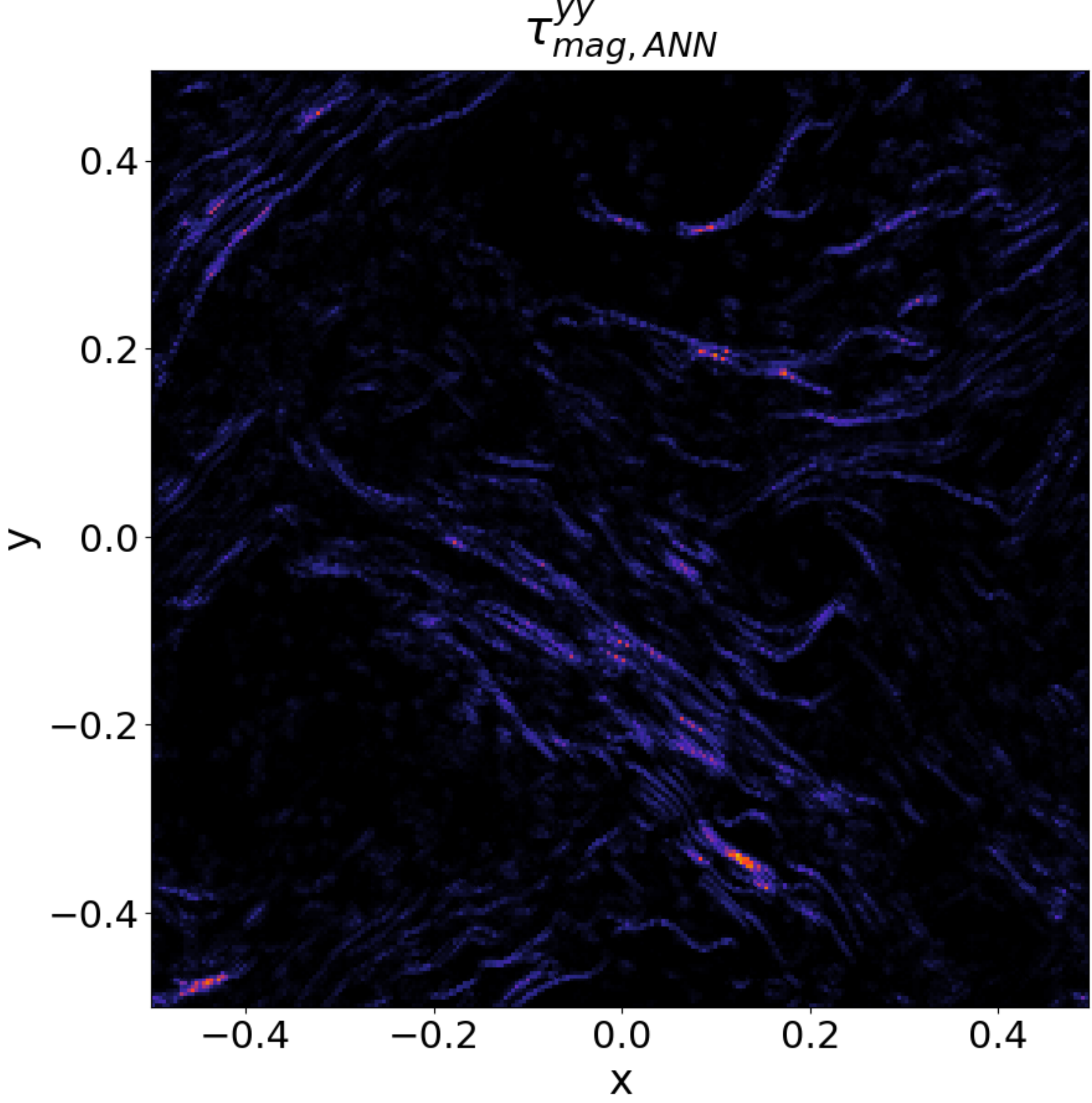}
\includegraphics[height=0.23\textheight]{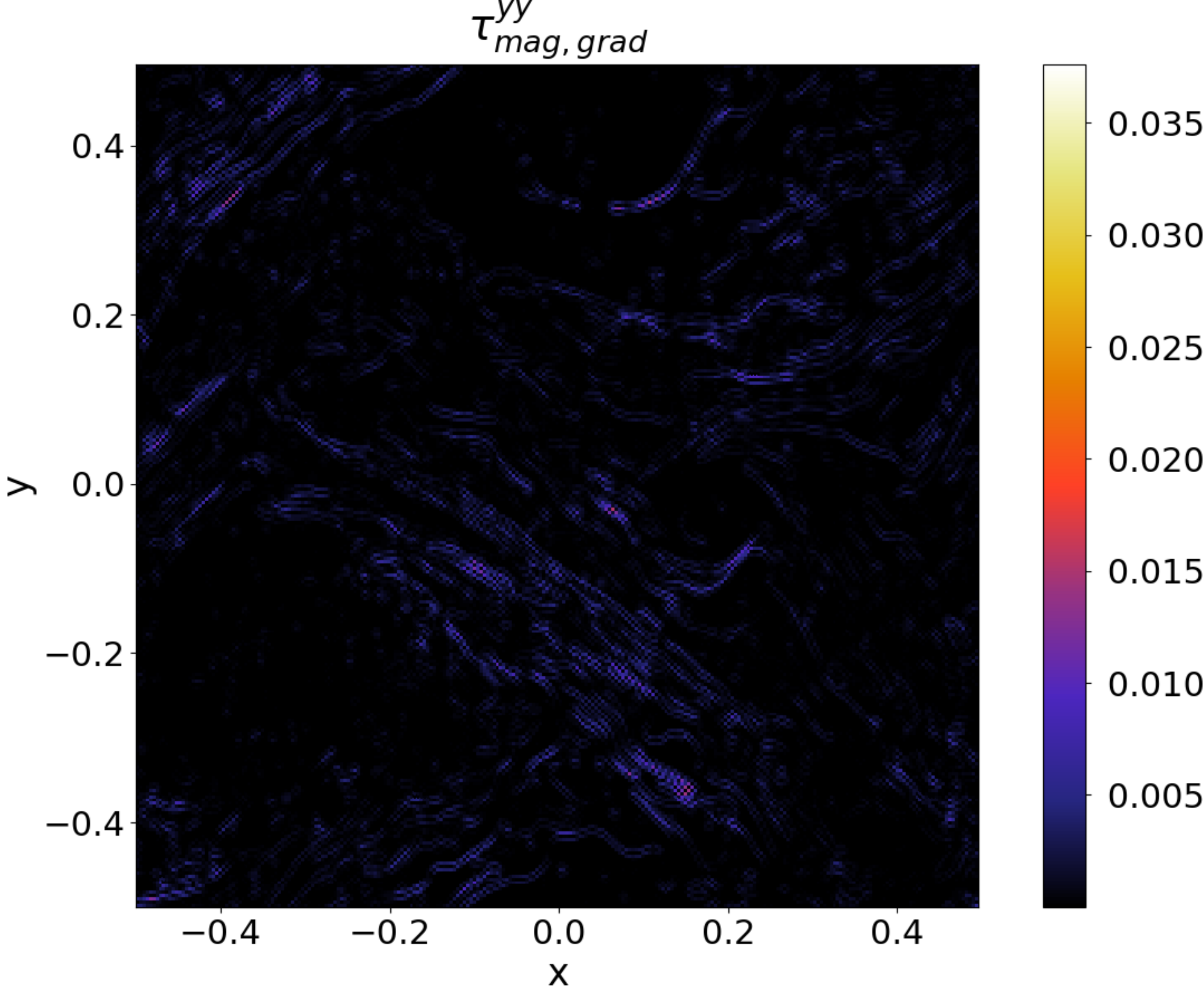}
\caption{Plots of the components of the $\tau_{mag}$ SGS tensor of the test dataset for the $N=2048^2$ resolution run with a filter size of $f=8$.  The columns depict the SGS tensor values of the exact DNS calculation, the ANN model reconstruction, and the gradient model reconstruction from left to right respectively.  The top, middle, and bottom rows display the $xx$, $xy$, and $yy$ components respectively.  We observe that while the ANN model appears to reproduce most of the visual features of the DNS calculation, the gradient model appears to struggle in regions with more detailed structure.}
\label{fig:tau_mag}
\end{figure*}

\subsection{Model Performance}
\label{sec:model_performace}

For all subgrid filter sizes $f$ and at all resolutions $N$, our findings indicate that our neural network model outperforms the gradient model when evaluated on the test data.  To show this, let us first take a look at the results of models with $N=2048^2$ at $f=8$, a case where the differences can be clearly observed between the two SGS models.

Fig.~\ref{fig:targets_vs_predictions} presents targets vs predictions of the SGS models for the test data at $N=2048^2$ and $f=8$.  We notice that both models show good performance when the magnitude of the SGS tensor is low.  However, at high SGS tensor magnitudes, the gradient model significantly underestimates the SGS quantities.  Compared to the gradient model, the ANN models predict more accurate values for those high magnitude targets, in particular for the components of $\tau_{mag}$ and $\tau_{ind}$ tensors.  We note that one reason for the poor performance of the gradient model for high SGS tensors may be due to it being a first order model and could potentially be improved using higher order corrections. 

In \Cref{fig:histogram}, we show a histogram the normalized distribution of the SGS tensors of the $N=2048^2$, $f=8$ test data as well as those predicted by the ANN and gradient models for this same dataset.  We find that overall, the ANN model's predictions more closely resemble the distribution calculated from the DNS dataset compared to those of the gradient model.  This improvement is especially noticeable for the $\tau_{mag}$ tensor components.  We also notice that the ANN model predicts some negative values for the diagonal components of the SGS tensors $\tau_{kin}$ and $\tau_{mag}$.  However, these predictions are unphysical as they violate the realizability constraint which requires $\tau_{ii} \geq 0$ \cite{Ghosal1999,Silvis2017}.  Such unphysical behavior can be resolved in ANN models by embedding the physical constraint in the loss function, which will be a subject of future work.

To understand exactly how models behave for a single time slice of data, \Cref{fig:tau_mag} shows the values of the components of the SGS tensor $\tau_{mag}$ for the actual DNS data $\tau_{DNS}$, the ANN model $\tau_{ANN}$, and the gradient model $\tau_{grad}$.   From these plots we observe that $\tau_{ANN}$ performs noticeable better in regions with significant small scale structure in $\tau_{DNS}$ compared to $\tau_{grad}$.  This effect is most prevalent in $\tau_{mag}$ in \Cref{fig:tau_mag}, though is visible for most of the other tensors.  The plots depicting the values of the other SGS tensors can be seen in \Cref{fig:tau_kin,fig:tau_ind,fig:tau_enth} in Appendix \ref{app:sgs_tensors}. 

\begin{figure*}
\centering
\includegraphics[width=0.33\linewidth]{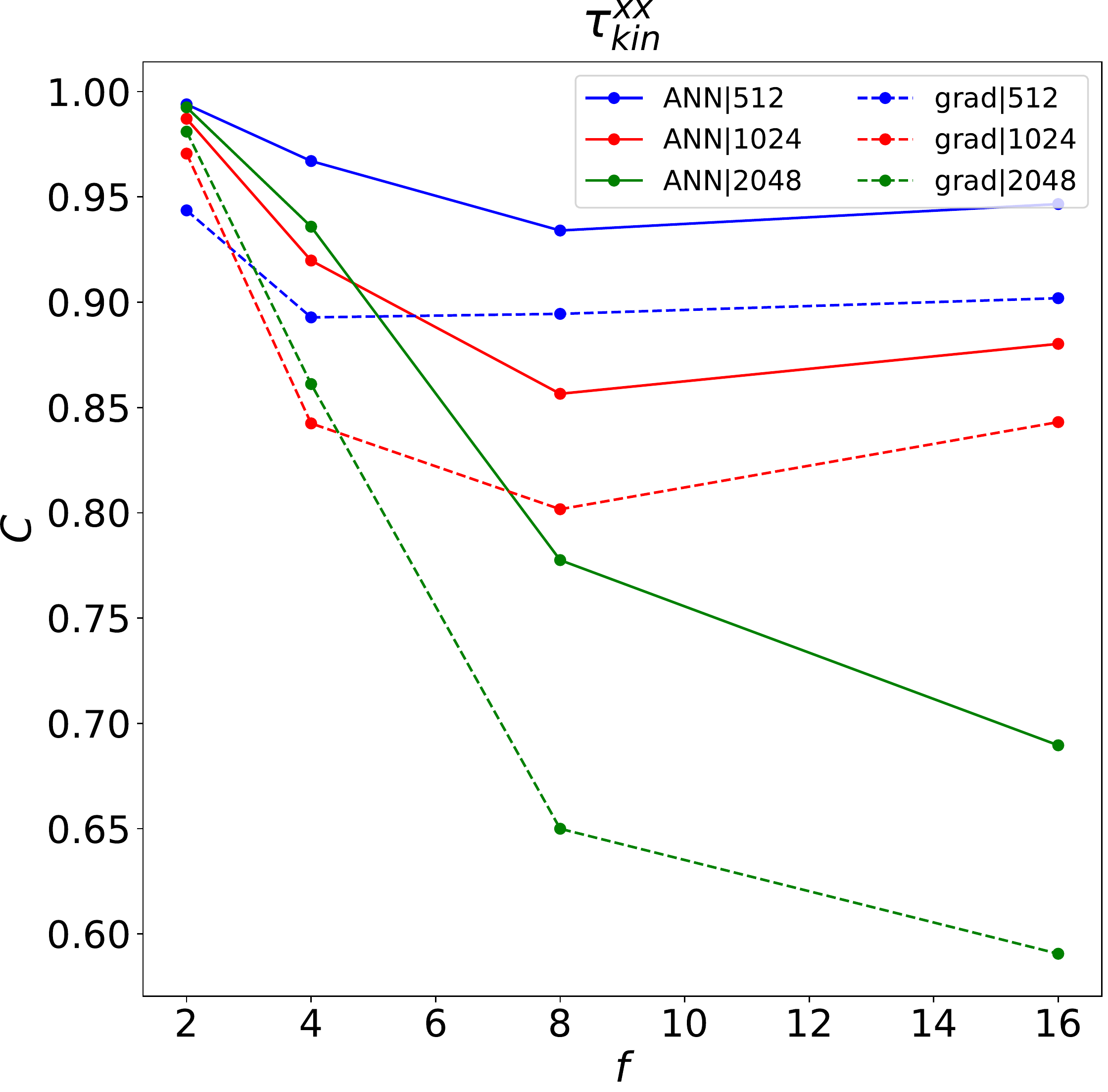}
\includegraphics[width=0.33\linewidth]{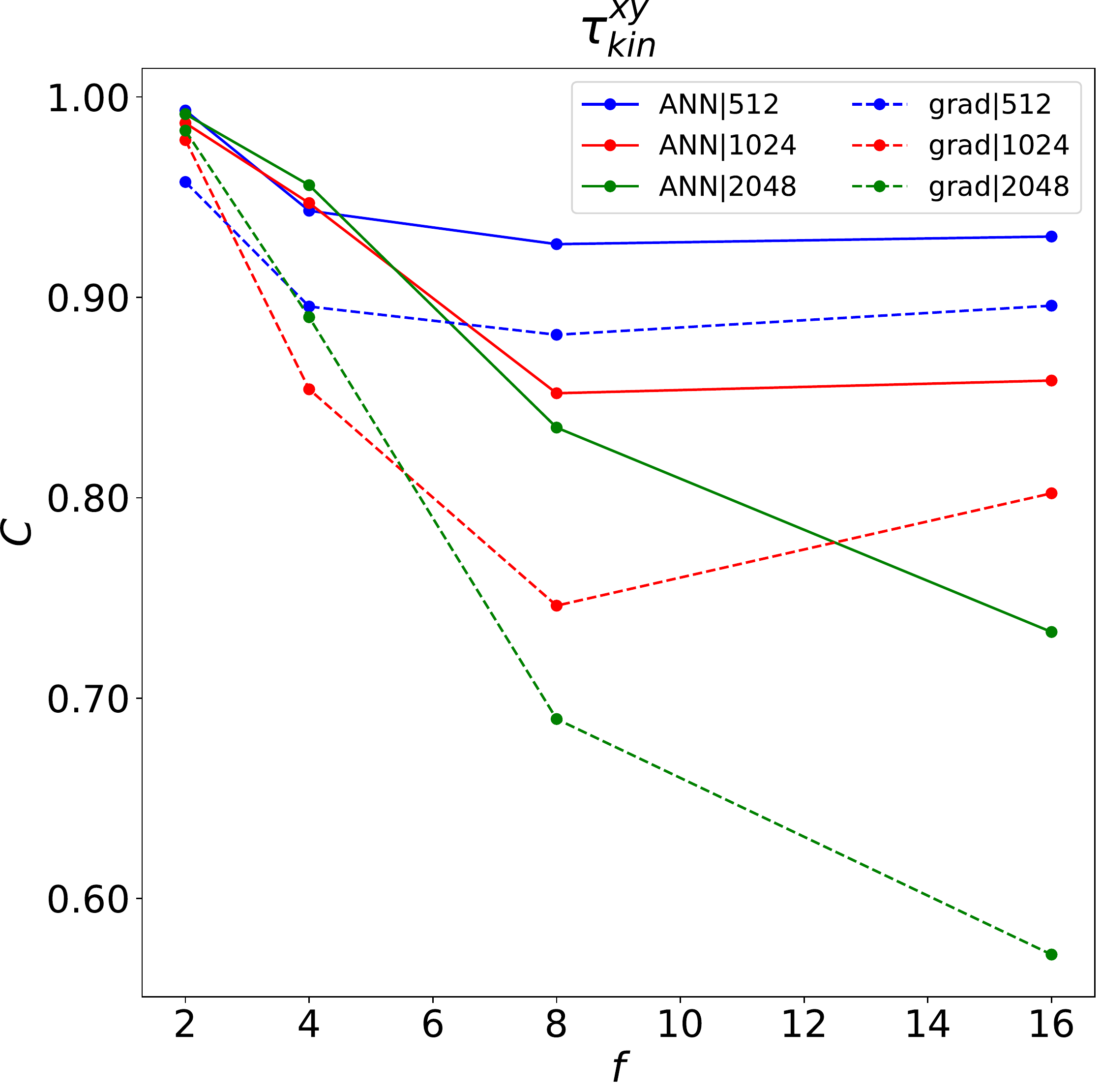}
\includegraphics[width=0.33\linewidth]{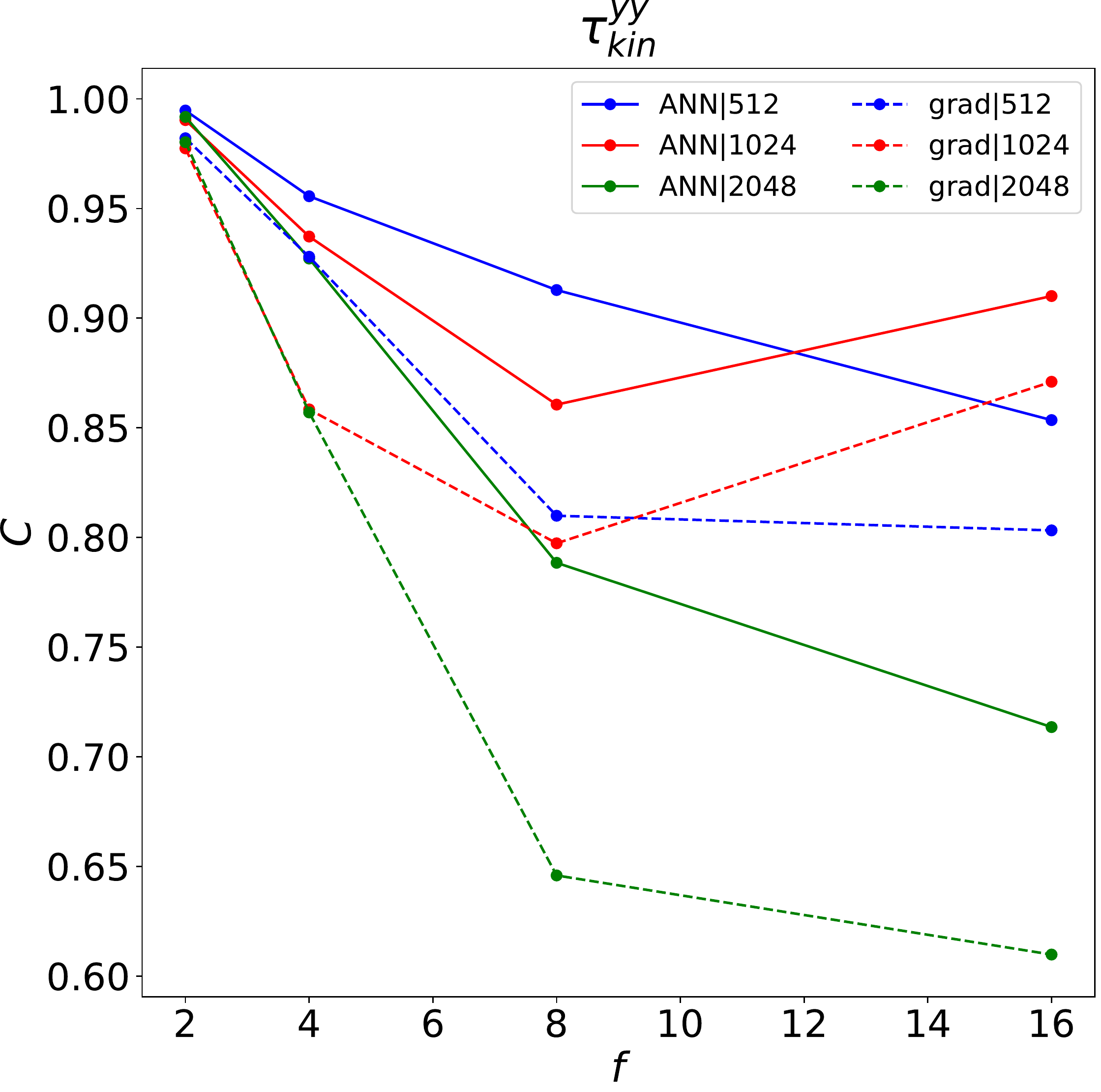}

\includegraphics[width=0.33\linewidth]{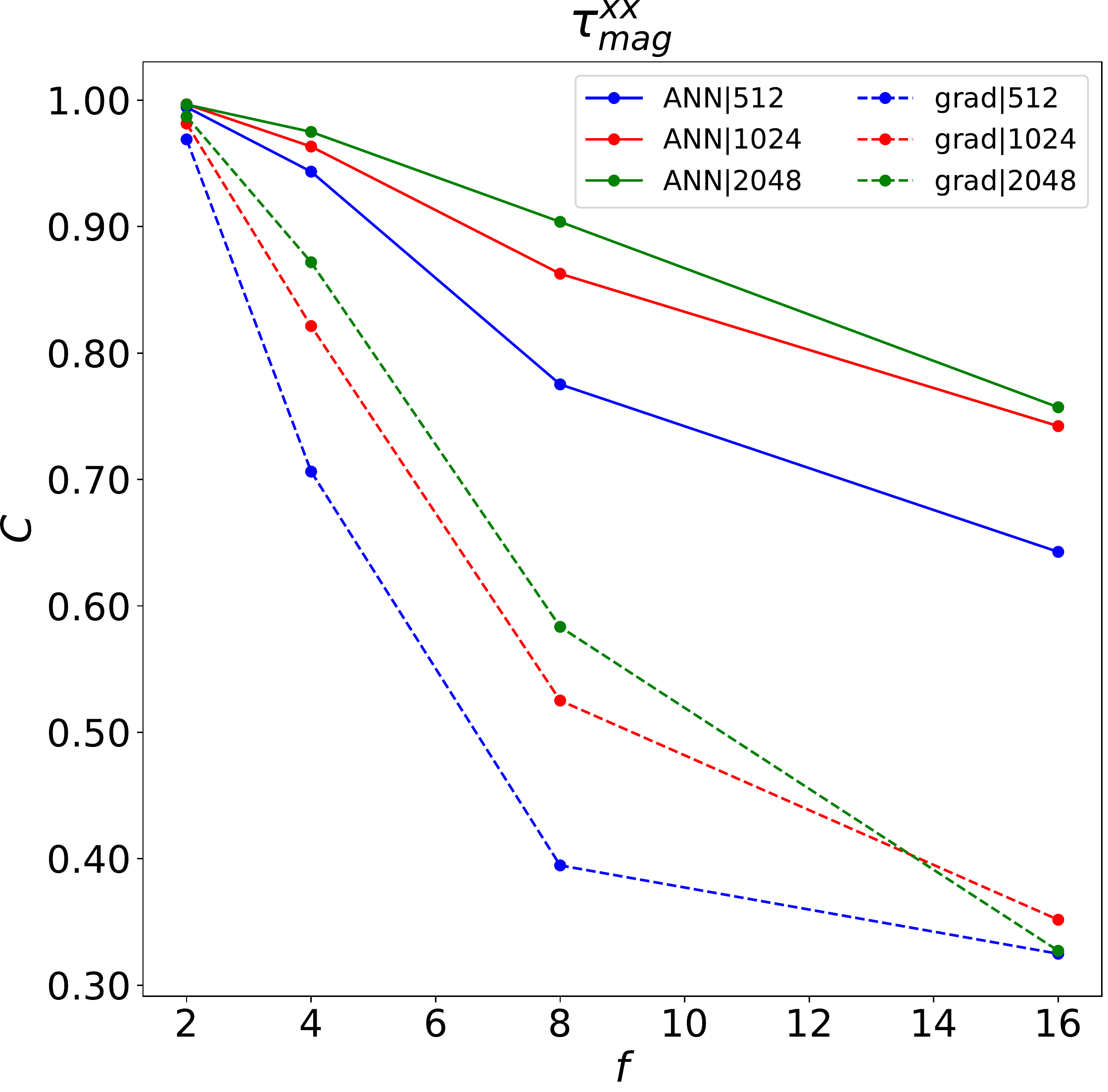}
\includegraphics[width=0.33\linewidth]{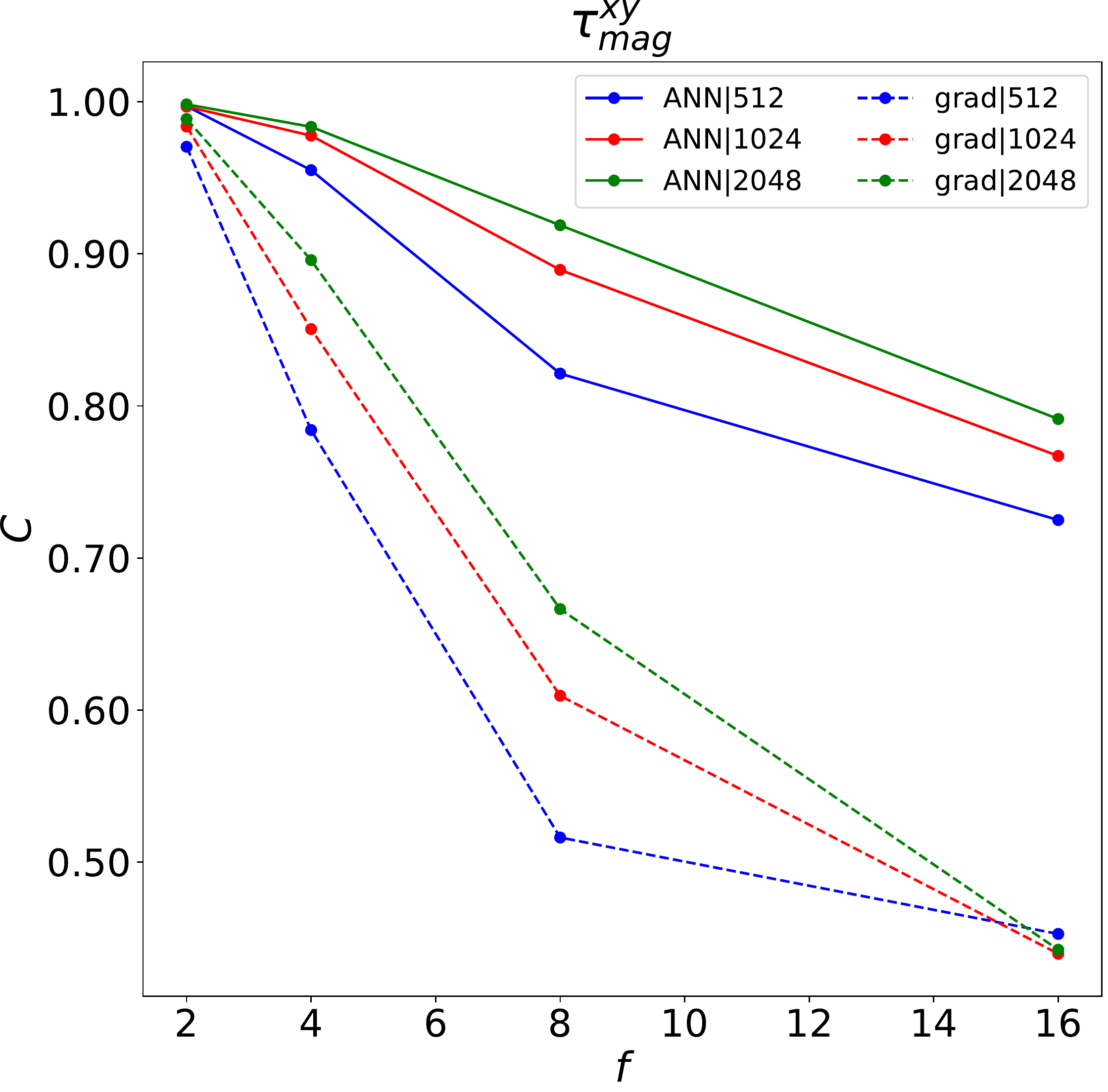}
\includegraphics[width=0.33\linewidth]{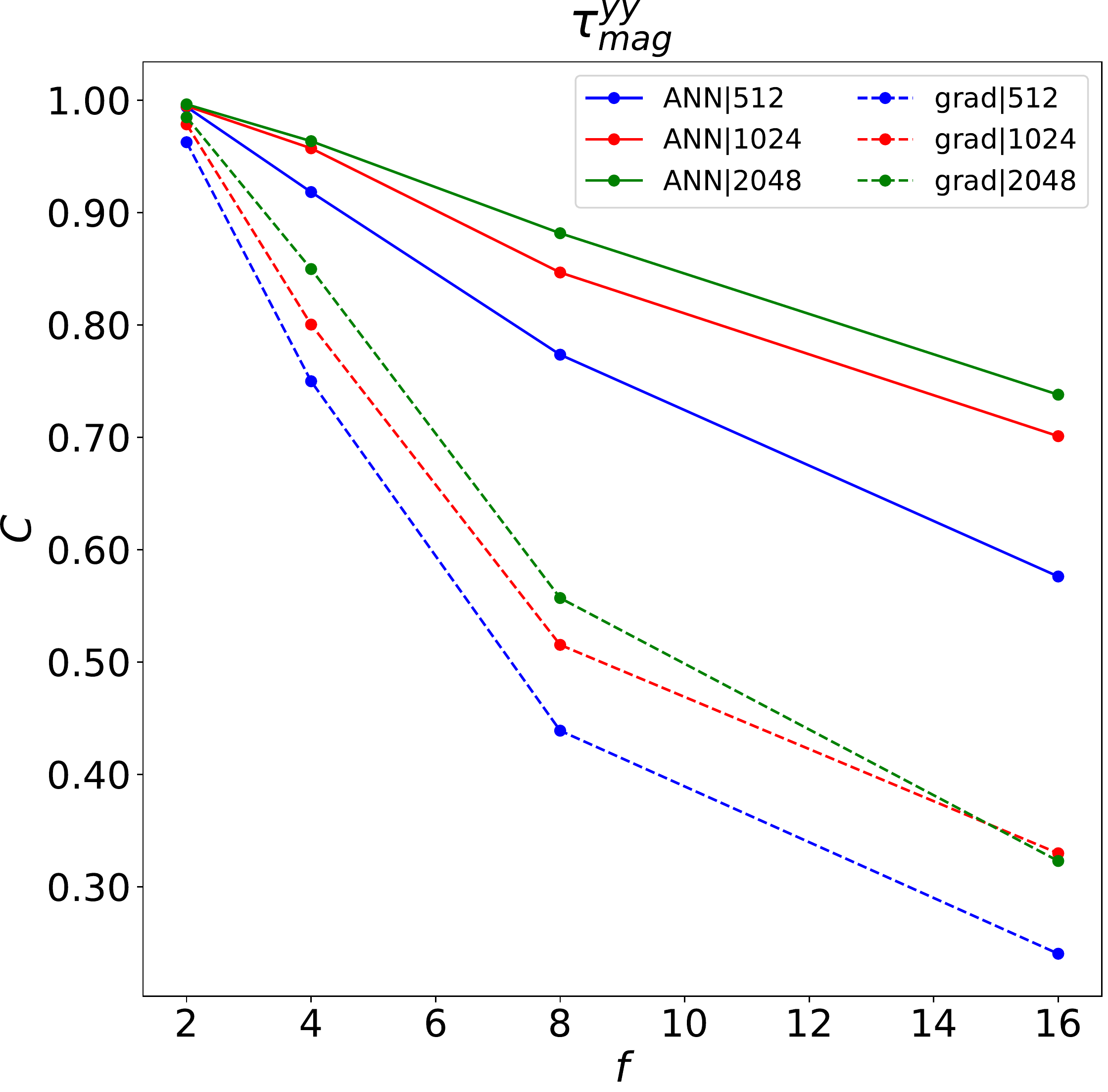}

\includegraphics[width=0.33\linewidth]{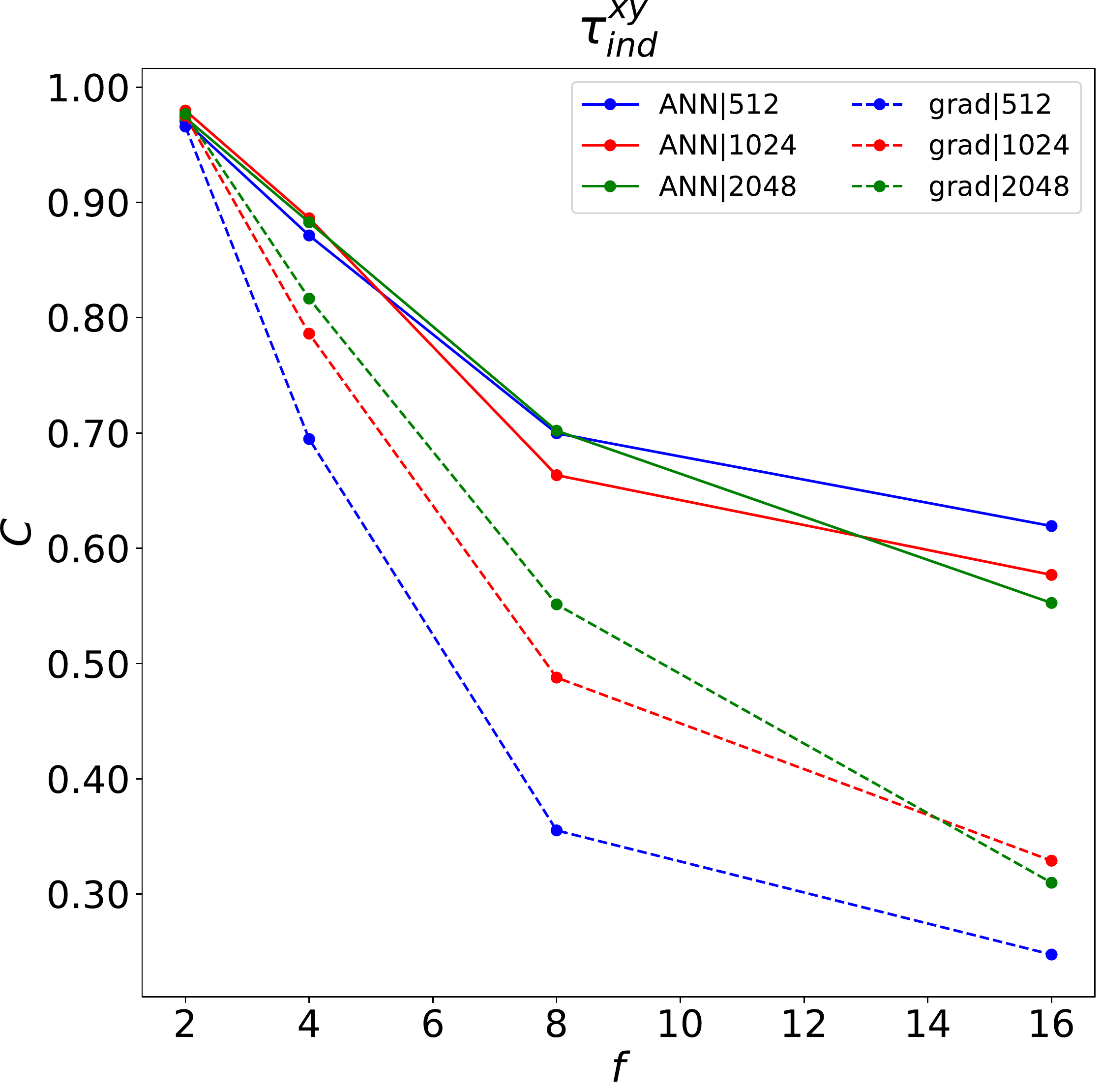}
\includegraphics[width=0.33\linewidth]{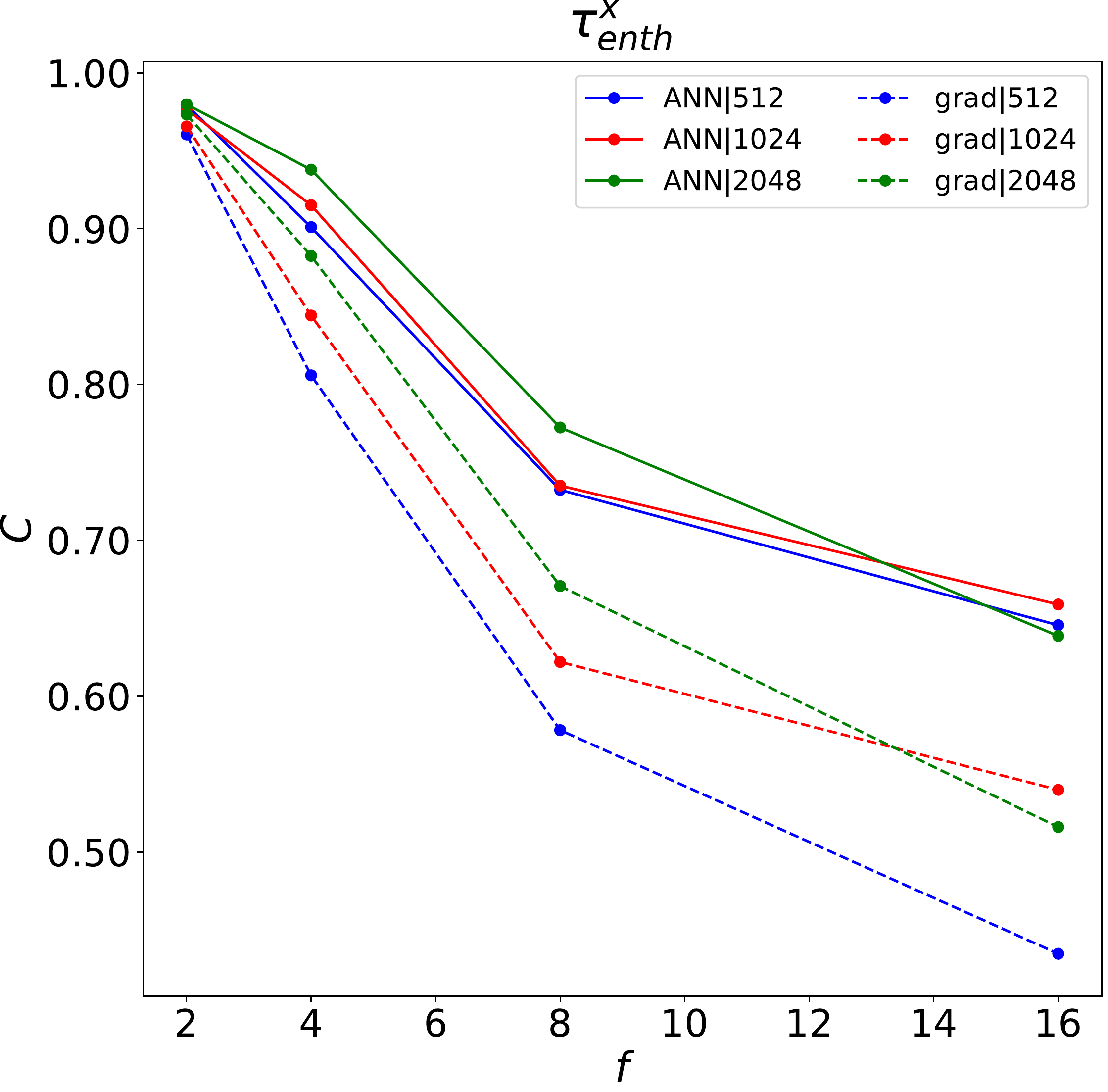}
\includegraphics[width=0.33\linewidth]{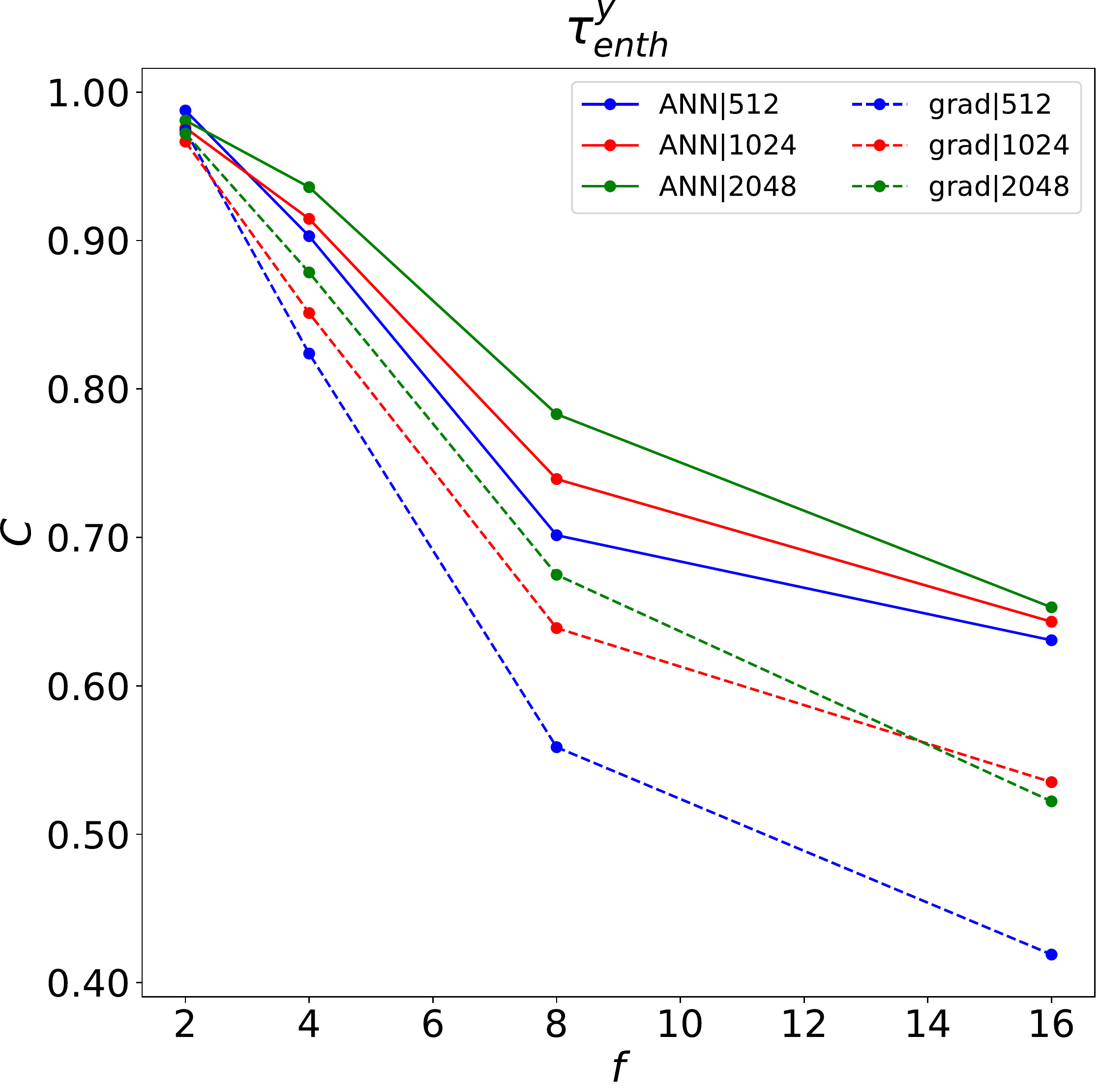}
\caption{Here we plot the correlation coefficient $C$ as a function of filter size $f$ for all resolutions $N$, SGS tensor components $\tau$, and SGS models.  The solid lines refer to the ANN model and the dashed lines refer to the gradient model.  The resolutions are given by the color of the line; blue represents the $N=512^2$ simulation, red represents the $N=1024^2$ simulation, and green represents the $N=2048^2$ simulation.  We observe that the ANN model has a higher correlation coefficient than the gradient model for all SGS tensor components at all resolutions.  We also note that $C$ generally decreases with increasing $f$, but this decay affects the gradient model more significantly.}
\label{fig:cor_coeff}
\end{figure*}

Having examined a specific SGS tensor qualitatively, we will now move towards a more general quantitative discussion of the behavior of the SGS tensors at different resolution and filter sizes for the models.  For this we will start by looking at the correlation coefficient $C$ presented in Fig.~\ref{fig:cor_coeff}.

In \Cref{fig:cor_coeff}, we show the plots of correlation coefficient $C$ vs the filter size $f$ for all resolutions simulated in this study.  Our findings show that all ANN models performed better than their gradient model counterparts for every SGS tensor component at the same $N$ and $f$.  The degree to which this improvement occurred was dependent primarily on the the filter size and the SGS tensor being analyzed.  The effect of the resolution is not entirely clear, but both models appear to follow similar trajectories on lines at the same resolution. In general, the value of $C$ decreased as $f$ increased.  This was particularly prevalent in the $\tau_{mag}$ and $\tau_{ind}$ tensors.  This decrease in $C$ for at high $f$ was much more significant in $\tau_{grad}$ than in $\tau_{ANN}$.  This indicates that $\tau_{ANN}$ performs better at higher filter sizes, implying that we would be able get accurate results from employing the ANN models at lower resolutions than we could from the gradient model.  

Moreover, the gradient model's difficulty calculating $\tau_{mag}$ and $\tau_{ind}$ at high filter sizes suggests that it is not able to reproduce the effects of turbulence on the magnetic fields at lower grid resolutions. In contrast, our results indicate that neural networks can address these limitations in an \textit{a posteriori} study. We observed that the purely hydrodynamical $\tau_{kin}$ SGS tensor was the easiest to compute accurately for both the ANN and gradient models.  Thus, the improvements in the ANN model's calculation of $\tau_{kin}$ should be considered less beneficial than those from $\tau_{mag}$ and $\tau_{ind}$. 

The energy SGS tensor $\tau_{enth}$ also receives a noticeable improvement from the use of the ANN model over the gradient model.  This effect is again most prevalent at high $f$ values, more than for $\tau_{kin}$ but not quite as significant as the $\tau_{mag}$ or $\tau_{ind}$ terms. We should note again that the gradient model is a leading order expansion of the filtering operator in grid spacing, which corresponds to filter size $f$.  If a higher order expansion of the filtering operator, we may see some improvement at high $f$.

\begin{figure*}
\centering
\includegraphics[width=0.33\linewidth]{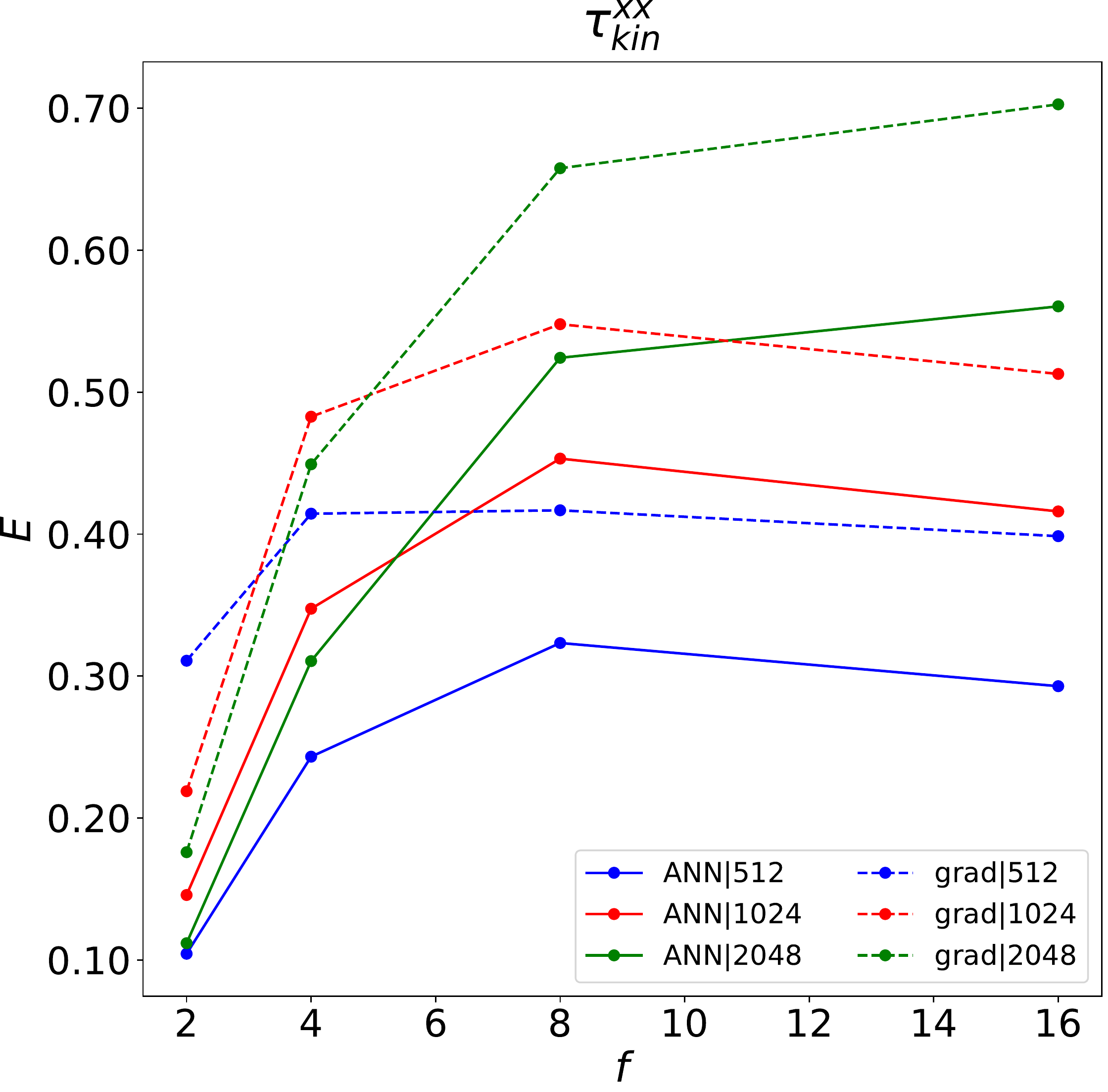}
\includegraphics[width=0.33\linewidth]{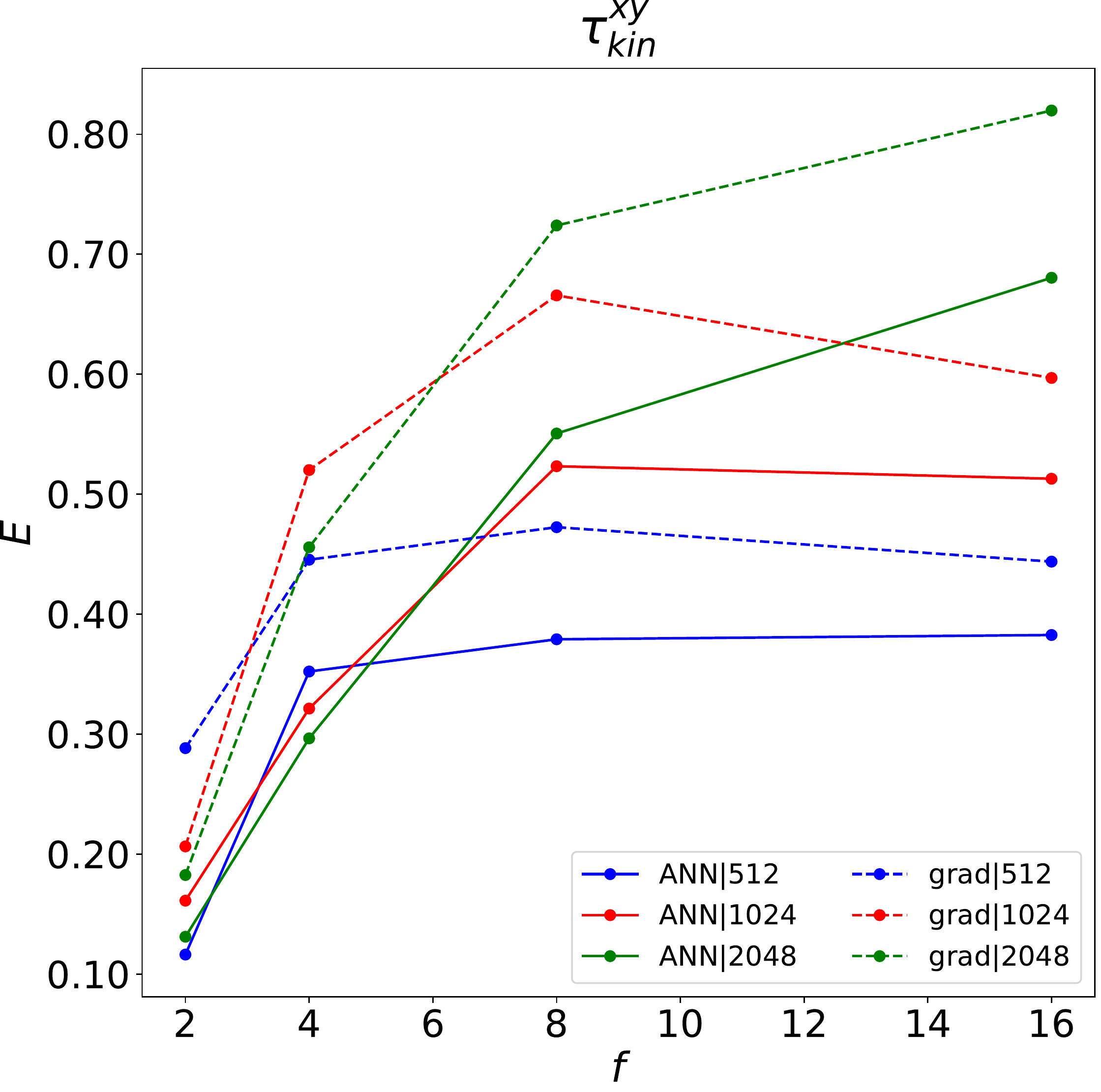}
\includegraphics[width=0.33\linewidth]{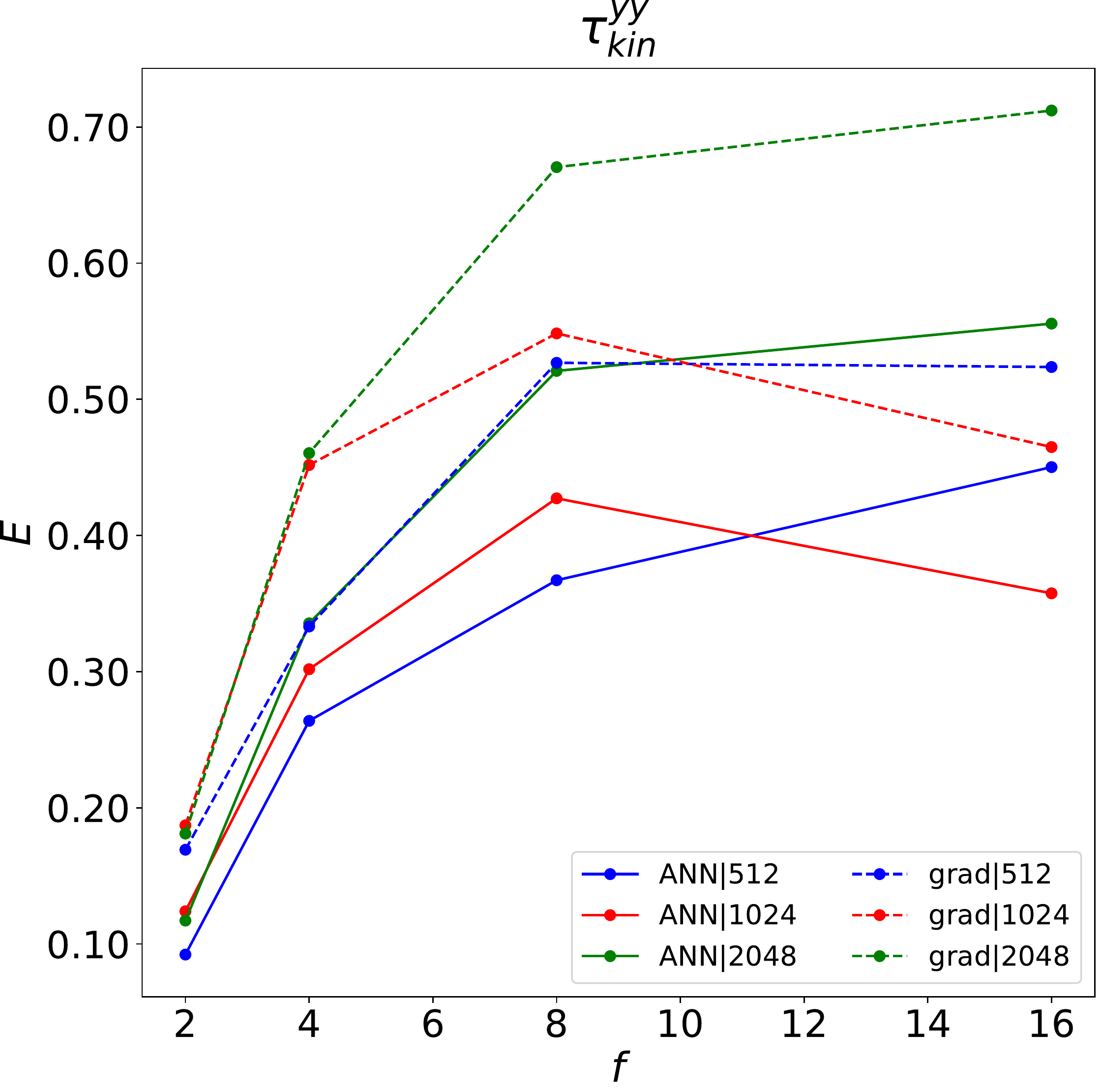}

\includegraphics[width=0.33\linewidth]{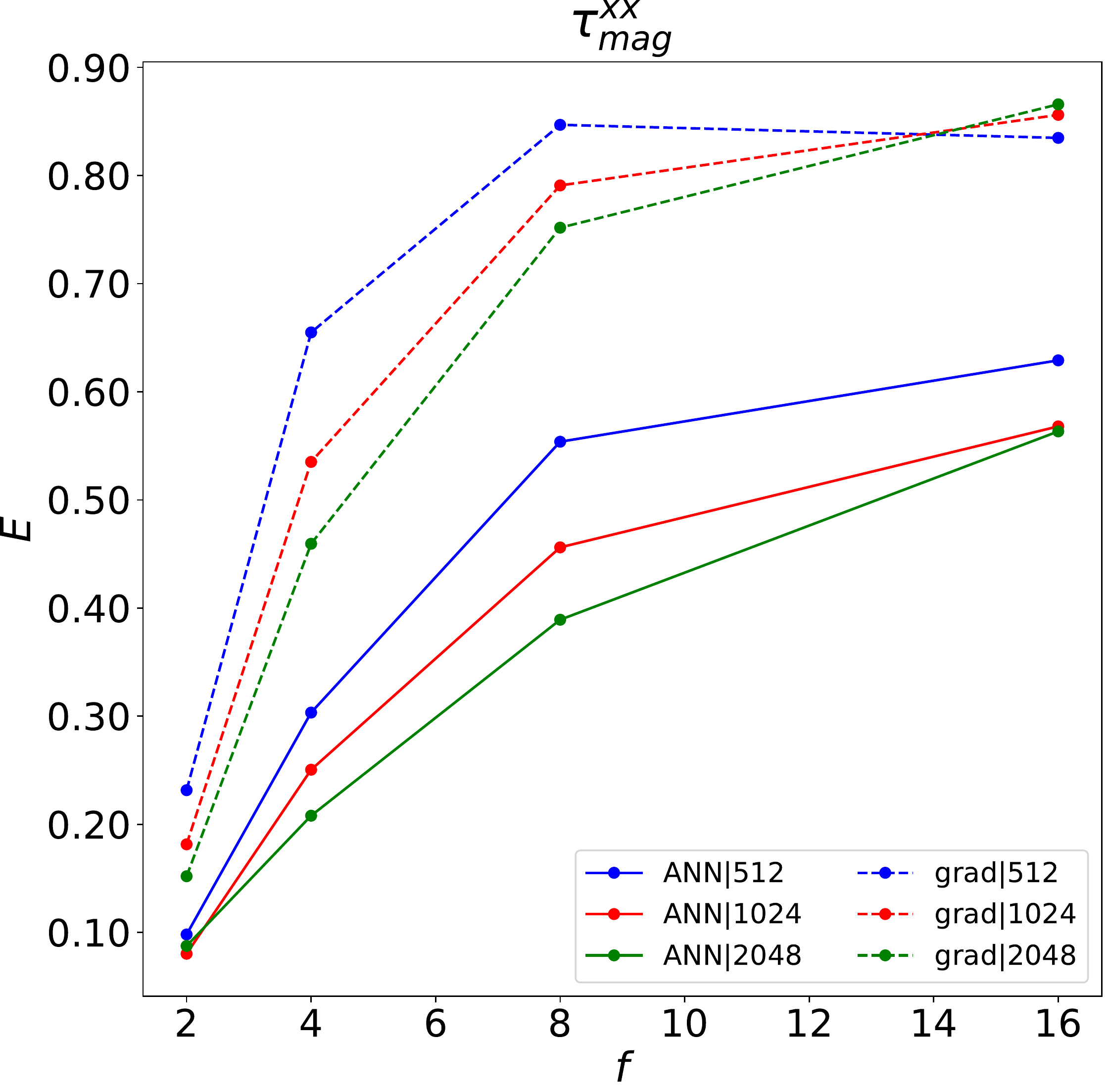}
\includegraphics[width=0.33\linewidth]{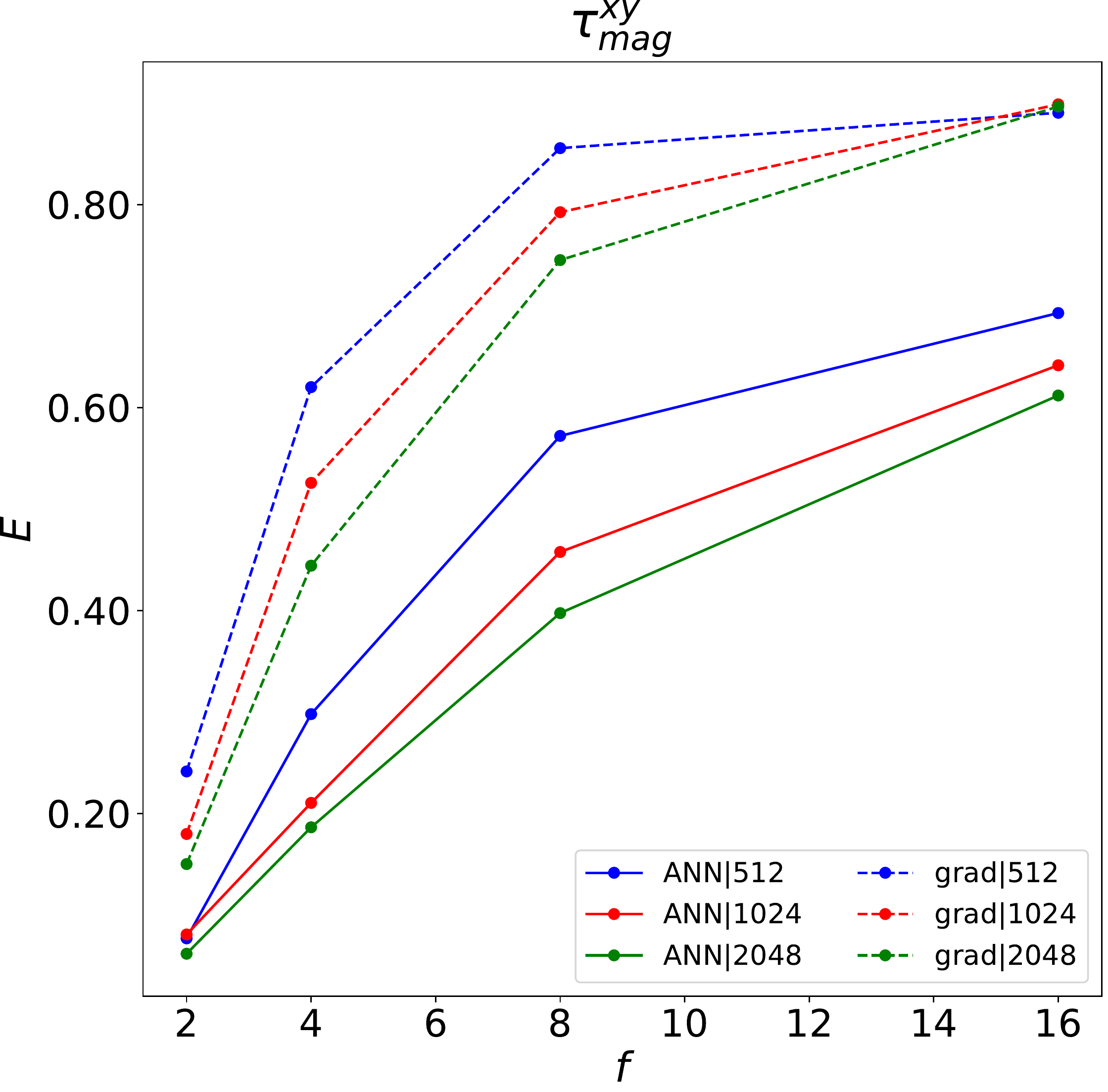}
\includegraphics[width=0.33\linewidth]{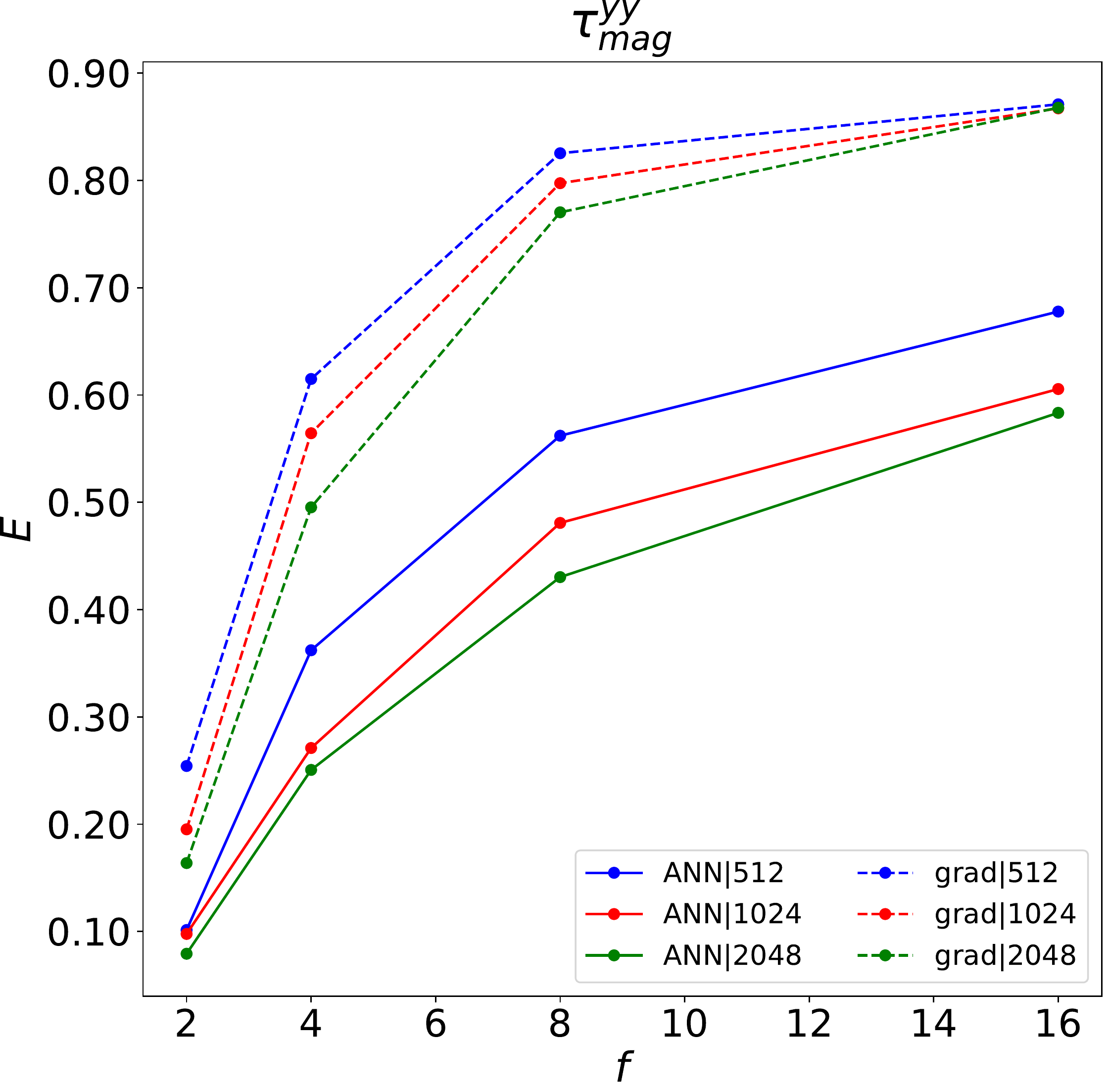}

\includegraphics[width=0.33\linewidth]{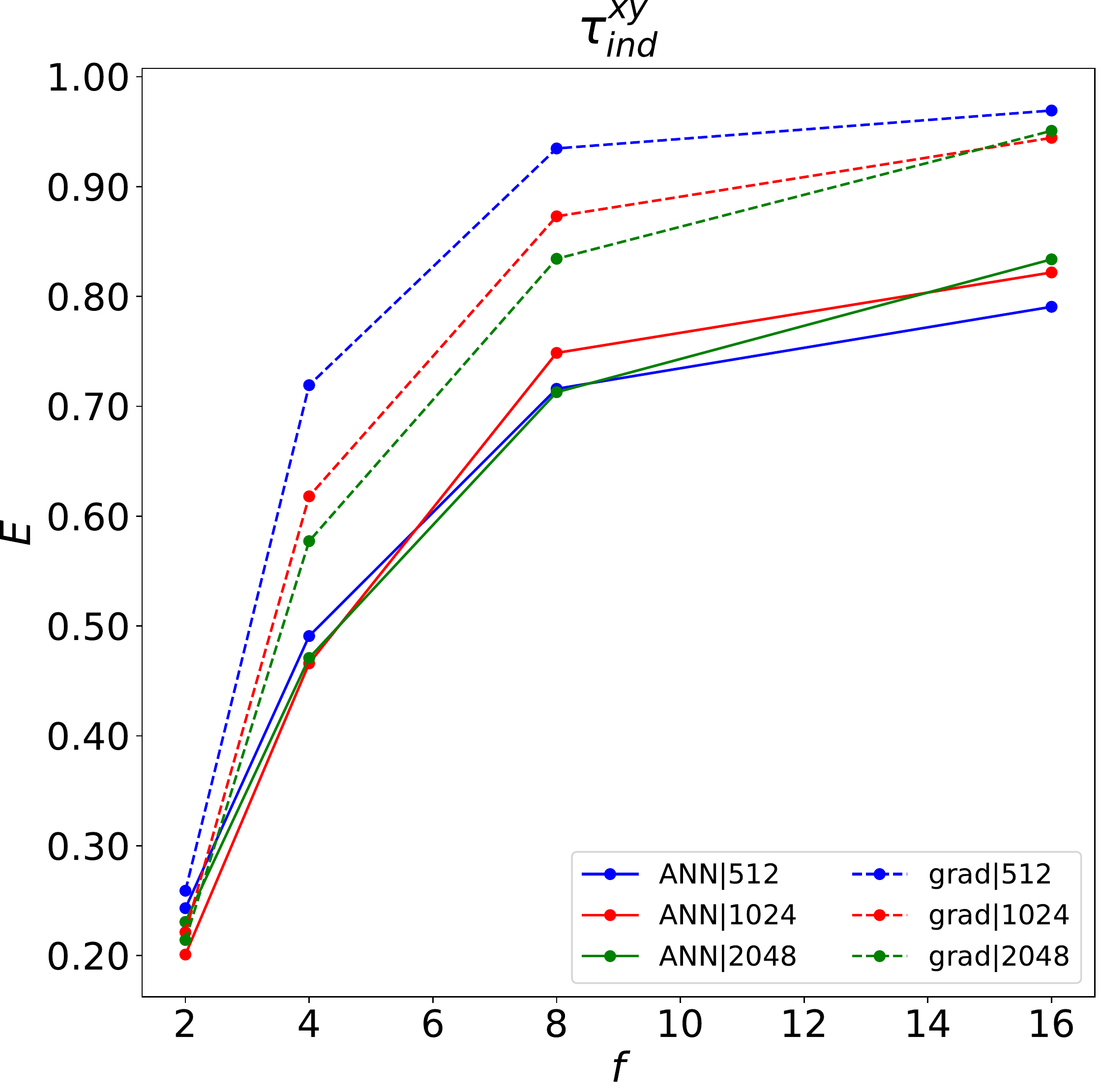}
\includegraphics[width=0.33\linewidth]{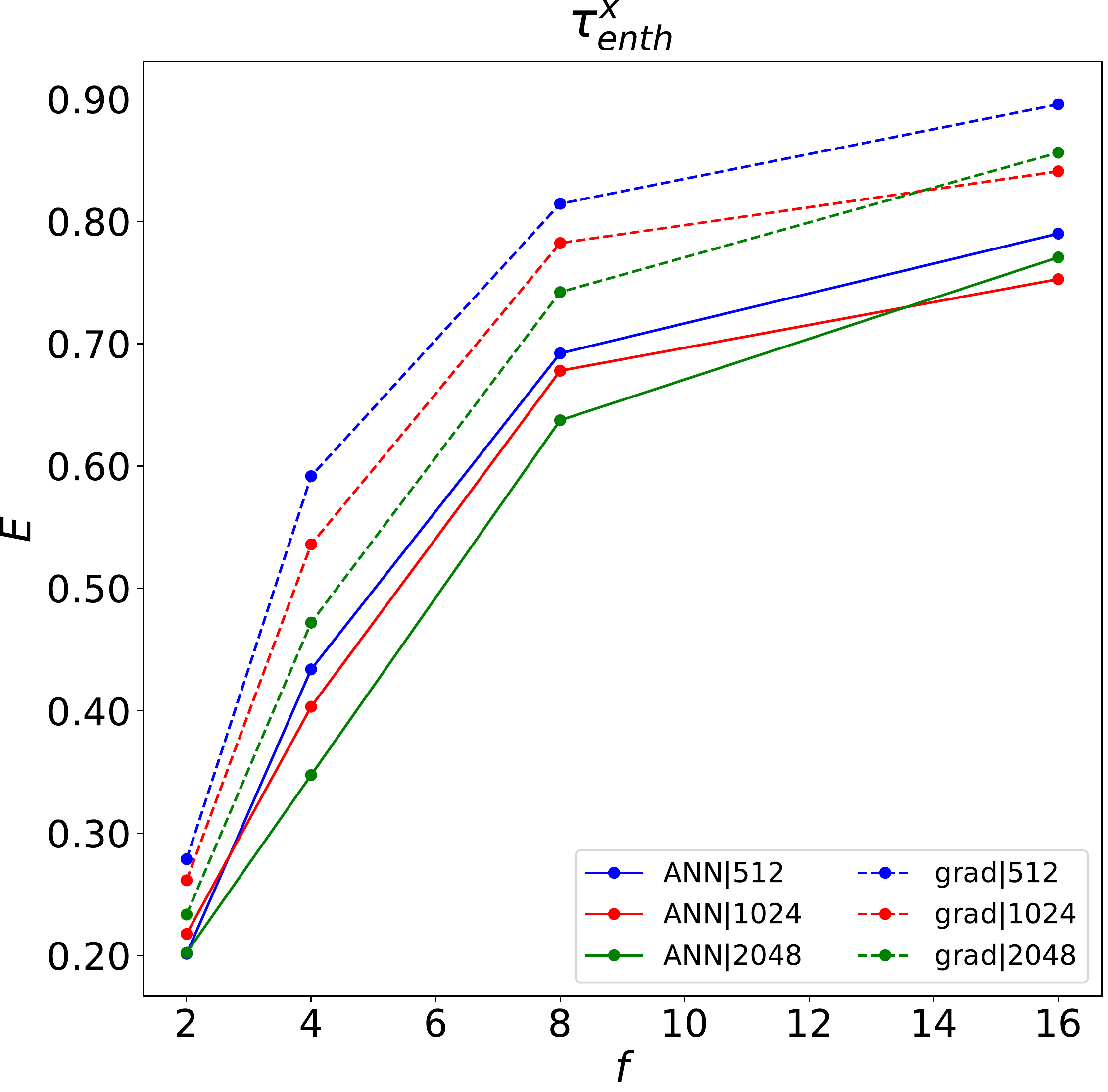}
\includegraphics[width=0.33\linewidth]{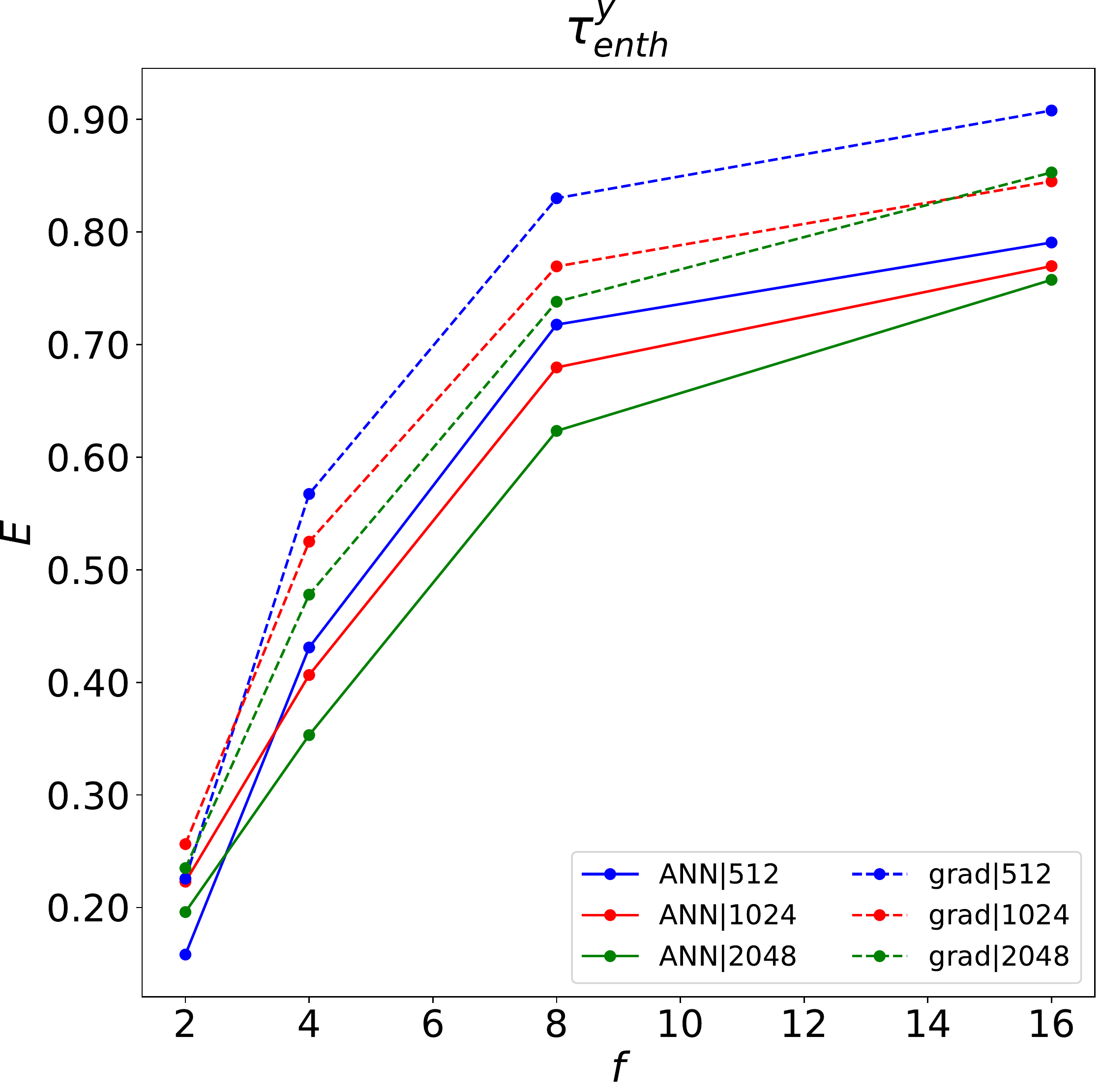}
\caption{Here we plot the relative error $E$ as a function of filter size $f$ for all resolutions $N$, SGS tensor components $\tau$, and SGS models.  The solid lines refer to the ANN model and the dashed lines refer to the gradient model.  The resolutions are given by the color of the line; blue represents the $N=512^2$ simulation, red represents the $N=1024^2$ simulation, and green represents the $N=2048^2$ simulation.  We observe that $E$ is lower for the ANN model than for the gradient model for all SGS tensor components at all resolutions.  We also note that $E$ generally increases with $f$, but is more severe for the gradient model than for the ANN model.}
\label{fig:relative_error}
\end{figure*}

\Cref{fig:relative_error} shows the relative error $E$ between the predictions of the gradient and ANN models compared to the DNS data.  The results mirror those discussed for the correlation coefficient $C$ in terms of $E$ increasing with $f$ more quickly for the gradient model compared to the ANN model. 

\begin{figure*}
\centering
\includegraphics[width=0.33\linewidth]{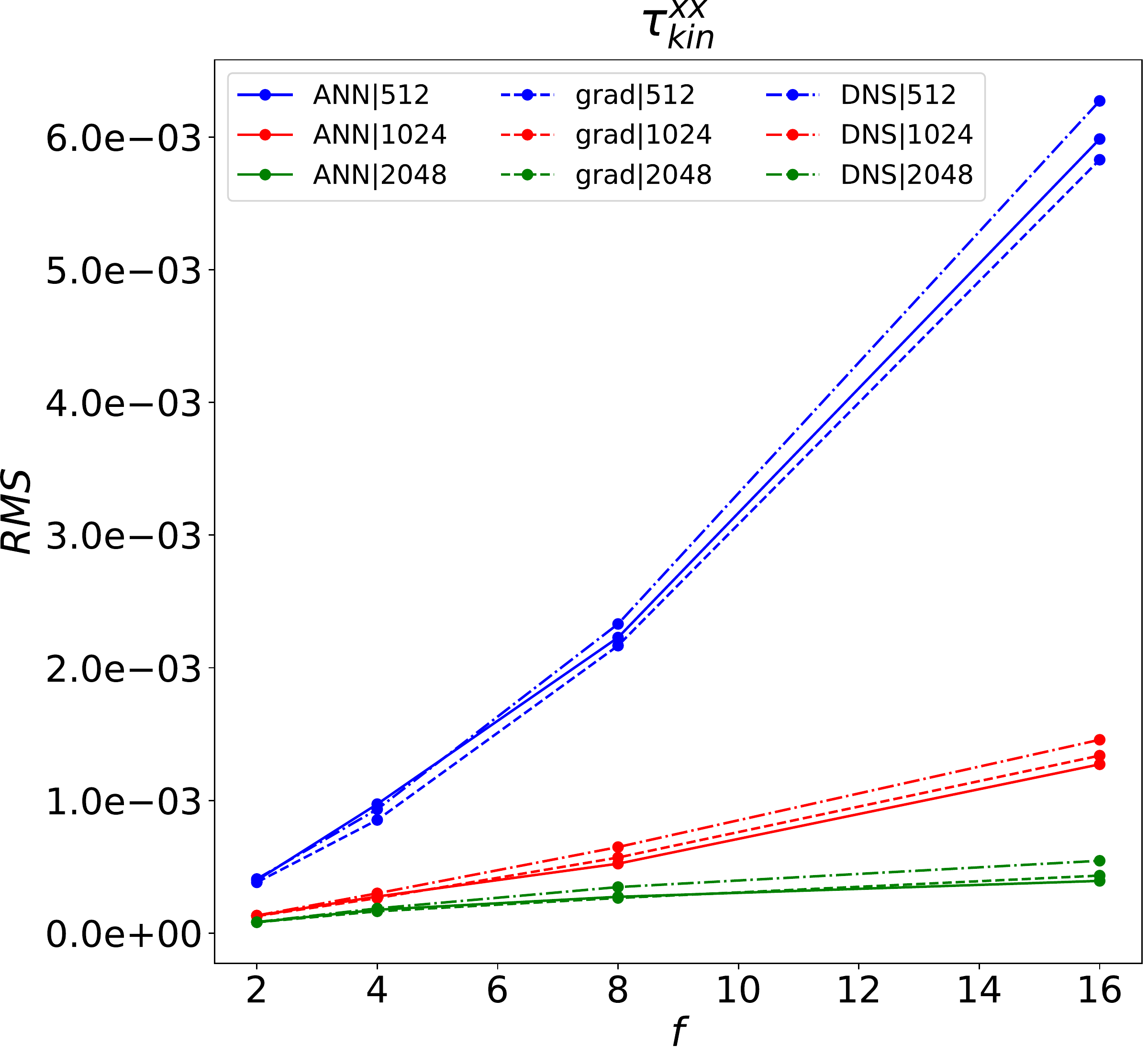}
\includegraphics[width=0.33\linewidth]{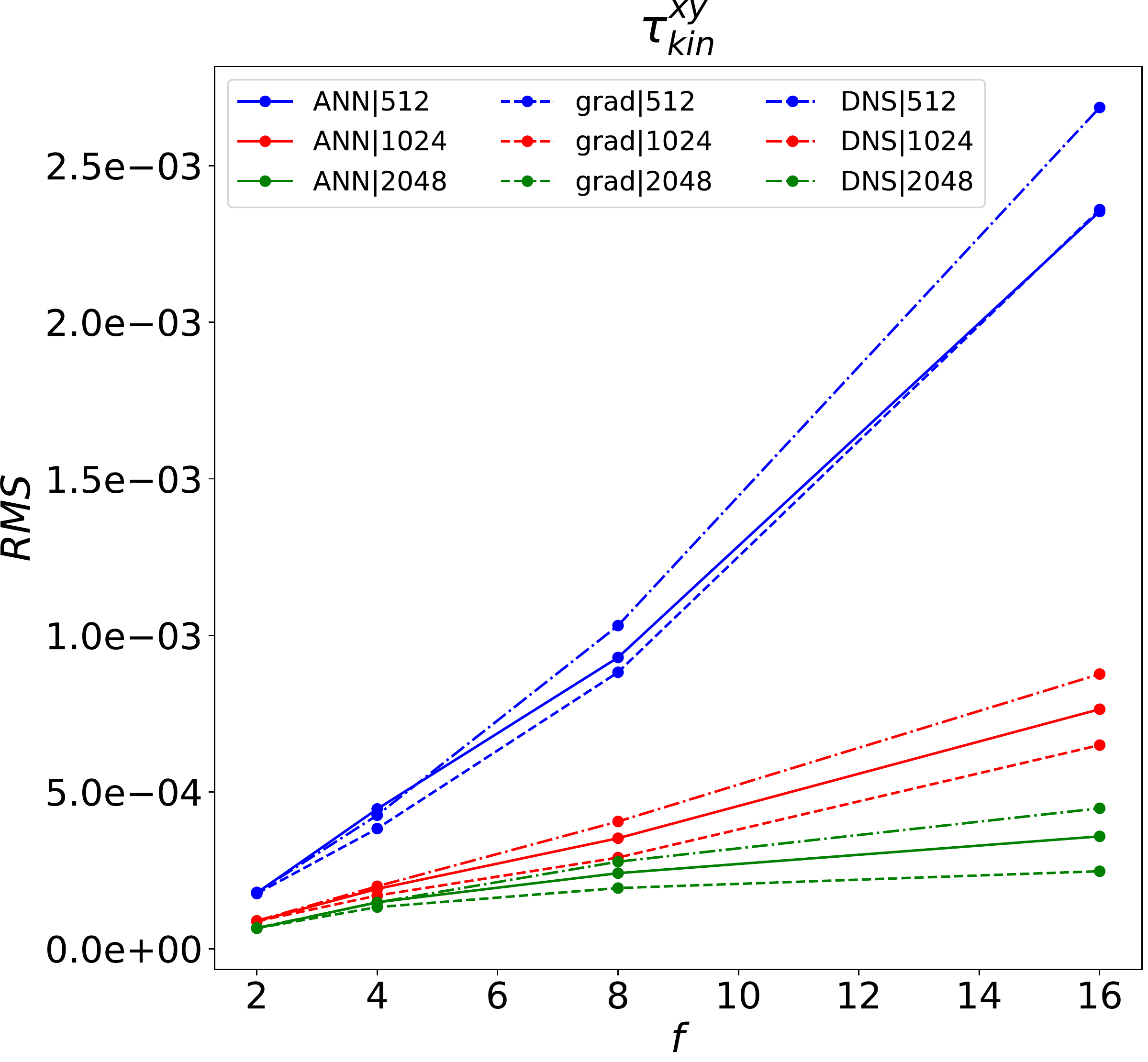}
\includegraphics[width=0.33\linewidth]{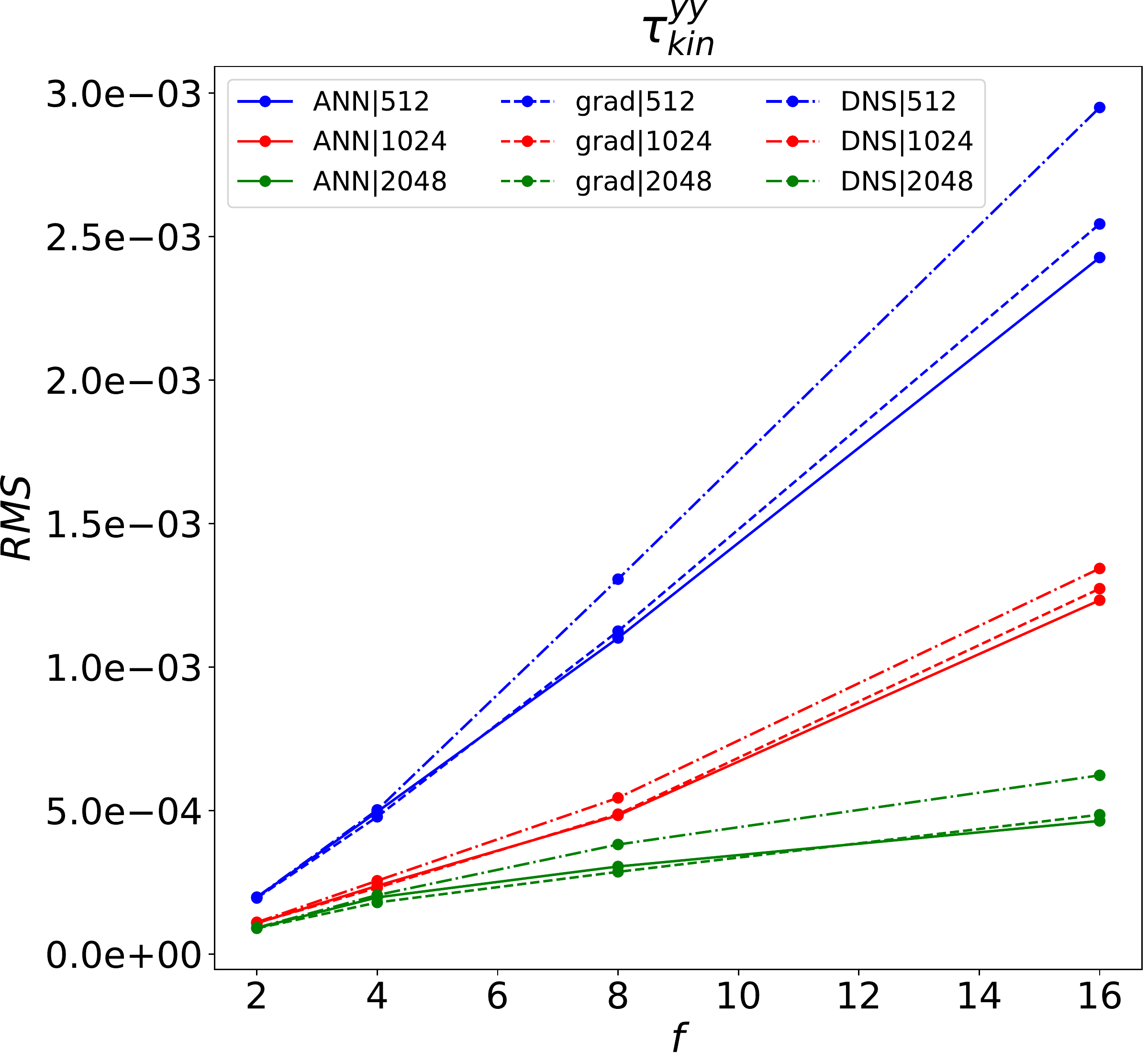}

\includegraphics[width=0.33\linewidth]{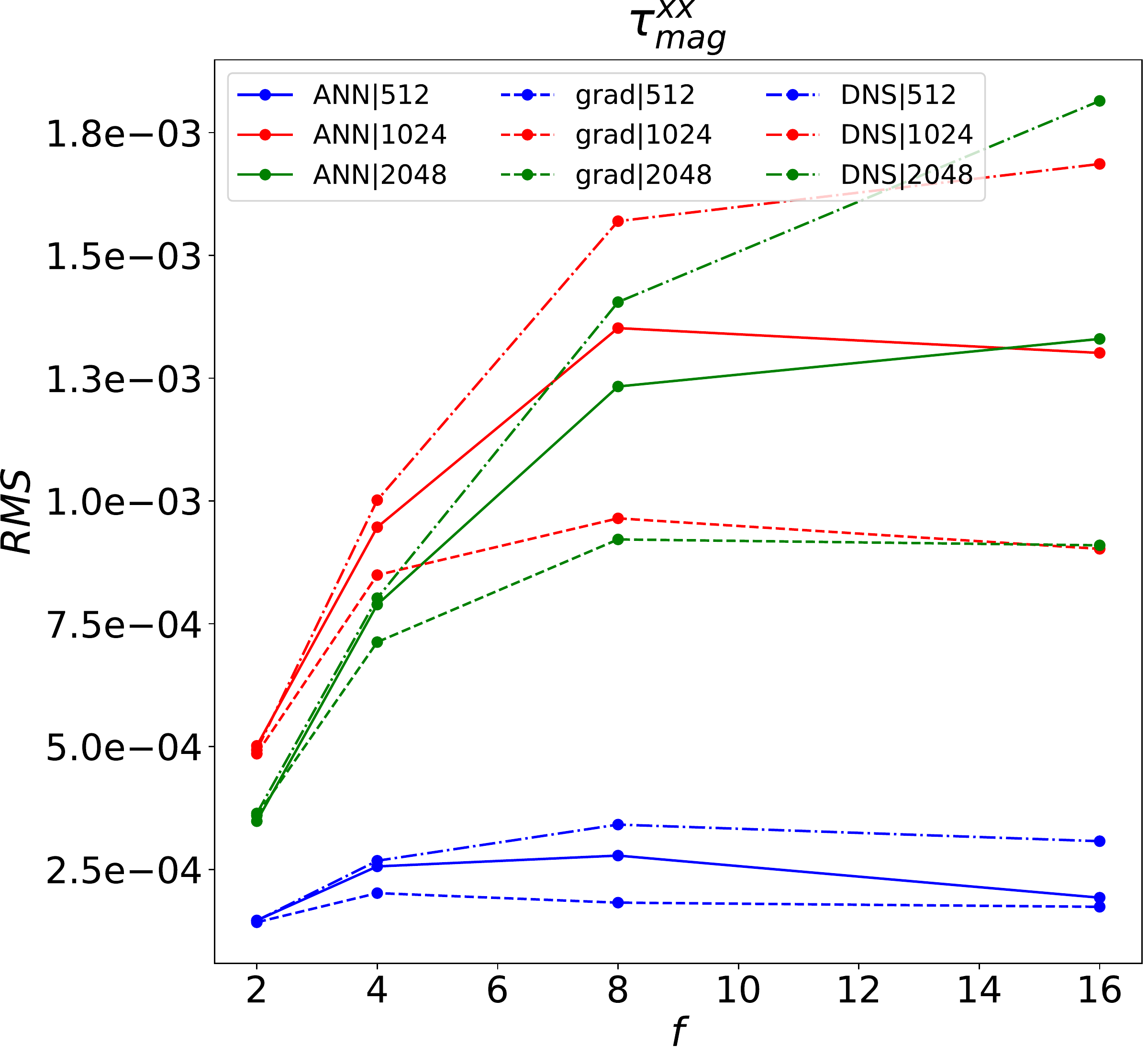}
\includegraphics[width=0.33\linewidth]{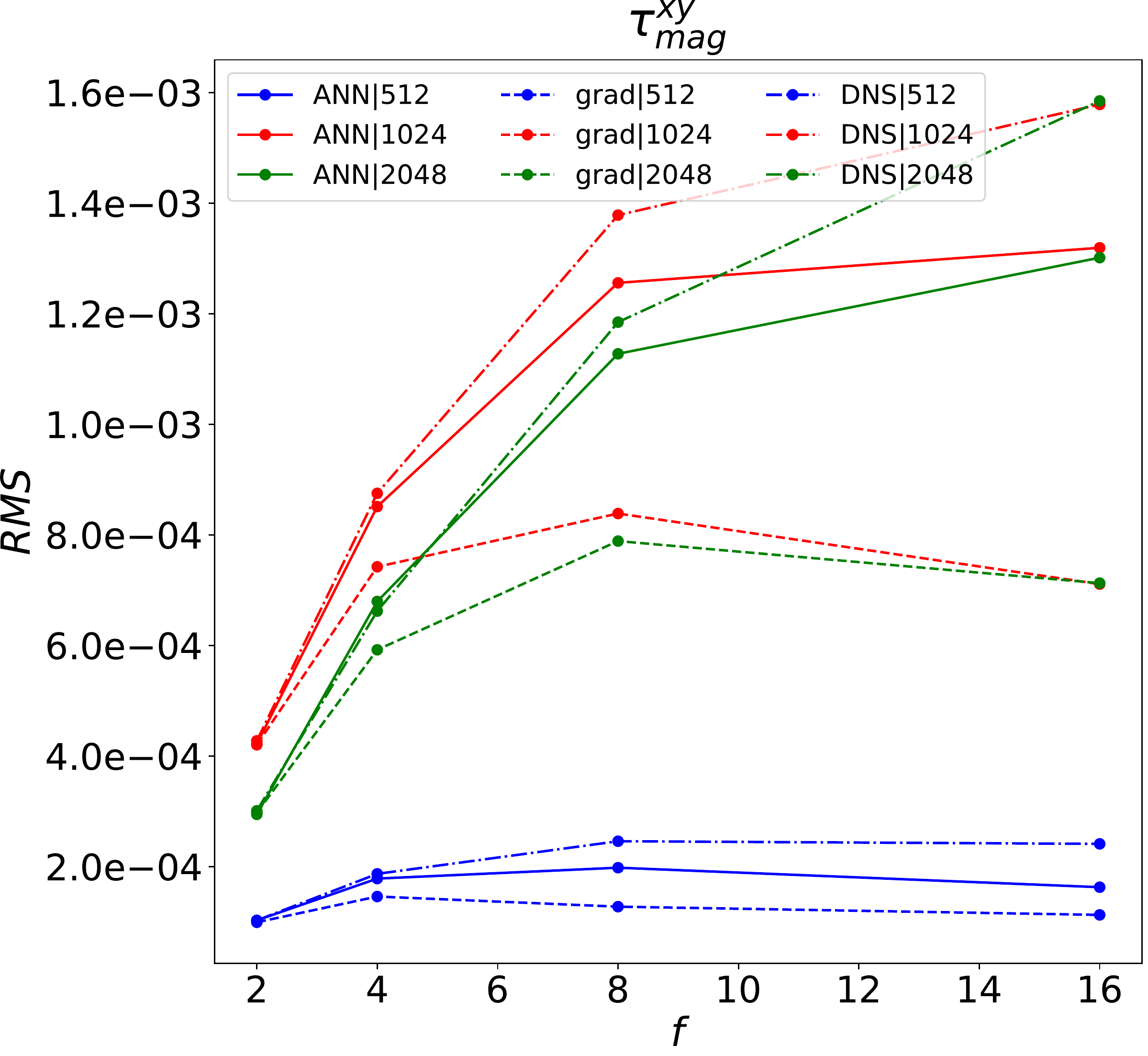}
\includegraphics[width=0.33\linewidth]{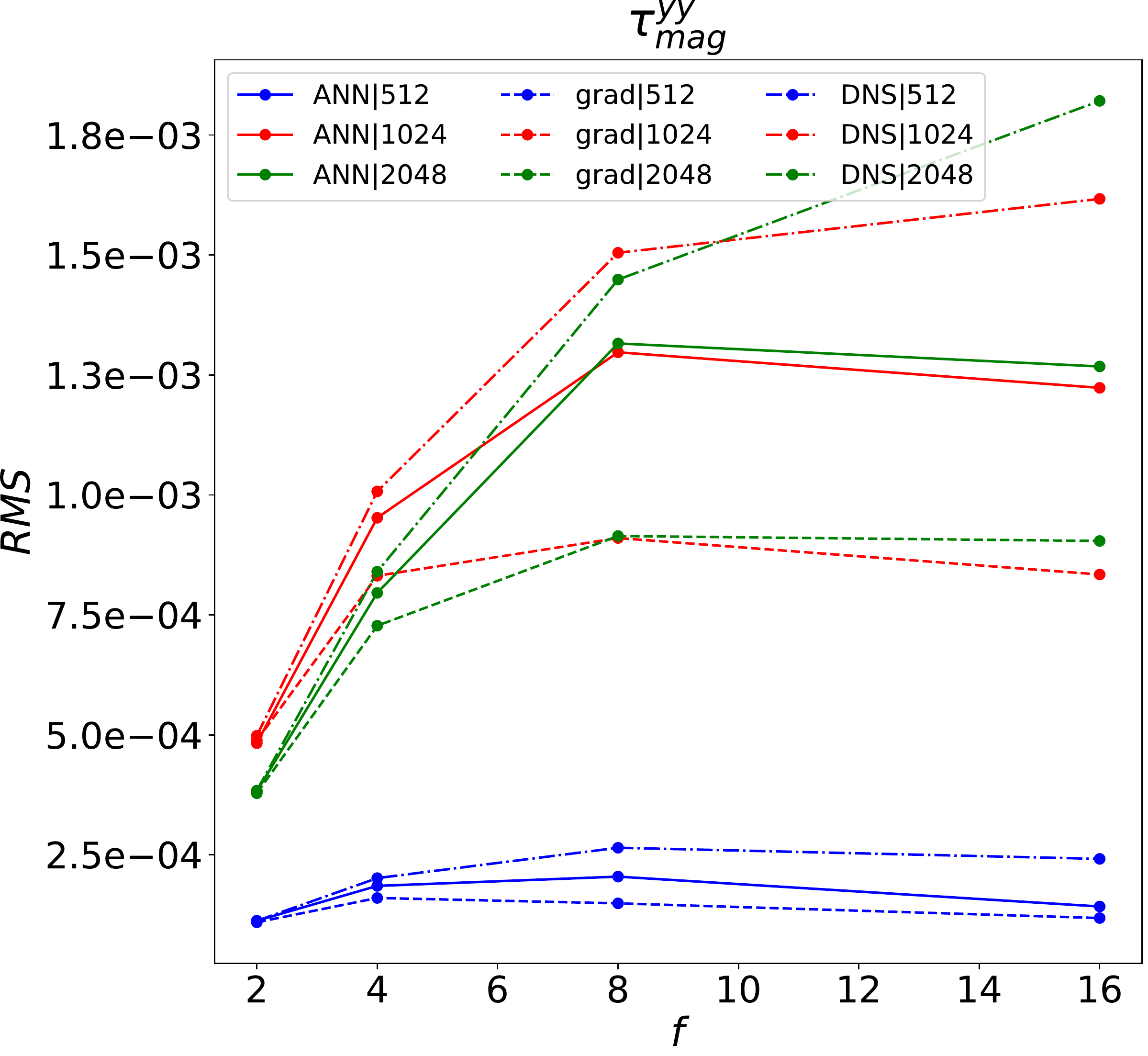}

\includegraphics[width=0.33\linewidth]{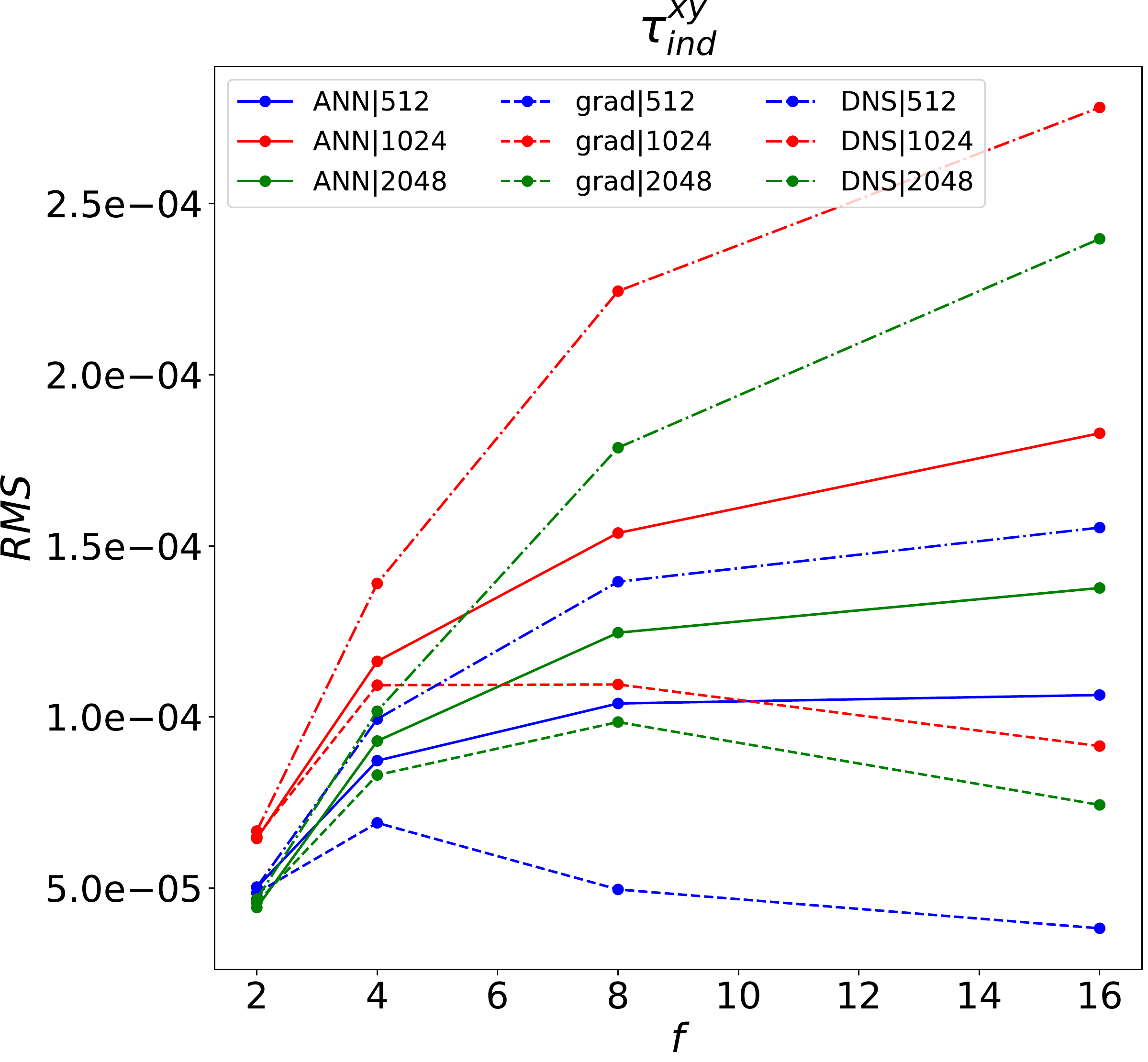}
\includegraphics[width=0.33\linewidth]{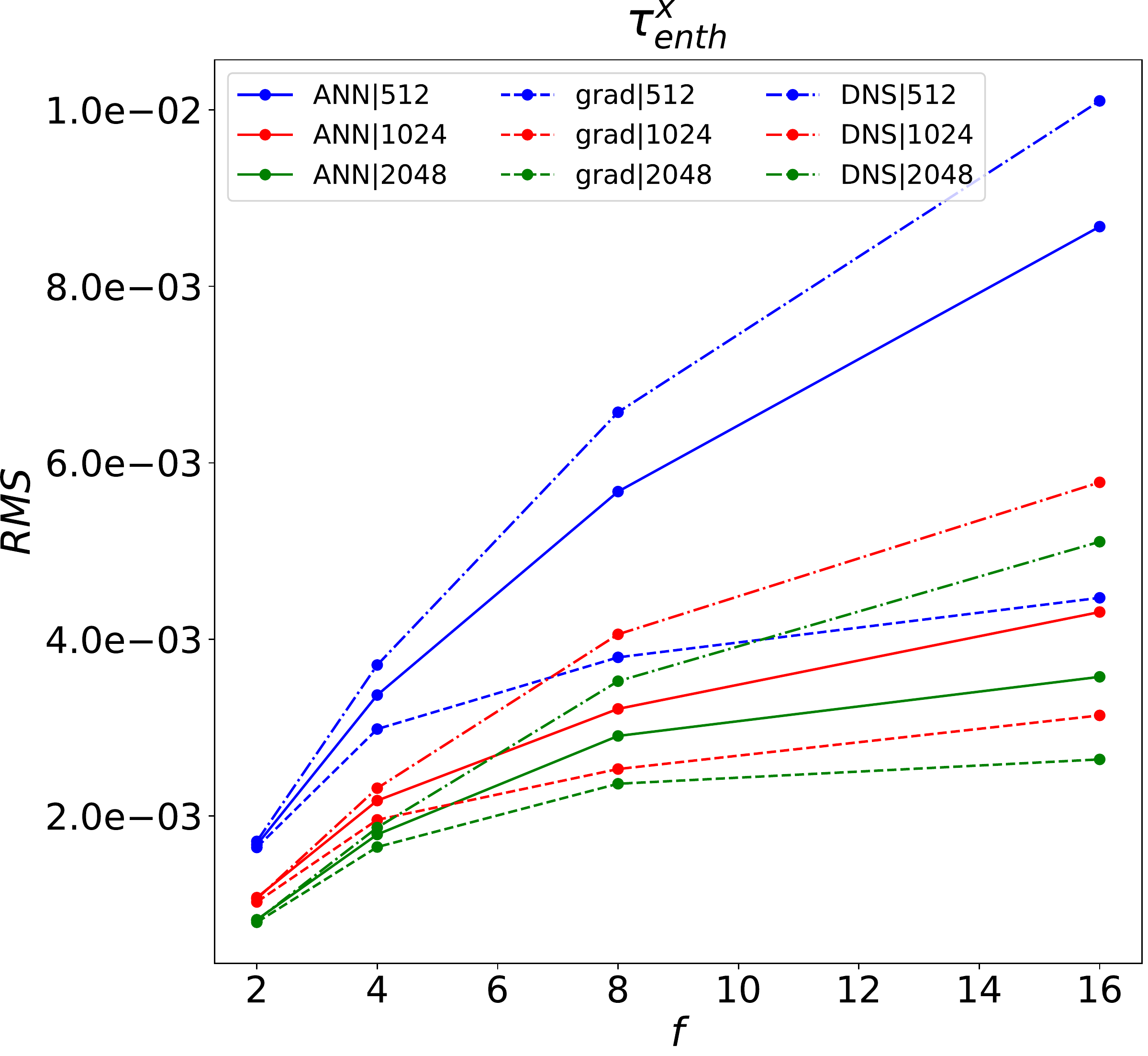}
\includegraphics[width=0.33\linewidth]{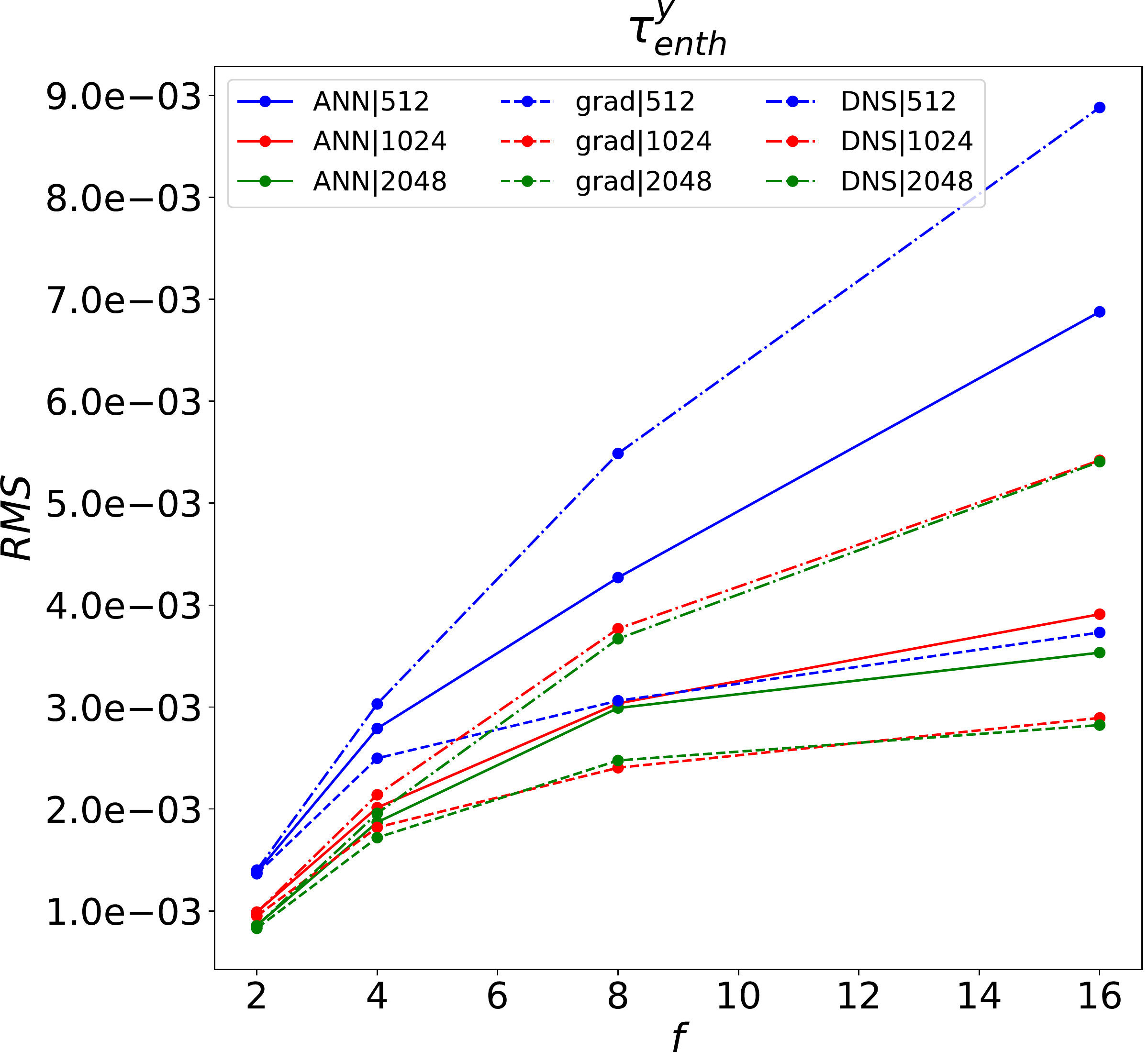}
\caption{Here we plot the $RMS$ value as a function of filter size $f$ for all resolutions $N$, SGS tensor components $\tau$, and SGS models including the DNS calculation.  The solid lines refer to the ANN model, the dashed lines refer to the gradient model, and the dash-dotted lines refer to the DNS calculation.  The resolutions are given by the color of the line; blue represents the $N=512^2$ simulation, red represents the $N=1024^2$ simulation, and green represents the $N=2048^2$ simulation.  We observe that the $RMS$ values are fairly similar for the $\tau_{kin}$ SGS tensor components (top row) for both SGS models and the DNS result.  For the SGS tensors $\tau_{mag}$ (middle row), $\tau_{ind}$ (bottom right), and $\tau_{enth}$ (bottom middle and left), the ANN model has a much closer $RMS$ value to the DNS result compared to the gradient model.}
\label{fig:rms}
\end{figure*}

In \Cref{fig:rms}, we show the $RMS$ of $\tau_{DNS}$, $\tau_{ANN}$, and $\tau_{grad}$ for all SGS tensor components at all resolutions.  For $\tau_{kin}$ we observe that the two models perform similarly in term of their proximity to the $RMS$ of $\tau_{DNS}$ and both slightly undershoot the true value for this SGS tensor.  We also notice that the value of the $RMS$ increases with filter size $f$, while the performance of both models decrease slightly for this metric at high $f$ for $\tau_{kin}$.  This makes sense as one would expect more SGS behavior at high $f$ as the grid increases we have more SGS phenomena, resulting in the SGS tensors being more difficult to model.  The $RMS$ of the low resolution data is greater than that of the high resolution simulations for $\tau_{kin}$.  Moreover, this rise in $RMS$ appears to be polynomial in $f$ and occurs much more prevalently at lower resolutions.  This would imply that there does not exist a preferred scale for $\tau_{kin}$ as the $RMS$ value appears the increase with the volume of the grid.  

The $RMS$ plots of $\tau_{mag}$ on the other hand demonstrate clear differences in the behavior of the models.  For example, the $RMS$ of the ANN model is considerably higher than that of the gradient model for all components of $\tau_{mag}$.  In turn, the $RMS$ of $\tau_{ANN}$ is much closer than $\tau_{grad}$ to the $RMS$ of $\tau_{DNS}$, which is greater than either model.  The difference between the $RMS$ of $\tau_{model}$ and the $RMS$ of $\tau_{DNS}$ is greatest at high values of $f$ for both the ANN and gradient models with the difference between the $RMS$ of $\tau_{DNS}$ and $\tau_{grad}$ being much more severe.  We also observe the $N=512^2$ resolution runs have a significantly lower $RMS$ value than the others, implying that this resolution was too low to capture much of the SGS behavior of the magnetic field.  Meanwhile, the RMS values of the $N=1024^2$ and the $N=2048^2$ simulations is fairly similar, which is not surprising given that the value of $E_{mag}$ is fairly similar for the two resolutions at $t=9.25$, the timeslice of the training data.  The $RMS$ value increases with $f$ for $\tau_{mag}$, but this increase slows down at high $f$.  In contrast, we recall the increase in $RMS$ of $\tau_{kin}$ accelerates at high values of $f$.  This behavior when taken in conjunction with the significantly lower $RMS$ value of the lowest resolution run implies that $\tau_{mag}$ prefers to act on small scales.  

The plots of the $RMS$ values of $\tau_{ind}$ share many of the same characteristics as those of $\tau_{mag}$ in terms of the superior performance of the ANN model over the gradient model, the $RMS$ of the lowest resolution run having the lowest value, and the deceleration of the increase in $RMS$ at high $f$ value.  However, we would like to emphasize that the $RMS$ of the $N=1024^2$ run is clearly greater than the $N=2048^2$ run unlike $\tau_{mag}$ where their values were fairly similar.  This phenomenon likely results from the quick acceleration of the increase of $E_{mag}$ of the test data slice at $t=9.25$ that was observed in \Cref{fig:energy} which is evidenced in $\tau_{ind}$ representing the turbulent amplification of the magnetic field.  We also observe that the gradient model performs particularly poorly for this SGS tensor at high $f$ where the $RMS$ of $\tau_{grad}$ actually decreases despite the $RMS$ of $\tau_{DNS}$ actually increasing albeit at a slower rate.

The $RMS$ plots of $\tau_{enth}$ show the lowest resolution having the highest $RMS$ value, followed by the middle resolution, then the high resolution as in the plots of $\tau_{kin}$.  However, we still observe a slower increase of $RMS$ at high $f$ like $\tau_{mag}$ and $\tau_{ind}$.  The ANN model clearly models the $RMS$ of $\tau_{enth}$ more closely than the gradient model as well.  As with all the previously mentioned SGS tensor components, the performance of the models in computing $RMS$ decreases at high $f$.

\section{Conclusions}
\label{sec:conclusions}

We performed a \textit{a priori} study to evaluate the accuracy of ANN models of SGS ideal MHD turbulence with high resolution 2-D simulations of the magnetized KHI.  This is the first such study of ANNs in MHD turbulence.  We compared the performance of the model to the gradient model that has been proposed in similar studies of modeling MHD turbulence in the LES framework.

In this study, we showed that the ANN performs significantly better than gradient model in reproducing the SGS tensors compared to the gradient model.  This improvement occurred at all resolutions, for all SGS tensors, and filter sizes.  However, the degree of improvement varied considerably with the SGS tensor and filter size.

In particular, $\tau_{mag}$, representing the turbulent effect of the magnetic field on the motion of the field, and $\tau_{ind}$, representing the turbulent amplification of the magnetic field, are modeled much more accurately than with the gradient model.  This allows ANNs to provide a better model of the turbulent effects of the magnetic field than any model in the MHD turbulence literature.

Moreover, we demonstrated that the gradient model's performance falls off significantly at high filter sizes.  However, the ANN is able to maintain a much higher correlation coefficient at high filter sizes.  This implies that ANNs may be able to reproduce the effect of turbulence more accurately than gradient models.

Having established the potential of these ANN models of MHD turbulence in an \textit{a priori} study, there are various pathways of future study for the use of ANN models.  The most obvious of which is \textit{a posteriori} study, where we deploy these models in an actual simulation, and quantify how well the SGS models reproduce the spectra.  One may also consider evaluating these models for more computationally intensive 3-D simulations, and eventually general relativistic MHD.  This work will also require the development of loss functions that incorporate physical constraints such as rotational invariance. These studies will be pursued in the near future. 

\section*{Acknowledgments}
\label{sec:acknowledgments}
EAH gratefully acknowledges National Science Foundation (NSF) awards OAC-1931561 and OAC-1934757. We are grateful to NVIDIA for donating several Tesla P100 and V100 GPUs that we used for our analysis, and the NSF grants NSF-1550514, NSF-1659702 and TG-PHY160053. We thank the \href{http://gravity.ncsa.illinois.edu}{NCSA grAvIty Group} for useful feedback.

\bibliography{LES_references}

\begin{thebibliography}{39}%
\makeatletter
\providecommand \@ifxundefined [1]{%
 \@ifx{#1\undefined}
}%
\providecommand \@ifnum [1]{%
 \ifnum #1\expandafter \@firstoftwo
 \else \expandafter \@secondoftwo
 \fi
}%
\providecommand \@ifx [1]{%
 \ifx #1\expandafter \@firstoftwo
 \else \expandafter \@secondoftwo
 \fi
}%
\providecommand \natexlab [1]{#1}%
\providecommand \enquote  [1]{``#1''}%
\providecommand \bibnamefont  [1]{#1}%
\providecommand \bibfnamefont [1]{#1}%
\providecommand \citenamefont [1]{#1}%
\providecommand \href@noop [0]{\@secondoftwo}%
\providecommand \href [0]{\begingroup \@sanitize@url \@href}%
\providecommand \@href[1]{\@@startlink{#1}\@@href}%
\providecommand \@@href[1]{\endgroup#1\@@endlink}%
\providecommand \@sanitize@url [0]{\catcode `\\12\catcode `\$12\catcode
  `\&12\catcode `\#12\catcode `\^12\catcode `\_12\catcode `\%12\relax}%
\providecommand \@@startlink[1]{}%
\providecommand \@@endlink[0]{}%
\providecommand \url  [0]{\begingroup\@sanitize@url \@url }%
\providecommand \@url [1]{\endgroup\@href {#1}{\urlprefix }}%
\providecommand \urlprefix  [0]{URL }%
\providecommand \Eprint [0]{\href }%
\providecommand \doibase [0]{http://dx.doi.org/}%
\providecommand \selectlanguage [0]{\@gobble}%
\providecommand \bibinfo  [0]{\@secondoftwo}%
\providecommand \bibfield  [0]{\@secondoftwo}%
\providecommand \translation [1]{[#1]}%
\providecommand \BibitemOpen [0]{}%
\providecommand \bibitemStop [0]{}%
\providecommand \bibitemNoStop [0]{.\EOS\space}%
\providecommand \EOS [0]{\spacefactor3000\relax}%
\providecommand \BibitemShut  [1]{\csname bibitem#1\endcsname}%
\let\auto@bib@innerbib\@empty
\bibitem [{\citenamefont {Schmidt}(2015)}]{Schmidt2015}%
  \BibitemOpen
  \bibfield  {author} {\bibinfo {author} {\bibfnamefont {W.}~\bibnamefont
  {Schmidt}},\ }\href {\doibase 10.1007/lrca-2015-2} {\bibfield  {journal}
  {\bibinfo  {journal} {Living Reviews in Computational Astrophysics}\ }\textbf
  {\bibinfo {volume} {1}},\ \bibinfo {pages} {2} (\bibinfo {year}
  {2015})}\BibitemShut {NoStop}%
\bibitem [{\citenamefont {{Kiuchi}}\ \emph {et~al.}(2015)\citenamefont
  {{Kiuchi}}, \citenamefont {{Cerd{\'a}-Dur{\'a}n}}, \citenamefont {{Kyutoku}},
  \citenamefont {{Sekiguchi}},\ and\ \citenamefont {{Shibata}}}]{Kiuchi2015a}%
  \BibitemOpen
  \bibfield  {author} {\bibinfo {author} {\bibfnamefont {K.}~\bibnamefont
  {{Kiuchi}}}, \bibinfo {author} {\bibfnamefont {P.}~\bibnamefont
  {{Cerd{\'a}-Dur{\'a}n}}}, \bibinfo {author} {\bibfnamefont {K.}~\bibnamefont
  {{Kyutoku}}}, \bibinfo {author} {\bibfnamefont {Y.}~\bibnamefont
  {{Sekiguchi}}}, \ and\ \bibinfo {author} {\bibfnamefont {M.}~\bibnamefont
  {{Shibata}}},\ }\href {\doibase 10.1103/PhysRevD.92.124034} {\bibfield
  {journal} {\bibinfo  {journal} {Phys. Rev. D}\ }\textbf {\bibinfo {volume}
  {92}},\ \bibinfo {eid} {124034} (\bibinfo {year} {2015})},\ \Eprint
  {http://arxiv.org/abs/1509.09205} {arXiv:1509.09205 [astro-ph.HE]}
  \BibitemShut {NoStop}%
\bibitem [{\citenamefont {{Baiotti}}\ and\ \citenamefont
  {{Rezzolla}}(2017)}]{Baiotti2017}%
  \BibitemOpen
  \bibfield  {author} {\bibinfo {author} {\bibfnamefont {L.}~\bibnamefont
  {{Baiotti}}}\ and\ \bibinfo {author} {\bibfnamefont {L.}~\bibnamefont
  {{Rezzolla}}},\ }\href {\doibase 10.1088/1361-6633/aa67bb} {\bibfield
  {journal} {\bibinfo  {journal} {Reports on Progress in Physics}\ }\textbf
  {\bibinfo {volume} {80}},\ \bibinfo {eid} {096901} (\bibinfo {year}
  {2017})},\ \Eprint {http://arxiv.org/abs/1607.03540} {arXiv:1607.03540
  [gr-qc]} \BibitemShut {NoStop}%
\bibitem [{\citenamefont {Müller}\ and\ \citenamefont
  {Carati}(2002{\natexlab{a}})}]{Muller2002}%
  \BibitemOpen
  \bibfield  {author} {\bibinfo {author} {\bibfnamefont {W.-C.}\ \bibnamefont
  {Müller}}\ and\ \bibinfo {author} {\bibfnamefont {D.}~\bibnamefont
  {Carati}},\ }\href {\doibase 10.1063/1.1448498} {\bibfield  {journal}
  {\bibinfo  {journal} {Physics of Plasmas}\ }\textbf {\bibinfo {volume} {9}},\
  \bibinfo {pages} {824} (\bibinfo {year} {2002}{\natexlab{a}})},\ \Eprint
  {http://arxiv.org/abs/https://doi.org/10.1063/1.1448498}
  {https://doi.org/10.1063/1.1448498} \BibitemShut {NoStop}%
\bibitem [{\citenamefont {Müller}\ and\ \citenamefont
  {Carati}(2002{\natexlab{b}})}]{Muller2002a}%
  \BibitemOpen
  \bibfield  {author} {\bibinfo {author} {\bibfnamefont {W.-C.}\ \bibnamefont
  {Müller}}\ and\ \bibinfo {author} {\bibfnamefont {D.}~\bibnamefont
  {Carati}},\ }\href {\doibase https://doi.org/10.1016/S0010-4655(02)00341-7}
  {\bibfield  {journal} {\bibinfo  {journal} {Computer Physics Communications}\
  }\textbf {\bibinfo {volume} {147}},\ \bibinfo {pages} {544 } (\bibinfo {year}
  {2002}{\natexlab{b}})},\ \bibinfo {note} {proceedings of the Europhysics
  Conference on Computational Physics Computational Modeling and Simulation of
  Complex Systems}\BibitemShut {NoStop}%
\bibitem [{\citenamefont {Miesch}\ \emph {et~al.}(2015)\citenamefont {Miesch},
  \citenamefont {Matthaeus}, \citenamefont {Brandenburg}, \citenamefont
  {Petrosyan}, \citenamefont {Pouquet}, \citenamefont {Cambon}, \citenamefont
  {Jenko}, \citenamefont {Uzdensky}, \citenamefont {Stone}, \citenamefont
  {Tobias}, \citenamefont {Toomre},\ and\ \citenamefont {Velli}}]{Miesch2015}%
  \BibitemOpen
  \bibfield  {author} {\bibinfo {author} {\bibfnamefont {M.}~\bibnamefont
  {Miesch}}, \bibinfo {author} {\bibfnamefont {W.}~\bibnamefont {Matthaeus}},
  \bibinfo {author} {\bibfnamefont {A.}~\bibnamefont {Brandenburg}}, \bibinfo
  {author} {\bibfnamefont {A.}~\bibnamefont {Petrosyan}}, \bibinfo {author}
  {\bibfnamefont {A.}~\bibnamefont {Pouquet}}, \bibinfo {author} {\bibfnamefont
  {C.}~\bibnamefont {Cambon}}, \bibinfo {author} {\bibfnamefont
  {F.}~\bibnamefont {Jenko}}, \bibinfo {author} {\bibfnamefont
  {D.}~\bibnamefont {Uzdensky}}, \bibinfo {author} {\bibfnamefont
  {J.}~\bibnamefont {Stone}}, \bibinfo {author} {\bibfnamefont
  {S.}~\bibnamefont {Tobias}}, \bibinfo {author} {\bibfnamefont
  {J.}~\bibnamefont {Toomre}}, \ and\ \bibinfo {author} {\bibfnamefont
  {M.}~\bibnamefont {Velli}},\ }\href {\doibase 10.1007/s11214-015-0190-7}
  {\bibfield  {journal} {\bibinfo  {journal} {Space Science Reviews}\ }\textbf
  {\bibinfo {volume} {194}},\ \bibinfo {pages} {97} (\bibinfo {year}
  {2015})}\BibitemShut {NoStop}%
\bibitem [{\citenamefont {Grete}\ \emph {et~al.}(2015)\citenamefont {Grete},
  \citenamefont {Vlaykov}, \citenamefont {Schmidt}, \citenamefont
  {Schleicher},\ and\ \citenamefont {Federrath}}]{Grete2015}%
  \BibitemOpen
  \bibfield  {author} {\bibinfo {author} {\bibfnamefont {P.}~\bibnamefont
  {Grete}}, \bibinfo {author} {\bibfnamefont {D.~G.}\ \bibnamefont {Vlaykov}},
  \bibinfo {author} {\bibfnamefont {W.}~\bibnamefont {Schmidt}}, \bibinfo
  {author} {\bibfnamefont {D.~R.~G.}\ \bibnamefont {Schleicher}}, \ and\
  \bibinfo {author} {\bibfnamefont {C.}~\bibnamefont {Federrath}},\ }\href
  {\doibase 10.1088/1367-2630/17/2/023070} {\bibfield  {journal} {\bibinfo
  {journal} {New Journal of Physics}\ }\textbf {\bibinfo {volume} {17}},\
  \bibinfo {pages} {023070} (\bibinfo {year} {2015})}\BibitemShut {NoStop}%
\bibitem [{\citenamefont {Grete}\ \emph {et~al.}(2016)\citenamefont {Grete},
  \citenamefont {Vlaykov}, \citenamefont {Schmidt},\ and\ \citenamefont
  {Schleicher}}]{Grete2016}%
  \BibitemOpen
  \bibfield  {author} {\bibinfo {author} {\bibfnamefont {P.}~\bibnamefont
  {Grete}}, \bibinfo {author} {\bibfnamefont {D.~G.}\ \bibnamefont {Vlaykov}},
  \bibinfo {author} {\bibfnamefont {W.}~\bibnamefont {Schmidt}}, \ and\
  \bibinfo {author} {\bibfnamefont {D.~R.~G.}\ \bibnamefont {Schleicher}},\
  }\href {\doibase 10.1063/1.4954304} {\bibfield  {journal} {\bibinfo
  {journal} {Physics of Plasmas}\ }\textbf {\bibinfo {volume} {23}},\ \bibinfo
  {pages} {062317} (\bibinfo {year} {2016})},\ \Eprint
  {http://arxiv.org/abs/https://doi.org/10.1063/1.4954304}
  {https://doi.org/10.1063/1.4954304} \BibitemShut {NoStop}%
\bibitem [{\citenamefont {Grete}\ \emph
  {et~al.}(2017{\natexlab{a}})\citenamefont {Grete}, \citenamefont {Vlaykov},
  \citenamefont {Schmidt},\ and\ \citenamefont {Schleicher}}]{Grete2017a}%
  \BibitemOpen
  \bibfield  {author} {\bibinfo {author} {\bibfnamefont {P.}~\bibnamefont
  {Grete}}, \bibinfo {author} {\bibfnamefont {D.~G.}\ \bibnamefont {Vlaykov}},
  \bibinfo {author} {\bibfnamefont {W.}~\bibnamefont {Schmidt}}, \ and\
  \bibinfo {author} {\bibfnamefont {D.~R.~G.}\ \bibnamefont {Schleicher}},\
  }\href {\doibase 10.1103/PhysRevE.95.033206} {\bibfield  {journal} {\bibinfo
  {journal} {Phys. Rev. E}\ }\textbf {\bibinfo {volume} {95}},\ \bibinfo
  {pages} {033206} (\bibinfo {year} {2017}{\natexlab{a}})}\BibitemShut
  {NoStop}%
\bibitem [{\citenamefont {Grete}\ \emph
  {et~al.}(2017{\natexlab{b}})\citenamefont {Grete}, \citenamefont {O'Shea},
  \citenamefont {Beckwith}, \citenamefont {Schmidt},\ and\ \citenamefont
  {Christlieb}}]{Grete2017b}%
  \BibitemOpen
  \bibfield  {author} {\bibinfo {author} {\bibfnamefont {P.}~\bibnamefont
  {Grete}}, \bibinfo {author} {\bibfnamefont {B.~W.}\ \bibnamefont {O'Shea}},
  \bibinfo {author} {\bibfnamefont {K.}~\bibnamefont {Beckwith}}, \bibinfo
  {author} {\bibfnamefont {W.}~\bibnamefont {Schmidt}}, \ and\ \bibinfo
  {author} {\bibfnamefont {A.}~\bibnamefont {Christlieb}},\ }\href {\doibase
  10.1063/1.4990613} {\bibfield  {journal} {\bibinfo  {journal} {Physics of
  Plasmas}\ }\textbf {\bibinfo {volume} {24}},\ \bibinfo {pages} {092311}
  (\bibinfo {year} {2017}{\natexlab{b}})},\ \Eprint
  {http://arxiv.org/abs/https://doi.org/10.1063/1.4990613}
  {https://doi.org/10.1063/1.4990613} \BibitemShut {NoStop}%
\bibitem [{\citenamefont {Kessar}\ \emph {et~al.}(2016)\citenamefont {Kessar},
  \citenamefont {Balarac},\ and\ \citenamefont {Plunian}}]{Kessar2016}%
  \BibitemOpen
  \bibfield  {author} {\bibinfo {author} {\bibfnamefont {M.}~\bibnamefont
  {Kessar}}, \bibinfo {author} {\bibfnamefont {G.}~\bibnamefont {Balarac}}, \
  and\ \bibinfo {author} {\bibfnamefont {F.}~\bibnamefont {Plunian}},\ }\href
  {\doibase 10.1063/1.4964782} {\bibfield  {journal} {\bibinfo  {journal}
  {Physics of Plasmas}\ }\textbf {\bibinfo {volume} {23}},\ \bibinfo {pages}
  {102305} (\bibinfo {year} {2016})},\ \Eprint
  {http://arxiv.org/abs/https://doi.org/10.1063/1.4964782}
  {https://doi.org/10.1063/1.4964782} \BibitemShut {NoStop}%
\bibitem [{\citenamefont {Vlaykov}\ \emph {et~al.}(2016)\citenamefont
  {Vlaykov}, \citenamefont {Grete}, \citenamefont {Schmidt},\ and\
  \citenamefont {Schleicher}}]{Vlaykov2016}%
  \BibitemOpen
  \bibfield  {author} {\bibinfo {author} {\bibfnamefont {D.~G.}\ \bibnamefont
  {Vlaykov}}, \bibinfo {author} {\bibfnamefont {P.}~\bibnamefont {Grete}},
  \bibinfo {author} {\bibfnamefont {W.}~\bibnamefont {Schmidt}}, \ and\
  \bibinfo {author} {\bibfnamefont {D.~R.~G.}\ \bibnamefont {Schleicher}},\
  }\href {\doibase 10.1063/1.4954303} {\bibfield  {journal} {\bibinfo
  {journal} {Physics of Plasmas}\ }\textbf {\bibinfo {volume} {23}},\ \bibinfo
  {pages} {062316} (\bibinfo {year} {2016})},\ \Eprint
  {http://arxiv.org/abs/https://doi.org/10.1063/1.4954303}
  {https://doi.org/10.1063/1.4954303} \BibitemShut {NoStop}%
\bibitem [{\citenamefont {{Vigan{\`o}}}\ \emph {et~al.}(2019)\citenamefont
  {{Vigan{\`o}}}, \citenamefont {{Aguilera-Miret}},\ and\ \citenamefont
  {{Palenzuela}}}]{Vigano2019a}%
  \BibitemOpen
  \bibfield  {author} {\bibinfo {author} {\bibfnamefont {D.}~\bibnamefont
  {{Vigan{\`o}}}}, \bibinfo {author} {\bibfnamefont {R.}~\bibnamefont
  {{Aguilera-Miret}}}, \ and\ \bibinfo {author} {\bibfnamefont
  {C.}~\bibnamefont {{Palenzuela}}},\ }\href@noop {} {\bibfield  {journal}
  {\bibinfo  {journal} {arXiv e-prints}\ ,\ \bibinfo {eid} {arXiv:1904.04099}}
  (\bibinfo {year} {2019})},\ \Eprint {http://arxiv.org/abs/1904.04099}
  {arXiv:1904.04099 [physics.flu-dyn]} \BibitemShut {NoStop}%
\bibitem [{\citenamefont {{Carrasco}}\ \emph {et~al.}(2019)\citenamefont
  {{Carrasco}}, \citenamefont {{Vigan{\`o}}},\ and\ \citenamefont
  {{Palenzuela}}}]{Carrasco2019a}%
  \BibitemOpen
  \bibfield  {author} {\bibinfo {author} {\bibfnamefont {F.}~\bibnamefont
  {{Carrasco}}}, \bibinfo {author} {\bibfnamefont {D.}~\bibnamefont
  {{Vigan{\`o}}}}, \ and\ \bibinfo {author} {\bibfnamefont {C.}~\bibnamefont
  {{Palenzuela}}},\ }\href@noop {} {\bibfield  {journal} {\bibinfo  {journal}
  {arXiv e-prints}\ ,\ \bibinfo {eid} {arXiv:1908.01419}} (\bibinfo {year}
  {2019})},\ \Eprint {http://arxiv.org/abs/1908.01419} {arXiv:1908.01419
  [astro-ph.HE]} \BibitemShut {NoStop}%
\bibitem [{\citenamefont {{Grete}}(2017)}]{Grete2017}%
  \BibitemOpen
  \bibfield  {author} {\bibinfo {author} {\bibfnamefont {P.}~\bibnamefont
  {{Grete}}},\ }\emph {\bibinfo {title} {{Large eddy simulations of
  compressible magnetohydrodynamic turbulence}}},\ \href@noop {} {Ph.D.
  thesis},\ \bibinfo  {school} {Max-Planck-Institut f{\"u}r
  Sonnensystemforschung} (\bibinfo {year} {2017})\BibitemShut {NoStop}%
\bibitem [{\citenamefont {Beresnyak}\ and\ \citenamefont
  {Lazarian}(2015)}]{Beresnyak2015}%
  \BibitemOpen
  \bibfield  {author} {\bibinfo {author} {\bibfnamefont {A.}~\bibnamefont
  {Beresnyak}}\ and\ \bibinfo {author} {\bibfnamefont {A.}~\bibnamefont
  {Lazarian}},\ }\enquote {\bibinfo {title} {Mhd turbulence, turbulent dynamo
  and applications},}\ in\ \href {\doibase 10.1007/978-3-662-44625-6_8} {\emph
  {\bibinfo {booktitle} {Magnetic Fields in Diffuse Media}}},\ \bibinfo
  {editor} {edited by\ \bibinfo {editor} {\bibfnamefont {A.}~\bibnamefont
  {Lazarian}}, \bibinfo {editor} {\bibfnamefont {E.~M.}\ \bibnamefont
  {de~Gouveia Dal~Pino}}, \ and\ \bibinfo {editor} {\bibfnamefont
  {C.}~\bibnamefont {Melioli}}}\ (\bibinfo  {publisher} {Springer Berlin
  Heidelberg},\ \bibinfo {address} {Berlin, Heidelberg},\ \bibinfo {year}
  {2015})\ pp.\ \bibinfo {pages} {163--226}\BibitemShut {NoStop}%
\bibitem [{\citenamefont {Beresnyak}(2019)}]{Beresnyak2019}%
  \BibitemOpen
  \bibfield  {author} {\bibinfo {author} {\bibfnamefont {A.}~\bibnamefont
  {Beresnyak}},\ }\href {\doibase 10.1007/s41115-019-0005-8} {\bibfield
  {journal} {\bibinfo  {journal} {Living Reviews in Computational
  Astrophysics}\ }\textbf {\bibinfo {volume} {5}},\ \bibinfo {pages} {2}
  (\bibinfo {year} {2019})}\BibitemShut {NoStop}%
\bibitem [{\citenamefont {{Ling}}\ \emph {et~al.}(2016)\citenamefont {{Ling}},
  \citenamefont {{Kurzawski}},\ and\ \citenamefont {{Templeton}}}]{Ling2016}%
  \BibitemOpen
  \bibfield  {author} {\bibinfo {author} {\bibfnamefont {J.}~\bibnamefont
  {{Ling}}}, \bibinfo {author} {\bibfnamefont {A.}~\bibnamefont {{Kurzawski}}},
  \ and\ \bibinfo {author} {\bibfnamefont {J.}~\bibnamefont {{Templeton}}},\
  }\href {\doibase 10.1017/jfm.2016.615} {\bibfield  {journal} {\bibinfo
  {journal} {Journal of Fluid Mechanics}\ }\textbf {\bibinfo {volume} {807}},\
  \bibinfo {pages} {155} (\bibinfo {year} {2016})}\BibitemShut {NoStop}%
\bibitem [{\citenamefont {{Maulik}}\ and\ \citenamefont
  {{San}}(2017)}]{Maulik2017}%
  \BibitemOpen
  \bibfield  {author} {\bibinfo {author} {\bibfnamefont {R.}~\bibnamefont
  {{Maulik}}}\ and\ \bibinfo {author} {\bibfnamefont {O.}~\bibnamefont
  {{San}}},\ }\href {\doibase 10.1017/jfm.2017.637} {\bibfield  {journal}
  {\bibinfo  {journal} {Journal of Fluid Mechanics}\ }\textbf {\bibinfo
  {volume} {831}},\ \bibinfo {pages} {151} (\bibinfo {year} {2017})},\ \Eprint
  {http://arxiv.org/abs/1706.00912} {arXiv:1706.00912 [physics.flu-dyn]}
  \BibitemShut {NoStop}%
\bibitem [{\citenamefont {{Fang}}\ \emph {et~al.}(2018)\citenamefont {{Fang}},
  \citenamefont {{Sondak}}, \citenamefont {{Protopapas}},\ and\ \citenamefont
  {{Succi}}}]{Fang2018}%
  \BibitemOpen
  \bibfield  {author} {\bibinfo {author} {\bibfnamefont {R.}~\bibnamefont
  {{Fang}}}, \bibinfo {author} {\bibfnamefont {D.}~\bibnamefont {{Sondak}}},
  \bibinfo {author} {\bibfnamefont {P.}~\bibnamefont {{Protopapas}}}, \ and\
  \bibinfo {author} {\bibfnamefont {S.}~\bibnamefont {{Succi}}},\ }\href@noop
  {} {\bibfield  {journal} {\bibinfo  {journal} {arXiv e-prints}\ ,\ \bibinfo
  {eid} {arXiv:1812.02241}} (\bibinfo {year} {2018})},\ \Eprint
  {http://arxiv.org/abs/1812.02241} {arXiv:1812.02241 [physics.flu-dyn]}
  \BibitemShut {NoStop}%
\bibitem [{\citenamefont {Wang}\ \emph {et~al.}(2018)\citenamefont {Wang},
  \citenamefont {Luo}, \citenamefont {Li}, \citenamefont {Tan},\ and\
  \citenamefont {Fan}}]{Wang2018}%
  \BibitemOpen
  \bibfield  {author} {\bibinfo {author} {\bibfnamefont {Z.}~\bibnamefont
  {Wang}}, \bibinfo {author} {\bibfnamefont {K.}~\bibnamefont {Luo}}, \bibinfo
  {author} {\bibfnamefont {D.}~\bibnamefont {Li}}, \bibinfo {author}
  {\bibfnamefont {J.}~\bibnamefont {Tan}}, \ and\ \bibinfo {author}
  {\bibfnamefont {J.}~\bibnamefont {Fan}},\ }\href {\doibase 10.1063/1.5054835}
  {\bibfield  {journal} {\bibinfo  {journal} {Physics of Fluids}\ }\textbf
  {\bibinfo {volume} {30}},\ \bibinfo {pages} {125101} (\bibinfo {year}
  {2018})},\ \Eprint {http://arxiv.org/abs/https://doi.org/10.1063/1.5054835}
  {https://doi.org/10.1063/1.5054835} \BibitemShut {NoStop}%
\bibitem [{\citenamefont {Xie}\ \emph {et~al.}(2019{\natexlab{a}})\citenamefont
  {Xie}, \citenamefont {Wang}, \citenamefont {Li},\ and\ \citenamefont
  {Ma}}]{Xie2019}%
  \BibitemOpen
  \bibfield  {author} {\bibinfo {author} {\bibfnamefont {C.}~\bibnamefont
  {Xie}}, \bibinfo {author} {\bibfnamefont {J.}~\bibnamefont {Wang}}, \bibinfo
  {author} {\bibfnamefont {K.}~\bibnamefont {Li}}, \ and\ \bibinfo {author}
  {\bibfnamefont {C.}~\bibnamefont {Ma}},\ }\href {\doibase
  10.1103/PhysRevE.99.053113} {\bibfield  {journal} {\bibinfo  {journal} {Phys.
  Rev. E}\ }\textbf {\bibinfo {volume} {99}},\ \bibinfo {pages} {053113}
  (\bibinfo {year} {2019}{\natexlab{a}})}\BibitemShut {NoStop}%
\bibitem [{\citenamefont {Xie}\ \emph {et~al.}(2019{\natexlab{b}})\citenamefont
  {Xie}, \citenamefont {Li}, \citenamefont {Ma},\ and\ \citenamefont
  {Wang}}]{Xie2019b}%
  \BibitemOpen
  \bibfield  {author} {\bibinfo {author} {\bibfnamefont {C.}~\bibnamefont
  {Xie}}, \bibinfo {author} {\bibfnamefont {K.}~\bibnamefont {Li}}, \bibinfo
  {author} {\bibfnamefont {C.}~\bibnamefont {Ma}}, \ and\ \bibinfo {author}
  {\bibfnamefont {J.}~\bibnamefont {Wang}},\ }\href {\doibase
  10.1103/PhysRevFluids.4.104605} {\bibfield  {journal} {\bibinfo  {journal}
  {Phys. Rev. Fluids}\ }\textbf {\bibinfo {volume} {4}},\ \bibinfo {pages}
  {104605} (\bibinfo {year} {2019}{\natexlab{b}})}\BibitemShut {NoStop}%
\bibitem [{\citenamefont {Xie}\ \emph {et~al.}(2020)\citenamefont {Xie},
  \citenamefont {Wang}, \citenamefont {Li}, \citenamefont {Wan},\ and\
  \citenamefont {Chen}}]{Xie2020a}%
  \BibitemOpen
  \bibfield  {author} {\bibinfo {author} {\bibfnamefont {C.~.}\ \bibnamefont
  {Xie}}, \bibinfo {author} {\bibfnamefont {J.~.}\ \bibnamefont {Wang}},
  \bibinfo {author} {\bibfnamefont {H.~.}\ \bibnamefont {Li}}, \bibinfo
  {author} {\bibfnamefont {M.~.}\ \bibnamefont {Wan}}, \ and\ \bibinfo {author}
  {\bibfnamefont {S.~.}\ \bibnamefont {Chen}},\ }\href {\doibase
  10.1063/1.5138681} {\bibfield  {journal} {\bibinfo  {journal} {AIP Advances}\
  }\textbf {\bibinfo {volume} {10}},\ \bibinfo {pages} {015044} (\bibinfo
  {year} {2020})},\ \Eprint
  {http://arxiv.org/abs/https://doi.org/10.1063/1.5138681}
  {https://doi.org/10.1063/1.5138681} \BibitemShut {NoStop}%
\bibitem [{\citenamefont {Pawar}\ \emph {et~al.}(2020)\citenamefont {Pawar},
  \citenamefont {San}, \citenamefont {Rasheed},\ and\ \citenamefont
  {Vedula}}]{Pawar2020}%
  \BibitemOpen
  \bibfield  {author} {\bibinfo {author} {\bibfnamefont {S.}~\bibnamefont
  {Pawar}}, \bibinfo {author} {\bibfnamefont {O.}~\bibnamefont {San}}, \bibinfo
  {author} {\bibfnamefont {A.}~\bibnamefont {Rasheed}}, \ and\ \bibinfo
  {author} {\bibfnamefont {P.}~\bibnamefont {Vedula}},\ }\href {\doibase
  10.1007/s00162-019-00512-z} {\bibfield  {journal} {\bibinfo  {journal}
  {Theoretical and Computational Fluid Dynamics}\ } (\bibinfo {year} {2020}),\
  10.1007/s00162-019-00512-z}\BibitemShut {NoStop}%
\bibitem [{\citenamefont {Brunton}\ \emph {et~al.}(2020)\citenamefont
  {Brunton}, \citenamefont {Noack},\ and\ \citenamefont
  {Koumoutsakos}}]{Brunton2020}%
  \BibitemOpen
  \bibfield  {author} {\bibinfo {author} {\bibfnamefont {S.~L.}\ \bibnamefont
  {Brunton}}, \bibinfo {author} {\bibfnamefont {B.~R.}\ \bibnamefont {Noack}},
  \ and\ \bibinfo {author} {\bibfnamefont {P.}~\bibnamefont {Koumoutsakos}},\
  }\href {\doibase 10.1146/annurev-fluid-010719-060214} {\bibfield  {journal}
  {\bibinfo  {journal} {Annual Review of Fluid Mechanics}\ }\textbf {\bibinfo
  {volume} {52}},\ \bibinfo {pages} {477} (\bibinfo {year} {2020})},\ \Eprint
  {http://arxiv.org/abs/https://doi.org/10.1146/annurev-fluid-010719-060214}
  {https://doi.org/10.1146/annurev-fluid-010719-060214} \BibitemShut {NoStop}%
\bibitem [{\citenamefont {Vollant}\ \emph {et~al.}(2016)\citenamefont
  {Vollant}, \citenamefont {Balarac},\ and\ \citenamefont
  {Corre}}]{Vollant2016}%
  \BibitemOpen
  \bibfield  {author} {\bibinfo {author} {\bibfnamefont {A.}~\bibnamefont
  {Vollant}}, \bibinfo {author} {\bibfnamefont {G.}~\bibnamefont {Balarac}}, \
  and\ \bibinfo {author} {\bibfnamefont {C.}~\bibnamefont {Corre}},\ }\href
  {\doibase 10.1063/1.4941781} {\bibfield  {journal} {\bibinfo  {journal}
  {Physics of Fluids}\ }\textbf {\bibinfo {volume} {28}},\ \bibinfo {pages}
  {025114} (\bibinfo {year} {2016})}\BibitemShut {NoStop}%
\bibitem [{\citenamefont {{Lecun}}\ \emph {et~al.}(2015)\citenamefont
  {{Lecun}}, \citenamefont {{Bengio}},\ and\ \citenamefont
  {{Hinton}}}]{LeCun:Nature}%
  \BibitemOpen
  \bibfield  {author} {\bibinfo {author} {\bibfnamefont {Y.}~\bibnamefont
  {{Lecun}}}, \bibinfo {author} {\bibfnamefont {Y.}~\bibnamefont {{Bengio}}}, \
  and\ \bibinfo {author} {\bibfnamefont {G.}~\bibnamefont {{Hinton}}},\ }\href
  {\doibase 10.1038/nature14539} {\bibfield  {journal} {\bibinfo  {journal}
  {\nat}\ }\textbf {\bibinfo {volume} {521}},\ \bibinfo {pages} {436} (\bibinfo
  {year} {2015})}\BibitemShut {NoStop}%
\bibitem [{\citenamefont {{Miotto}}\ \emph {et~al.}(2018)\citenamefont
  {{Miotto}}, \citenamefont {{Wang}}, \citenamefont {{Wang}}, \citenamefont
  {{Jiang}},\ and\ \citenamefont {{Dudley}}}]{miotto}%
  \BibitemOpen
  \bibfield  {author} {\bibinfo {author} {\bibfnamefont {R.}~\bibnamefont
  {{Miotto}}}, \bibinfo {author} {\bibfnamefont {F.}~\bibnamefont {{Wang}}},
  \bibinfo {author} {\bibfnamefont {S.}~\bibnamefont {{Wang}}}, \bibinfo
  {author} {\bibfnamefont {X.}~\bibnamefont {{Jiang}}}, \ and\ \bibinfo
  {author} {\bibfnamefont {J.}~\bibnamefont {{Dudley}}},\ }\href {\doibase
  10.1093/bib/bbx044.} {\bibfield  {journal} {\bibinfo  {journal} {Briefings in
  Bioinformatics}\ } (\bibinfo {year} {2018}),\
  10.1093/bib/bbx044.}\BibitemShut {Stop}%
\bibitem [{\citenamefont {Ismail~Fawaz}\ \emph {et~al.}(2019)\citenamefont
  {Ismail~Fawaz}, \citenamefont {Forestier}, \citenamefont {Weber},
  \citenamefont {Idoumghar},\ and\ \citenamefont {Muller}}]{IsmailFawaz2019}%
  \BibitemOpen
  \bibfield  {author} {\bibinfo {author} {\bibfnamefont {H.}~\bibnamefont
  {Ismail~Fawaz}}, \bibinfo {author} {\bibfnamefont {G.}~\bibnamefont
  {Forestier}}, \bibinfo {author} {\bibfnamefont {J.}~\bibnamefont {Weber}},
  \bibinfo {author} {\bibfnamefont {L.}~\bibnamefont {Idoumghar}}, \ and\
  \bibinfo {author} {\bibfnamefont {P.-A.}\ \bibnamefont {Muller}},\ }\href
  {\doibase 10.1007/s10618-019-00619-1} {\bibfield  {journal} {\bibinfo
  {journal} {Data Mining and Knowledge Discovery}\ }\textbf {\bibinfo {volume}
  {33}},\ \bibinfo {pages} {917} (\bibinfo {year} {2019})}\BibitemShut
  {NoStop}%
\bibitem [{\citenamefont {Schmidhuber}(2015)}]{SCHMIDHUBER201585}%
  \BibitemOpen
  \bibfield  {author} {\bibinfo {author} {\bibfnamefont {J.}~\bibnamefont
  {Schmidhuber}},\ }\href {\doibase
  https://doi.org/10.1016/j.neunet.2014.09.003} {\bibfield  {journal} {\bibinfo
   {journal} {Neural Networks}\ }\textbf {\bibinfo {volume} {61}},\ \bibinfo
  {pages} {85 } (\bibinfo {year} {2015})}\BibitemShut {NoStop}%
\bibitem [{\citenamefont {{Huerta}}\ \emph {et~al.}(2019)\citenamefont
  {{Huerta}}, \citenamefont {{Allen}}, \citenamefont {{Andreoni}},
  \citenamefont {{Antelis}}, \citenamefont {{Bachelet}}, \citenamefont
  {{Berriman}}, \citenamefont {{Bianco}}, \citenamefont {{Biswas}},
  \citenamefont {{Carrasco{\^A} Kind}}, \citenamefont {{Chard}}, \citenamefont
  {{Cho}}, \citenamefont {{Cowperthwaite}}, \citenamefont {{Etienne}},
  \citenamefont {{Fishbach}}, \citenamefont {{Forster}}, \citenamefont
  {{George}}, \citenamefont {{Gibbs}}, \citenamefont {{Graham}}, \citenamefont
  {{Gropp}}, \citenamefont {{Gruendl}}, \citenamefont {{Gupta}}, \citenamefont
  {{Haas}}, \citenamefont {{Habib}}, \citenamefont {{Jennings}}, \citenamefont
  {{Johnson}}, \citenamefont {{Katsavounidis}}, \citenamefont {{Katz}},
  \citenamefont {{Khan}}, \citenamefont {{Kindratenko}}, \citenamefont
  {{Kramer}}, \citenamefont {{Liu}}, \citenamefont {{Mahabal}}, \citenamefont
  {{Marka}}, \citenamefont {{McHenry}}, \citenamefont {{Miller}}, \citenamefont
  {{Moreno}}, \citenamefont {{Neubauer}}, \citenamefont {{Oberlin}},
  \citenamefont {{Olivas}}, \citenamefont {{Petravick}}, \citenamefont
  {{Rebei}}, \citenamefont {{Rosofsky}}, \citenamefont {{Ruiz}}, \citenamefont
  {{Saxton}}, \citenamefont {{Schutz}}, \citenamefont {{Schwing}},
  \citenamefont {{Seidel}}, \citenamefont {{Shapiro}}, \citenamefont {{Shen}},
  \citenamefont {{Shen}}, \citenamefont {{Singer}}, \citenamefont {{Sipocz}},
  \citenamefont {{Sun}}, \citenamefont {{Towns}}, \citenamefont {{Tsokaros}},
  \citenamefont {{Wei}}, \citenamefont {{Wells}}, \citenamefont {{Williams}},
  \citenamefont {{Xiong}},\ and\ \citenamefont {{Zhao}}}]{2019NatRP6}%
  \BibitemOpen
  \bibfield  {author} {\bibinfo {author} {\bibfnamefont {E.~A.}\ \bibnamefont
  {{Huerta}}}, \bibinfo {author} {\bibfnamefont {G.}~\bibnamefont {{Allen}}},
  \bibinfo {author} {\bibfnamefont {I.}~\bibnamefont {{Andreoni}}}, \bibinfo
  {author} {\bibfnamefont {J.~M.}\ \bibnamefont {{Antelis}}}, \bibinfo {author}
  {\bibfnamefont {E.}~\bibnamefont {{Bachelet}}}, \bibinfo {author}
  {\bibfnamefont {G.~B.}\ \bibnamefont {{Berriman}}}, \bibinfo {author}
  {\bibfnamefont {F.~B.}\ \bibnamefont {{Bianco}}}, \bibinfo {author}
  {\bibfnamefont {R.}~\bibnamefont {{Biswas}}}, \bibinfo {author}
  {\bibfnamefont {M.}~\bibnamefont {{Carrasco{\^A} Kind}}}, \bibinfo {author}
  {\bibfnamefont {K.}~\bibnamefont {{Chard}}}, \bibinfo {author} {\bibfnamefont
  {M.}~\bibnamefont {{Cho}}}, \bibinfo {author} {\bibfnamefont {P.~S.}\
  \bibnamefont {{Cowperthwaite}}}, \bibinfo {author} {\bibfnamefont {Z.~B.}\
  \bibnamefont {{Etienne}}}, \bibinfo {author} {\bibfnamefont {M.}~\bibnamefont
  {{Fishbach}}}, \bibinfo {author} {\bibfnamefont {F.}~\bibnamefont
  {{Forster}}}, \bibinfo {author} {\bibfnamefont {D.}~\bibnamefont {{George}}},
  \bibinfo {author} {\bibfnamefont {T.}~\bibnamefont {{Gibbs}}}, \bibinfo
  {author} {\bibfnamefont {M.}~\bibnamefont {{Graham}}}, \bibinfo {author}
  {\bibfnamefont {W.}~\bibnamefont {{Gropp}}}, \bibinfo {author} {\bibfnamefont
  {R.}~\bibnamefont {{Gruendl}}}, \bibinfo {author} {\bibfnamefont
  {A.}~\bibnamefont {{Gupta}}}, \bibinfo {author} {\bibfnamefont
  {R.}~\bibnamefont {{Haas}}}, \bibinfo {author} {\bibfnamefont
  {S.}~\bibnamefont {{Habib}}}, \bibinfo {author} {\bibfnamefont
  {E.}~\bibnamefont {{Jennings}}}, \bibinfo {author} {\bibfnamefont {M.~W.~G.}\
  \bibnamefont {{Johnson}}}, \bibinfo {author} {\bibfnamefont {E.}~\bibnamefont
  {{Katsavounidis}}}, \bibinfo {author} {\bibfnamefont {D.~S.}\ \bibnamefont
  {{Katz}}}, \bibinfo {author} {\bibfnamefont {A.}~\bibnamefont {{Khan}}},
  \bibinfo {author} {\bibfnamefont {V.}~\bibnamefont {{Kindratenko}}}, \bibinfo
  {author} {\bibfnamefont {W.~T.~C.}\ \bibnamefont {{Kramer}}}, \bibinfo
  {author} {\bibfnamefont {X.}~\bibnamefont {{Liu}}}, \bibinfo {author}
  {\bibfnamefont {A.}~\bibnamefont {{Mahabal}}}, \bibinfo {author}
  {\bibfnamefont {Z.}~\bibnamefont {{Marka}}}, \bibinfo {author} {\bibfnamefont
  {K.}~\bibnamefont {{McHenry}}}, \bibinfo {author} {\bibfnamefont {J.~M.}\
  \bibnamefont {{Miller}}}, \bibinfo {author} {\bibfnamefont {C.}~\bibnamefont
  {{Moreno}}}, \bibinfo {author} {\bibfnamefont {M.~S.}\ \bibnamefont
  {{Neubauer}}}, \bibinfo {author} {\bibfnamefont {S.}~\bibnamefont
  {{Oberlin}}}, \bibinfo {author} {\bibfnamefont {A.~R.}\ \bibnamefont
  {{Olivas}}}, \bibinfo {author} {\bibfnamefont {D.}~\bibnamefont
  {{Petravick}}}, \bibinfo {author} {\bibfnamefont {A.}~\bibnamefont
  {{Rebei}}}, \bibinfo {author} {\bibfnamefont {S.}~\bibnamefont {{Rosofsky}}},
  \bibinfo {author} {\bibfnamefont {M.}~\bibnamefont {{Ruiz}}}, \bibinfo
  {author} {\bibfnamefont {A.}~\bibnamefont {{Saxton}}}, \bibinfo {author}
  {\bibfnamefont {B.~F.}\ \bibnamefont {{Schutz}}}, \bibinfo {author}
  {\bibfnamefont {A.}~\bibnamefont {{Schwing}}}, \bibinfo {author}
  {\bibfnamefont {E.}~\bibnamefont {{Seidel}}}, \bibinfo {author}
  {\bibfnamefont {S.~L.}\ \bibnamefont {{Shapiro}}}, \bibinfo {author}
  {\bibfnamefont {H.}~\bibnamefont {{Shen}}}, \bibinfo {author} {\bibfnamefont
  {Y.}~\bibnamefont {{Shen}}}, \bibinfo {author} {\bibfnamefont {L.~P.}\
  \bibnamefont {{Singer}}}, \bibinfo {author} {\bibfnamefont {B.~M.}\
  \bibnamefont {{Sipocz}}}, \bibinfo {author} {\bibfnamefont {L.}~\bibnamefont
  {{Sun}}}, \bibinfo {author} {\bibfnamefont {J.}~\bibnamefont {{Towns}}},
  \bibinfo {author} {\bibfnamefont {A.}~\bibnamefont {{Tsokaros}}}, \bibinfo
  {author} {\bibfnamefont {W.}~\bibnamefont {{Wei}}}, \bibinfo {author}
  {\bibfnamefont {J.}~\bibnamefont {{Wells}}}, \bibinfo {author} {\bibfnamefont
  {T.~J.}\ \bibnamefont {{Williams}}}, \bibinfo {author} {\bibfnamefont
  {J.}~\bibnamefont {{Xiong}}}, \ and\ \bibinfo {author} {\bibfnamefont
  {Z.}~\bibnamefont {{Zhao}}},\ }\href {\doibase 10.1038/s42254-019-0097-4}
  {\bibfield  {journal} {\bibinfo  {journal} {Nature Reviews Physics}\ }\textbf
  {\bibinfo {volume} {1}},\ \bibinfo {pages} {600} (\bibinfo {year} {2019})},\
  \Eprint {http://arxiv.org/abs/1911.11779} {arXiv:1911.11779 [gr-qc]}
  \BibitemShut {NoStop}%
\bibitem [{\citenamefont {Arbona}\ \emph {et~al.}(2013)\citenamefont {Arbona},
  \citenamefont {Artigues}, \citenamefont {Bona-Casas}, \citenamefont
  {Mass{\'o}}, \citenamefont {Mi{\~n}ano}, \citenamefont {Rigo},\ and\
  \citenamefont {Trias}}]{simflowny1}%
  \BibitemOpen
  \bibfield  {author} {\bibinfo {author} {\bibfnamefont {A.}~\bibnamefont
  {Arbona}}, \bibinfo {author} {\bibfnamefont {A.}~\bibnamefont {Artigues}},
  \bibinfo {author} {\bibfnamefont {C.}~\bibnamefont {Bona-Casas}}, \bibinfo
  {author} {\bibfnamefont {J.}~\bibnamefont {Mass{\'o}}}, \bibinfo {author}
  {\bibfnamefont {B.}~\bibnamefont {Mi{\~n}ano}}, \bibinfo {author}
  {\bibfnamefont {A.}~\bibnamefont {Rigo}}, \ and\ \bibinfo {author}
  {\bibfnamefont {M.}~\bibnamefont {Trias}},\ }\href@noop {} {\bibfield
  {journal} {\bibinfo  {journal} {Computer Physics Communications}\ }\textbf
  {\bibinfo {volume} {184}},\ \bibinfo {pages} {2321} (\bibinfo {year}
  {2013})}\BibitemShut {NoStop}%
\bibitem [{\citenamefont {Arbona}\ \emph {et~al.}(2018)\citenamefont {Arbona},
  \citenamefont {Miñano}, \citenamefont {Rigo}, \citenamefont {Bona},
  \citenamefont {Palenzuela}, \citenamefont {Artigues}, \citenamefont
  {Bona-Casas},\ and\ \citenamefont {Massó}}]{simflowny2}%
  \BibitemOpen
  \bibfield  {author} {\bibinfo {author} {\bibfnamefont {A.}~\bibnamefont
  {Arbona}}, \bibinfo {author} {\bibfnamefont {B.}~\bibnamefont {Miñano}},
  \bibinfo {author} {\bibfnamefont {A.}~\bibnamefont {Rigo}}, \bibinfo {author}
  {\bibfnamefont {C.}~\bibnamefont {Bona}}, \bibinfo {author} {\bibfnamefont
  {C.}~\bibnamefont {Palenzuela}}, \bibinfo {author} {\bibfnamefont
  {A.}~\bibnamefont {Artigues}}, \bibinfo {author} {\bibfnamefont
  {C.}~\bibnamefont {Bona-Casas}}, \ and\ \bibinfo {author} {\bibfnamefont
  {J.}~\bibnamefont {Massó}},\ }\href {\doibase
  https://doi.org/10.1016/j.cpc.2018.03.015} {\bibfield  {journal} {\bibinfo
  {journal} {Computer Physics Communications}\ }\textbf {\bibinfo {volume}
  {229}},\ \bibinfo {pages} {170 } (\bibinfo {year} {2018})}\BibitemShut
  {NoStop}%
\bibitem [{\citenamefont {Dedner}\ \emph {et~al.}(2002)\citenamefont {Dedner},
  \citenamefont {Kemm}, \citenamefont {Kröner}, \citenamefont {Munz},
  \citenamefont {Schnitzer},\ and\ \citenamefont {Wesenberg}}]{Dedner2002}%
  \BibitemOpen
  \bibfield  {author} {\bibinfo {author} {\bibfnamefont {A.}~\bibnamefont
  {Dedner}}, \bibinfo {author} {\bibfnamefont {F.}~\bibnamefont {Kemm}},
  \bibinfo {author} {\bibfnamefont {D.}~\bibnamefont {Kröner}}, \bibinfo
  {author} {\bibfnamefont {C.-D.}\ \bibnamefont {Munz}}, \bibinfo {author}
  {\bibfnamefont {T.}~\bibnamefont {Schnitzer}}, \ and\ \bibinfo {author}
  {\bibfnamefont {M.}~\bibnamefont {Wesenberg}},\ }\href {\doibase
  https://doi.org/10.1006/jcph.2001.6961} {\bibfield  {journal} {\bibinfo
  {journal} {Journal of Computational Physics}\ }\textbf {\bibinfo {volume}
  {175}},\ \bibinfo {pages} {645 } (\bibinfo {year} {2002})}\BibitemShut
  {NoStop}%
\bibitem [{\citenamefont {Abadi}\ \emph {et~al.}(2015)\citenamefont {Abadi},
  \citenamefont {Agarwal}, \citenamefont {Barham}, \citenamefont {Brevdo},
  \citenamefont {Chen}, \citenamefont {Citro}, \citenamefont {Corrado},
  \citenamefont {Davis}, \citenamefont {Dean}, \citenamefont {Devin},
  \citenamefont {Ghemawat}, \citenamefont {Goodfellow}, \citenamefont {Harp},
  \citenamefont {Irving}, \citenamefont {Isard}, \citenamefont {Jia},
  \citenamefont {Jozefowicz}, \citenamefont {Kaiser}, \citenamefont {Kudlur},
  \citenamefont {Levenberg}, \citenamefont {Man\'{e}}, \citenamefont {Monga},
  \citenamefont {Moore}, \citenamefont {Murray}, \citenamefont {Olah},
  \citenamefont {Schuster}, \citenamefont {Shlens}, \citenamefont {Steiner},
  \citenamefont {Sutskever}, \citenamefont {Talwar}, \citenamefont {Tucker},
  \citenamefont {Vanhoucke}, \citenamefont {Vasudevan}, \citenamefont
  {Vi\'{e}gas}, \citenamefont {Vinyals}, \citenamefont {Warden}, \citenamefont
  {Wattenberg}, \citenamefont {Wicke}, \citenamefont {Yu},\ and\ \citenamefont
  {Zheng}}]{tensorflow2015-whitepaper}%
  \BibitemOpen
  \bibfield  {author} {\bibinfo {author} {\bibfnamefont {M.}~\bibnamefont
  {Abadi}}, \bibinfo {author} {\bibfnamefont {A.}~\bibnamefont {Agarwal}},
  \bibinfo {author} {\bibfnamefont {P.}~\bibnamefont {Barham}}, \bibinfo
  {author} {\bibfnamefont {E.}~\bibnamefont {Brevdo}}, \bibinfo {author}
  {\bibfnamefont {Z.}~\bibnamefont {Chen}}, \bibinfo {author} {\bibfnamefont
  {C.}~\bibnamefont {Citro}}, \bibinfo {author} {\bibfnamefont {G.~S.}\
  \bibnamefont {Corrado}}, \bibinfo {author} {\bibfnamefont {A.}~\bibnamefont
  {Davis}}, \bibinfo {author} {\bibfnamefont {J.}~\bibnamefont {Dean}},
  \bibinfo {author} {\bibfnamefont {M.}~\bibnamefont {Devin}}, \bibinfo
  {author} {\bibfnamefont {S.}~\bibnamefont {Ghemawat}}, \bibinfo {author}
  {\bibfnamefont {I.}~\bibnamefont {Goodfellow}}, \bibinfo {author}
  {\bibfnamefont {A.}~\bibnamefont {Harp}}, \bibinfo {author} {\bibfnamefont
  {G.}~\bibnamefont {Irving}}, \bibinfo {author} {\bibfnamefont
  {M.}~\bibnamefont {Isard}}, \bibinfo {author} {\bibfnamefont
  {Y.}~\bibnamefont {Jia}}, \bibinfo {author} {\bibfnamefont {R.}~\bibnamefont
  {Jozefowicz}}, \bibinfo {author} {\bibfnamefont {L.}~\bibnamefont {Kaiser}},
  \bibinfo {author} {\bibfnamefont {M.}~\bibnamefont {Kudlur}}, \bibinfo
  {author} {\bibfnamefont {J.}~\bibnamefont {Levenberg}}, \bibinfo {author}
  {\bibfnamefont {D.}~\bibnamefont {Man\'{e}}}, \bibinfo {author}
  {\bibfnamefont {R.}~\bibnamefont {Monga}}, \bibinfo {author} {\bibfnamefont
  {S.}~\bibnamefont {Moore}}, \bibinfo {author} {\bibfnamefont
  {D.}~\bibnamefont {Murray}}, \bibinfo {author} {\bibfnamefont
  {C.}~\bibnamefont {Olah}}, \bibinfo {author} {\bibfnamefont {M.}~\bibnamefont
  {Schuster}}, \bibinfo {author} {\bibfnamefont {J.}~\bibnamefont {Shlens}},
  \bibinfo {author} {\bibfnamefont {B.}~\bibnamefont {Steiner}}, \bibinfo
  {author} {\bibfnamefont {I.}~\bibnamefont {Sutskever}}, \bibinfo {author}
  {\bibfnamefont {K.}~\bibnamefont {Talwar}}, \bibinfo {author} {\bibfnamefont
  {P.}~\bibnamefont {Tucker}}, \bibinfo {author} {\bibfnamefont
  {V.}~\bibnamefont {Vanhoucke}}, \bibinfo {author} {\bibfnamefont
  {V.}~\bibnamefont {Vasudevan}}, \bibinfo {author} {\bibfnamefont
  {F.}~\bibnamefont {Vi\'{e}gas}}, \bibinfo {author} {\bibfnamefont
  {O.}~\bibnamefont {Vinyals}}, \bibinfo {author} {\bibfnamefont
  {P.}~\bibnamefont {Warden}}, \bibinfo {author} {\bibfnamefont
  {M.}~\bibnamefont {Wattenberg}}, \bibinfo {author} {\bibfnamefont
  {M.}~\bibnamefont {Wicke}}, \bibinfo {author} {\bibfnamefont
  {Y.}~\bibnamefont {Yu}}, \ and\ \bibinfo {author} {\bibfnamefont
  {X.}~\bibnamefont {Zheng}},\ }\href {http://tensorflow.org/} {\enquote
  {\bibinfo {title} {{TensorFlow}: Large-scale machine learning on
  heterogeneous systems},}\ } (\bibinfo {year} {2015}),\ \bibinfo {note}
  {software available from tensorflow.org}\BibitemShut {NoStop}%
\bibitem [{\citenamefont {{Kingma}}\ and\ \citenamefont
  {{Ba}}(2014)}]{Kingma2014}%
  \BibitemOpen
  \bibfield  {author} {\bibinfo {author} {\bibfnamefont {D.~P.}\ \bibnamefont
  {{Kingma}}}\ and\ \bibinfo {author} {\bibfnamefont {J.}~\bibnamefont
  {{Ba}}},\ }\href@noop {} {\bibfield  {journal} {\bibinfo  {journal} {arXiv
  e-prints}\ ,\ \bibinfo {eid} {arXiv:1412.6980}} (\bibinfo {year} {2014})},\
  \Eprint {http://arxiv.org/abs/1412.6980} {arXiv:1412.6980 [cs.LG]}
  \BibitemShut {NoStop}%
\bibitem [{\citenamefont {Ghosal}(1999)}]{Ghosal1999}%
  \BibitemOpen
  \bibfield  {author} {\bibinfo {author} {\bibfnamefont {S.}~\bibnamefont
  {Ghosal}},\ }\href {\doibase 10.2514/2.752} {\bibfield  {journal} {\bibinfo
  {journal} {AIAA Journal}\ }\textbf {\bibinfo {volume} {37}},\ \bibinfo
  {pages} {425} (\bibinfo {year} {1999})},\ \Eprint
  {http://arxiv.org/abs/https://doi.org/10.2514/2.752}
  {https://doi.org/10.2514/2.752} \BibitemShut {NoStop}%
\bibitem [{\citenamefont {Silvis}\ \emph {et~al.}(2017)\citenamefont {Silvis},
  \citenamefont {Remmerswaal},\ and\ \citenamefont {Verstappen}}]{Silvis2017}%
  \BibitemOpen
  \bibfield  {author} {\bibinfo {author} {\bibfnamefont {M.~H.}\ \bibnamefont
  {Silvis}}, \bibinfo {author} {\bibfnamefont {R.~A.}\ \bibnamefont
  {Remmerswaal}}, \ and\ \bibinfo {author} {\bibfnamefont {R.}~\bibnamefont
  {Verstappen}},\ }\href {\doibase 10.1063/1.4974093} {\bibfield  {journal}
  {\bibinfo  {journal} {Physics of Fluids}\ }\textbf {\bibinfo {volume} {29}},\
  \bibinfo {pages} {015105} (\bibinfo {year} {2017})},\ \Eprint
  {http://arxiv.org/abs/https://doi.org/10.1063/1.4974093}
  {https://doi.org/10.1063/1.4974093} \BibitemShut {NoStop}%
\end{thebibliography}%
\bibliographystyle{apsrev4-1}

\appendix

\section{ANN Model Inputs}
\label{app:ann_inputs}

Here we explicitly note the inputs to each SGS tensor of $\tau_{ANN}$ for clarity.

\begin{align}
I_{\tau_{kin,net}} =&\
\Bigl\{
\widetilde{v}_{i}^{m, n},
\widetilde{v}_{i}^{m \pm 1, n},
\widetilde{v}_{i}^{m, n \pm 1}, 
\partial_{p} \widetilde{v}_{i}^{m, n},
\partial_{p} \widetilde{v}_{i}^{m \pm 1, n}, 
\partial_{p} \widetilde{v}_{i}^{m, n \pm 1}, 
\partial_{p} \partial_{q} \widetilde{v}_{i}^{m, n}, 
\partial_{p} \partial_{q} \widetilde{v}_{i}^{m \pm 1, n}, 
\partial_{p} \partial_{q} \widetilde{v}_{i}^{m, n \pm 1} 
\Bigr\} 
\label{eq:inputs_tau_kin}   
\\
I_{\tau_{mag,net}} =&\
\Bigl\{
\overline{B}_{i}^{m, n},
\overline{B}_{i}^{m \pm 1, n},
\overline{B}_{i}^{m, n \pm 1}, 
\partial_{p} \overline{B}_{i}^{m, n},
\partial_{p} \overline{B}_{i}^{m \pm 1, n}, 
\partial_{p} \overline{B}_{i}^{m, n \pm 1},
\partial_{p} \partial_{q} \overline{B}_{i}^{m, n}, 
\partial_{p} \partial_{q} \overline{B}_{i}^{m \pm 1, n}, 
\partial_{p} \partial_{q} \overline{B}_{i}^{m, n \pm 1}  
\Bigr\} 
\label{eq:inputs_tau_mag}
\\
I_{\tau_{ind,net}} =&\
\Bigl\{
\overline{\rho}^{m, n},
\overline{\rho}^{m \pm 1, n},
\overline{\rho}^{m, n \pm 1}, 
\partial_{p} \overline{\rho}^{m, n},
\partial_{p} \overline{\rho}^{m \pm 1, n}, 
\partial_{p} \overline{\rho}^{m, n \pm 1}, 
\partial_{p} \partial_{q} \overline{\rho}^{m, n}, 
\partial_{p} \partial_{q} \overline{\rho}^{m \pm 1, n}, 
\partial_{p} \partial_{q} \overline{\rho}^{m, n \pm 1}, 
\notag \\
&\widetilde{v}_{i}^{m, n},
\widetilde{v}_{i}^{m \pm 1, n},
\widetilde{v}_{i}^{m, n \pm 1}, 
\partial_{p} \widetilde{v}_{i}^{m, n},
\partial_{p} \widetilde{v}_{i}^{m \pm 1, n}, 
\partial_{p} \widetilde{v}_{i}^{m, n \pm 1},
\partial_{p} \partial_{q} \widetilde{v}_{i}^{m, n}, 
\partial_{p} \partial_{q} \widetilde{v}_{i}^{m \pm 1, n}, 
\partial_{p} \partial_{q} \widetilde{v}_{i}^{m, n \pm 1},  
\notag \\
&\overline{B}_{i}^{m, n},
\overline{B}_{i}^{m \pm 1, n},
\overline{B}_{i}^{m, n \pm 1}, 
\partial_{p} \overline{B}_{i}^{m, n},
\partial_{p} \overline{B}_{i}^{m \pm 1, n}, 
\partial_{p} \overline{B}_{i}^{m, n \pm 1},
\partial_{p} \partial_{q} \overline{B}_{i}^{m, n}, 
\partial_{p} \partial_{q} \overline{B}_{i}^{m \pm 1, n}, 
\partial_{p} \partial_{q} \overline{B}_{i}^{m, n \pm 1}  
\Bigr\}
\label{eq:inputs_tau_ind}
\\
I_{\tau_{enth,net}} = &\
\Bigl\{
\overline{\rho}^{m, n},
\overline{\rho}^{m \pm 1, n},
\overline{\rho}^{m, n \pm 1},
\partial_{p} \overline{\rho}^{m, n},
\partial_{p} \overline{\rho}^{m \pm 1, n}, 
\partial_{p} \overline{\rho}^{m, n \pm 1}, 
\partial_{p} \partial_{q} \overline{\rho}^{m, n}, 
\partial_{p} \partial_{q} \overline{\rho}^{m \pm 1, n}, 
\partial_{p} \partial_{q} \overline{\rho}^{m, n \pm 1},  
\notag \\
&\widetilde{v}_{i}^{m, n},
\widetilde{v}_{i}^{m \pm 1, n},
\widetilde{v}_{i}^{m, n \pm 1}, 
\partial_{p} \widetilde{v}_{i}^{m, n},
\partial_{p} \widetilde{v}_{i}^{m \pm 1, n}, 
\partial_{p} \widetilde{v}_{i}^{m, n \pm 1},
\partial_{p} \partial_{q} \widetilde{v}_{i}^{m, n}, 
\partial_{p} \partial_{q} \widetilde{v}_{i}^{m \pm 1, n}, 
\partial_{p} \partial_{q} \widetilde{v}_{i}^{m, n \pm 1},  
\notag \\
&\widetilde{h}^{m, n},
\widetilde{h}^{m \pm 1, n},
\widetilde{h}^{m, n \pm 1},
\partial_{p} \widetilde{h}^{m, n},
\partial_{p} \widetilde{h}^{m \pm 1, n}, 
\partial_{p} \widetilde{h}^{m, n \pm 1},
\partial_{p} \partial_{q} \widetilde{h}^{m, n}, 
\partial_{p} \partial_{q} \widetilde{h}^{m \pm 1, n}, 
\partial_{p} \partial_{q} \widetilde{h}^{m, n \pm 1}  
\Bigr\} 
\label{eq:inputs_tau_enth}
\end{align}

\noindent where the index $i=1,2$ are the components of the vector, the indices $m$ and $n$ correspond to the discrete spatial location on the grid after filtering, and the indices $p,q=1,2$ represent the spatial indices along which we are taking derivatives.  The $(m,n\pm 1)$ index refers to the value of the quantities in the cells located at $(m,n-1)$ and $m,n+1$, while $(m\pm 1,n)$ refers to cell designated by $(m-1,n)$ and $(m+1,n)$. 
\section{Spectra Calculation Details}
\label{app:specta_calc}
The spectra of the simulation $\Ekin(k)$ and $\Emag(k)$ were computed in Fourier space with a 2D shifted Fast Fourier Transform (FFT) such that the FFT is centered at $k=0$ under the assumption of periodic boundary conditions that were used in the simulation.  We first calculated the 2D wavenumber $k_{2D}(k_x,k_y) = \sqrt{k_x^2+k_y^2}$, where $k_x={2\pi n_x}/{L_x}$ and $k_y={2\pi n_y}/{L_y}$ and $L_x=L_y=1$ is the length in the $x$ and $y$ directions respectively.  Here $n_x \in [-N_x/2,N_x/2-1], n_y \in [-N_y/2,N_y/2-1]$ are integers and $N_x,N_y$ are the number of grid points in the $x$ and $y$ directions respectively.  We then calculated the energy spectra associated with each of these wave numbers $k_x,k_y$ for the 2D kinetic energy and magnetic energy as

\begin{gather}
\mathcal{E}_{kin,2D}\left( k_x,k_y \right) = \frac{\widehat{\sqrt{\rho}v_x} \widehat{\sqrt{\rho}v_x}^*+\widehat{\sqrt{\rho}v_y} \widehat{\sqrt{\rho}v_y}^*}{N_x^2 N_y^2} \\ 
\mathcal{E}_{mag,2D}\left( k_x,k_y \right) = \frac{\widehat{B_x}\widehat{B_x}^* + \widehat{B_y}\widehat{B_y}^*}{N_x^2 N_y^2} 
\end{gather}
where $\widehat{x}$ is the 2D shifted FFT of $x$ rendering it a function of $k_x,k_y$ and ${x}^*$ is the complex conjugate of $x$.

$k_{2D}$ was then resampled over as $k=n \Delta k$ where $\Delta k=\sqrt{\Delta k_x^2 + \Delta k_y^2}$, $\Delta k_x ={\pi}/{L_x}$,  $\Delta k_x ={\pi}/{L_y}$, and $n\in [1,N]$ is an integer.  To resample, we computed $k_{diff}(k,k_x,k_y)=\abs{k-k_{2D}(k_x,k_y)}$ for every value of $k_x,k_y$ looping over values of $k$.  Then, for each value of $k$ we compute $\Ekin(k)$ and $\Emag(k)$ as

\begin{align}
\Ekin(k) &= \sum_{k_x}\sum_{k_y} \left\{\begin{array}{ll}{\mathcal{E}_{kin,2D}(k_x,k_y)} & {\abs{k_{diff}(k,k_x,k_y)}}<\frac{\Delta k}{2}\,, \\ {0} & {\text { otherwise }}\end{array}\right.\\
\Emag(k) &= \sum_{k_x}\sum_{k_y} \left\{\begin{array}{ll}{\mathcal{E}_{mag,2D}(k_x,k_y)} & {\abs{k_{diff}(k,k_x,k_y)}}<\frac{\Delta k}{2}\,, \\ {0} & {\text { otherwise }}\end{array}\right.
\end{align}

\section{SGS Tensors}
\label{app:sgs_tensors}

\begin{figure}[h]
\centering
\includegraphics[height=0.23\textheight]{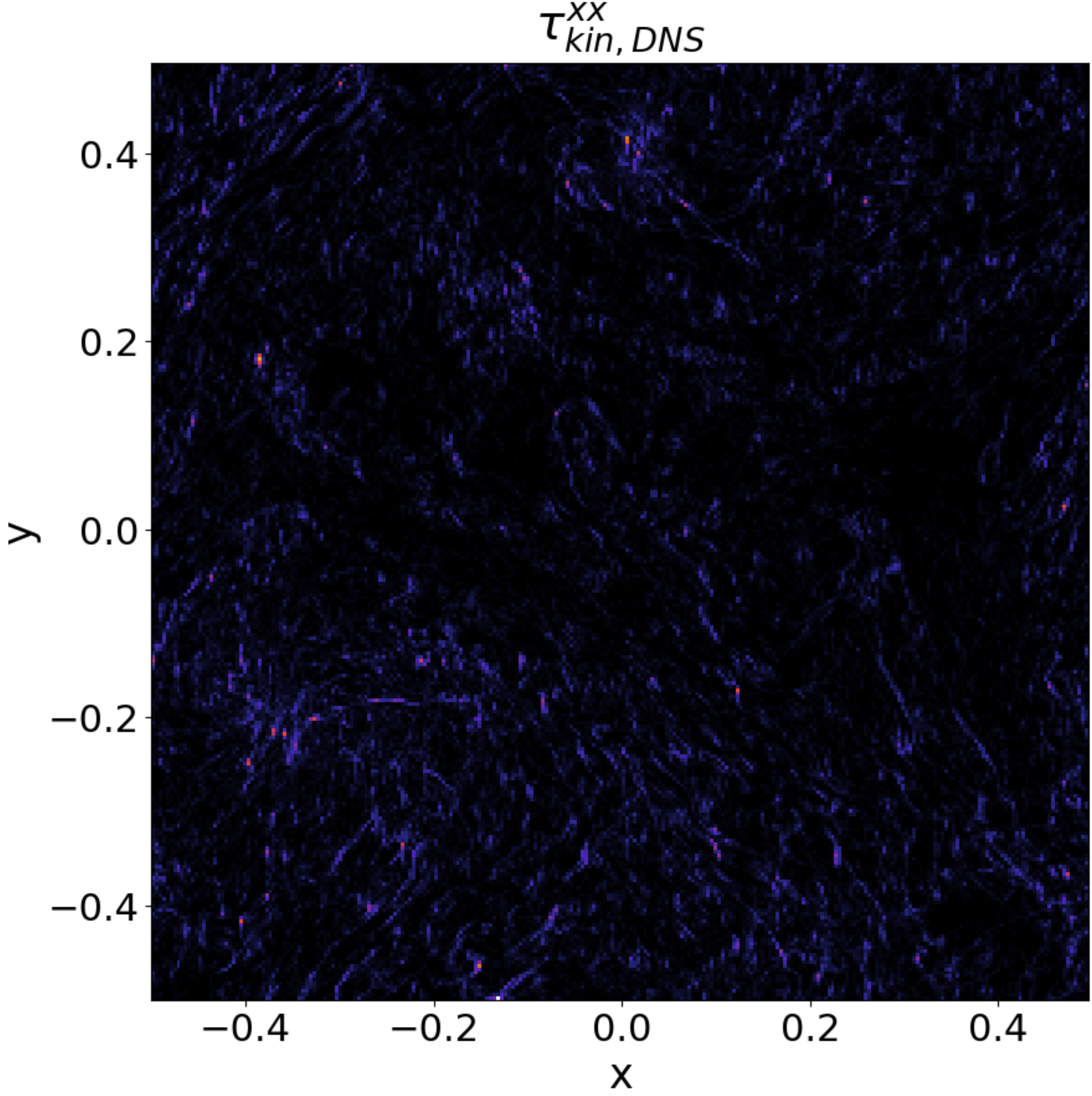}
\includegraphics[height=0.23\textheight]{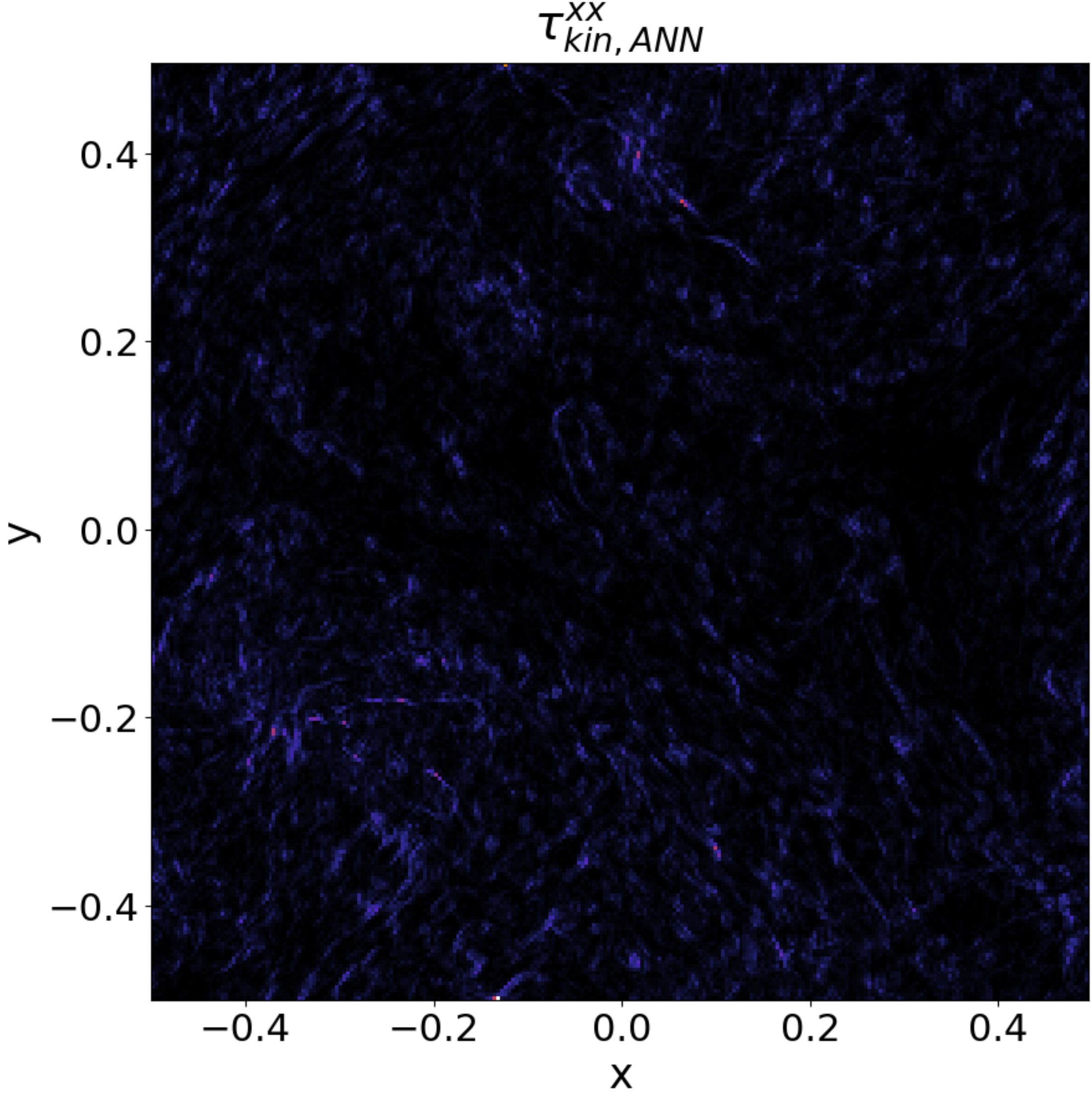}
\includegraphics[height=0.23\textheight]{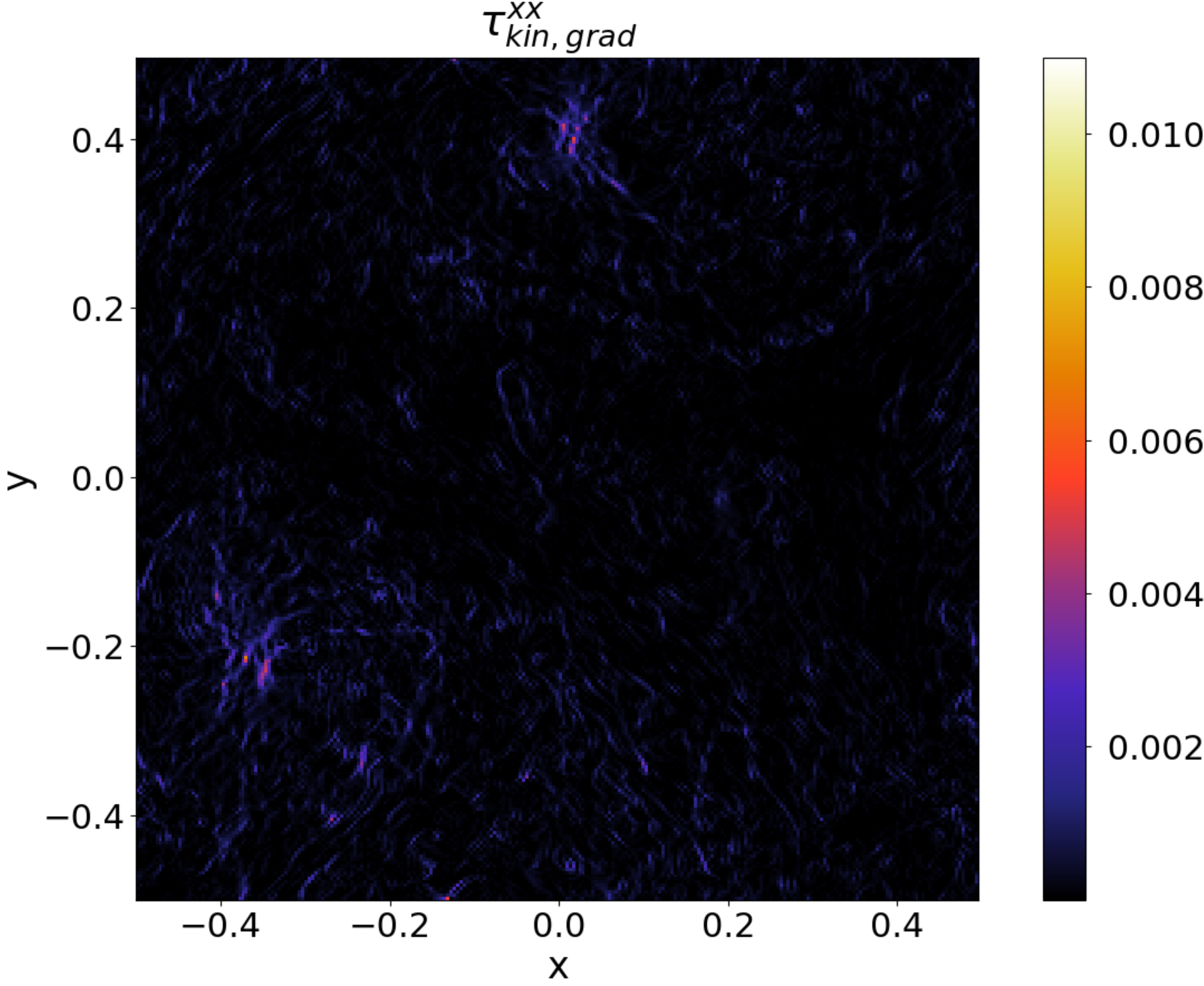}

\includegraphics[height=0.23\textheight]{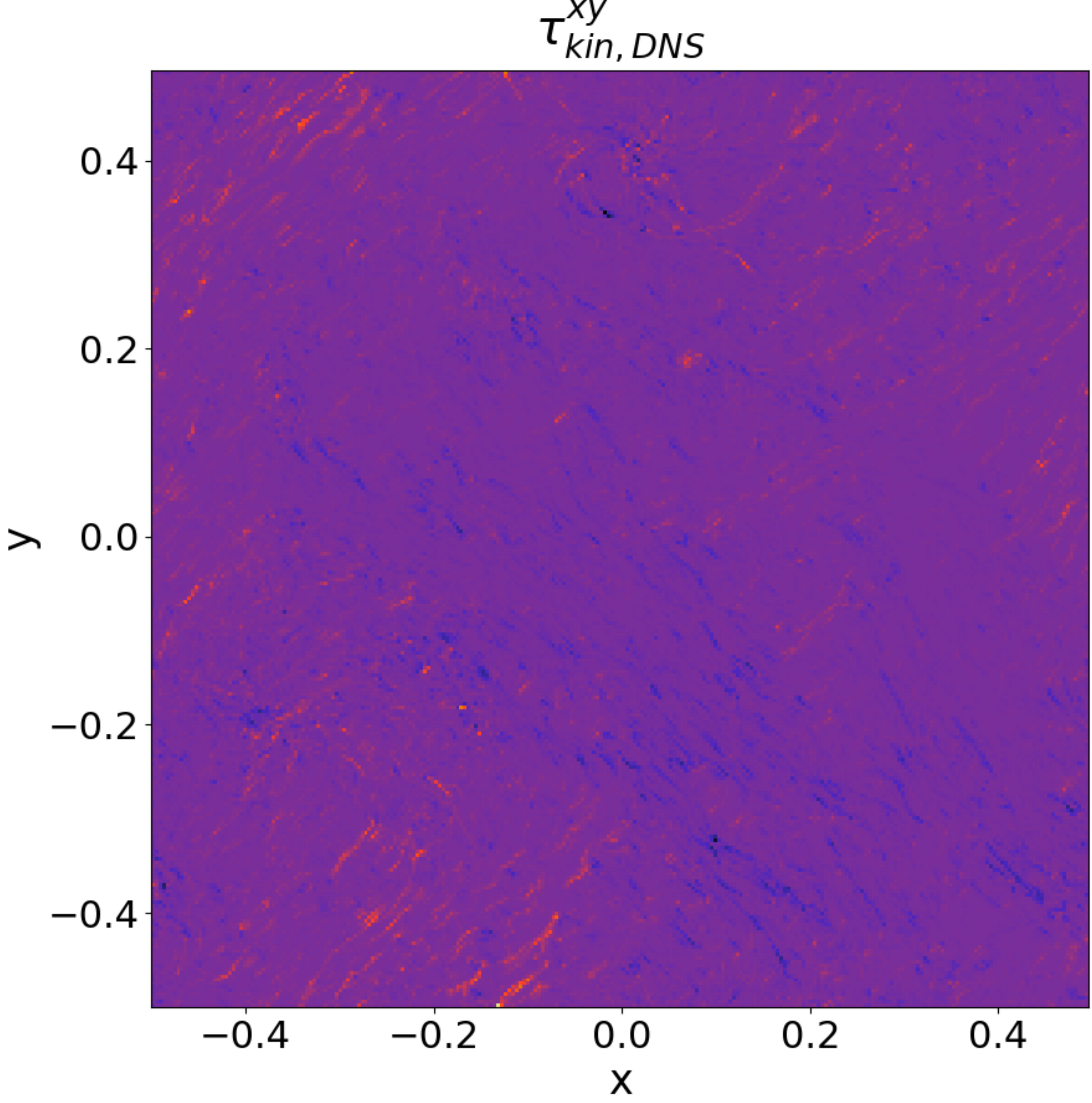}
\includegraphics[height=0.23\textheight]{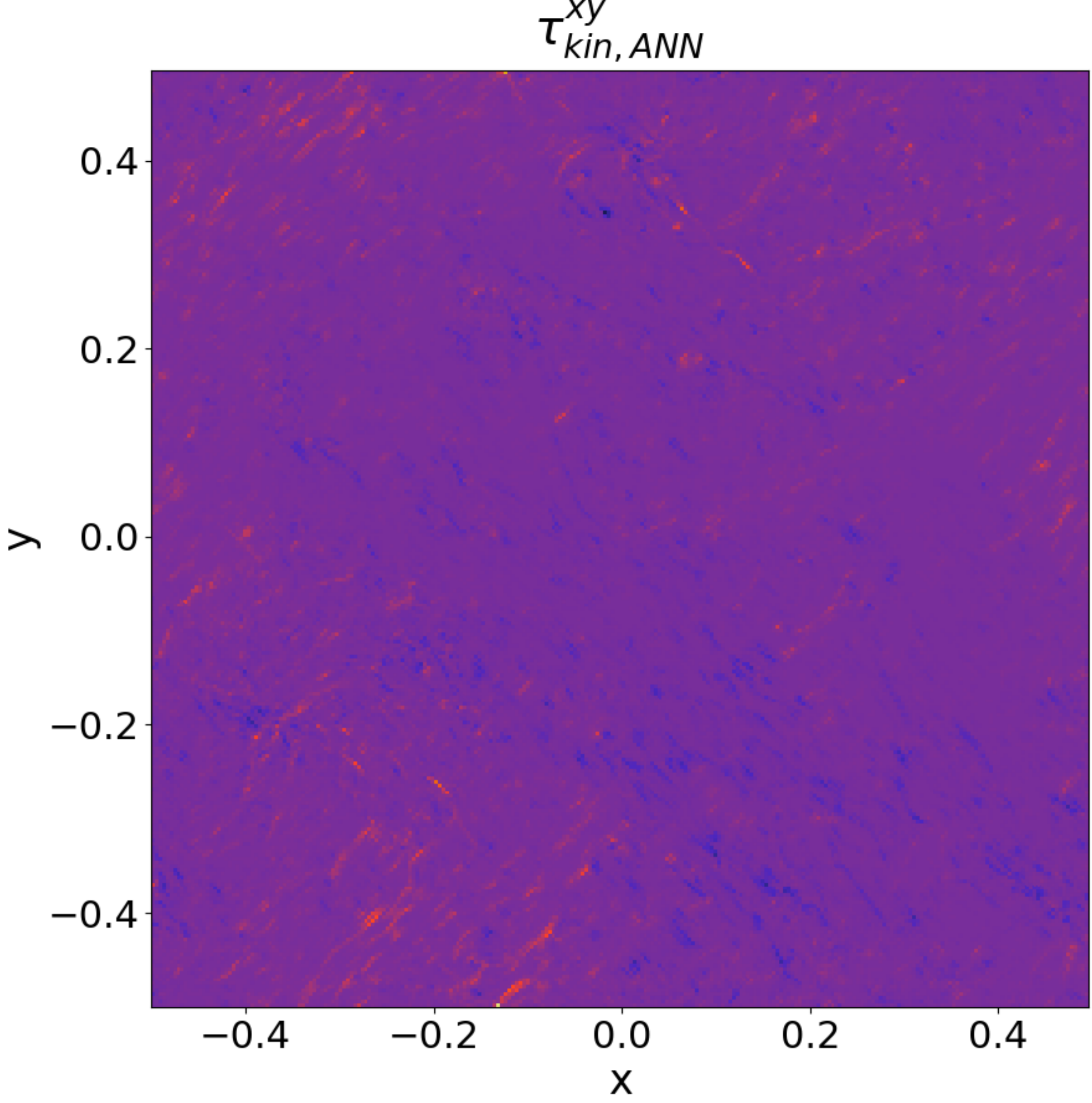}
\includegraphics[height=0.23\textheight]{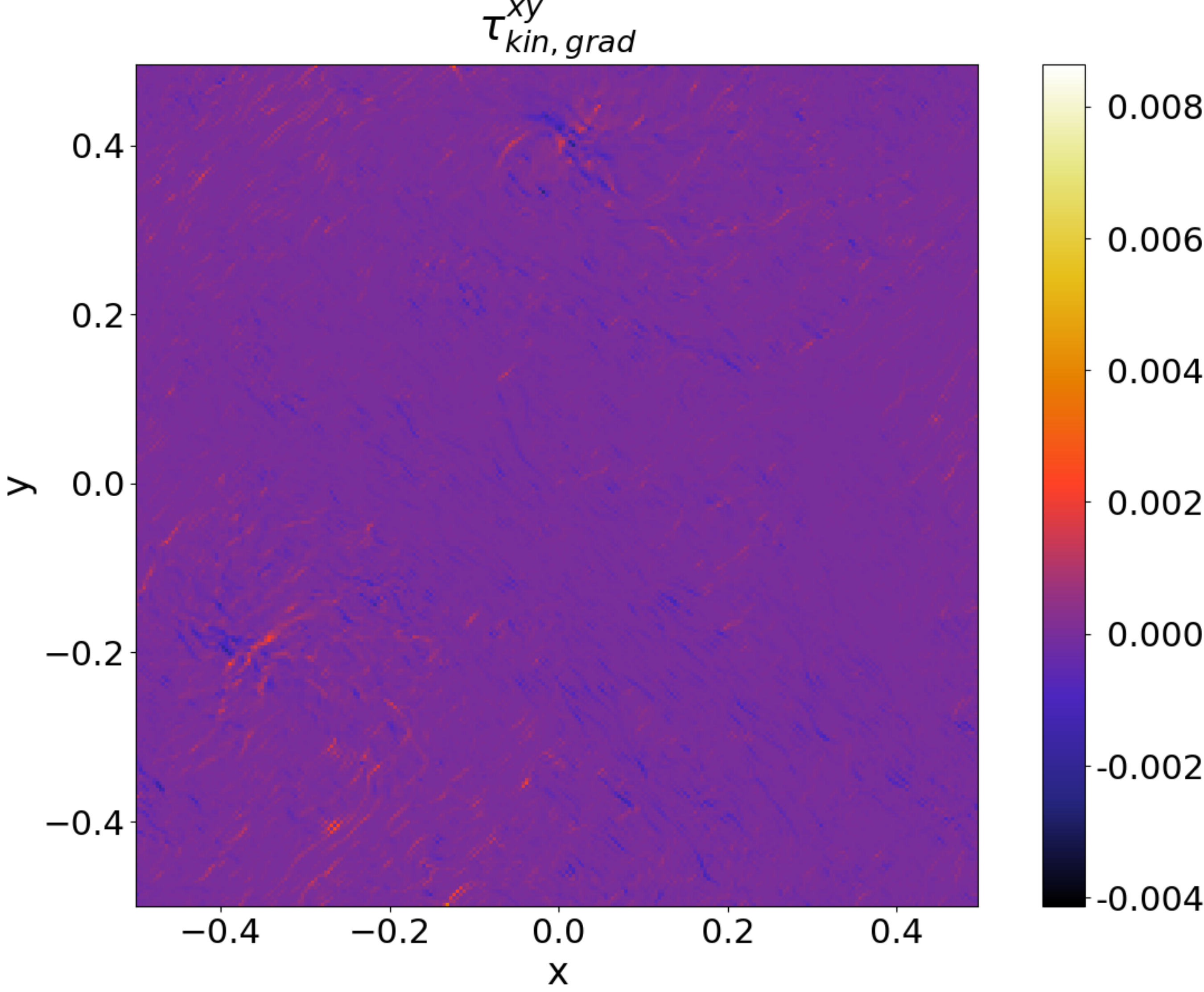}

\includegraphics[height=0.23\textheight]{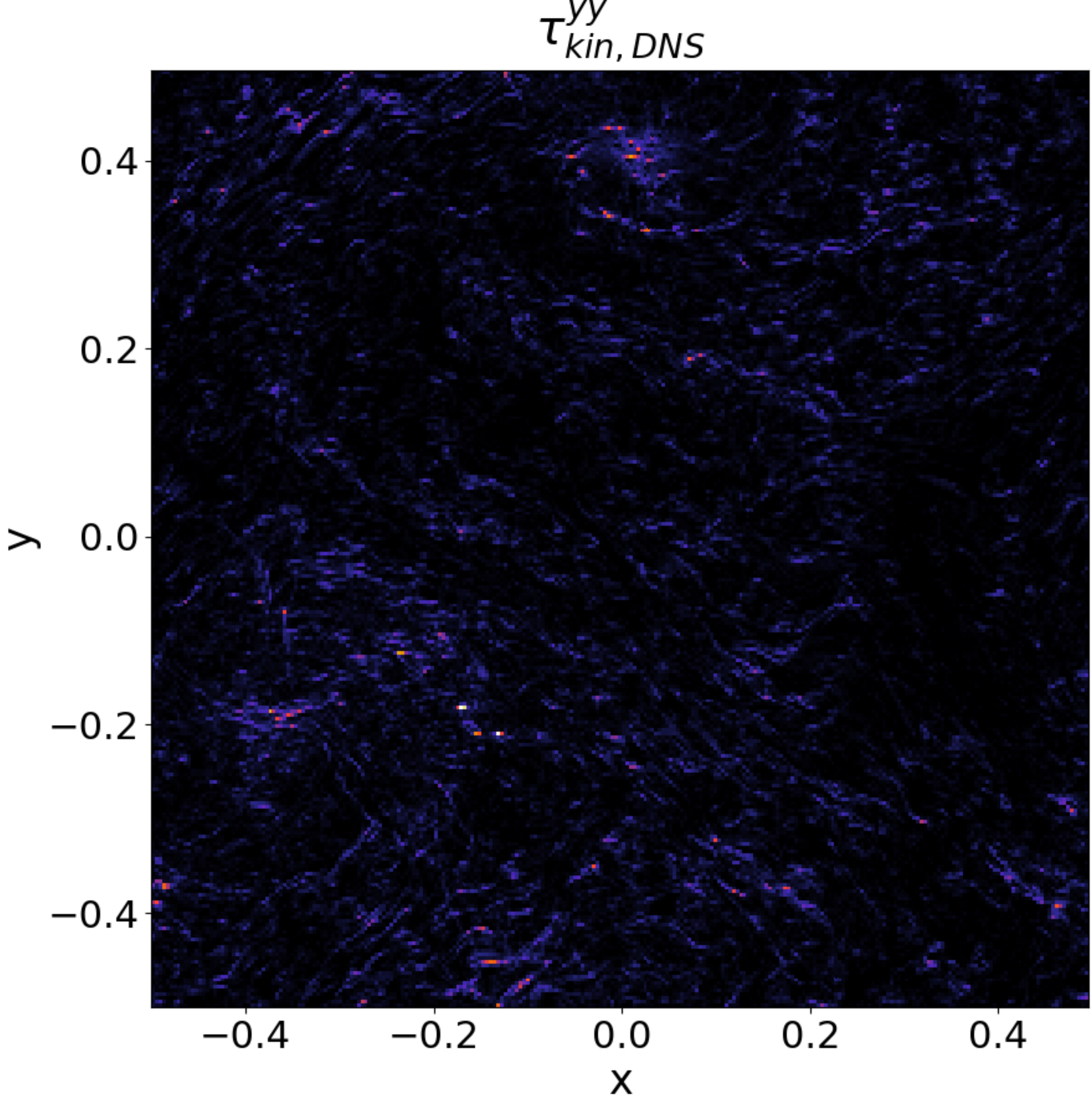}
\includegraphics[height=0.23\textheight]{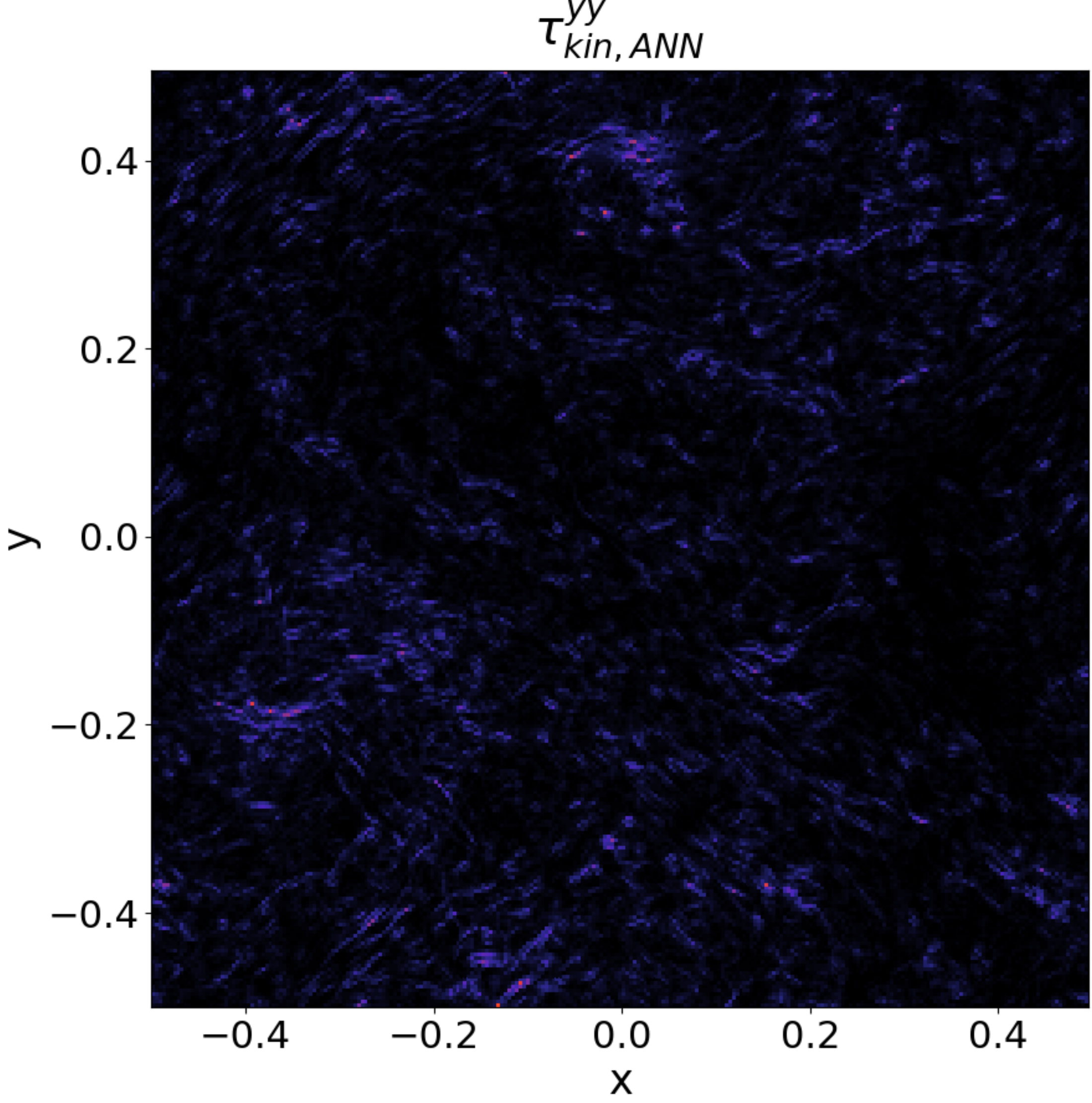}
\includegraphics[height=0.23\textheight]{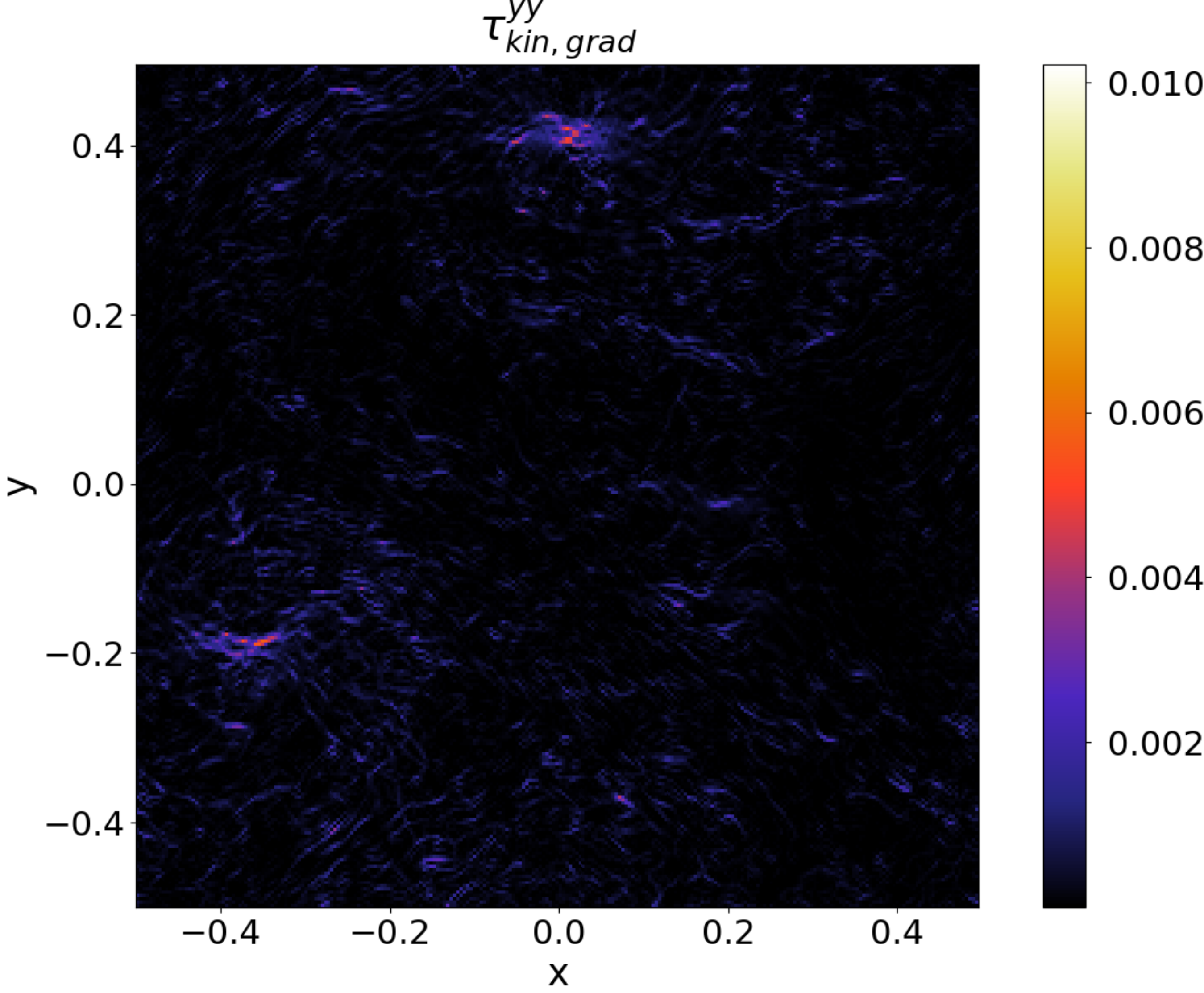}
\caption{Same as \Cref{fig:tau_mag} for the SGS tensor components of $\tau_{kin}$.}
\label{fig:tau_kin}
\end{figure}

\begin{figure}[h]
\centering
\includegraphics[height=0.23\textheight]{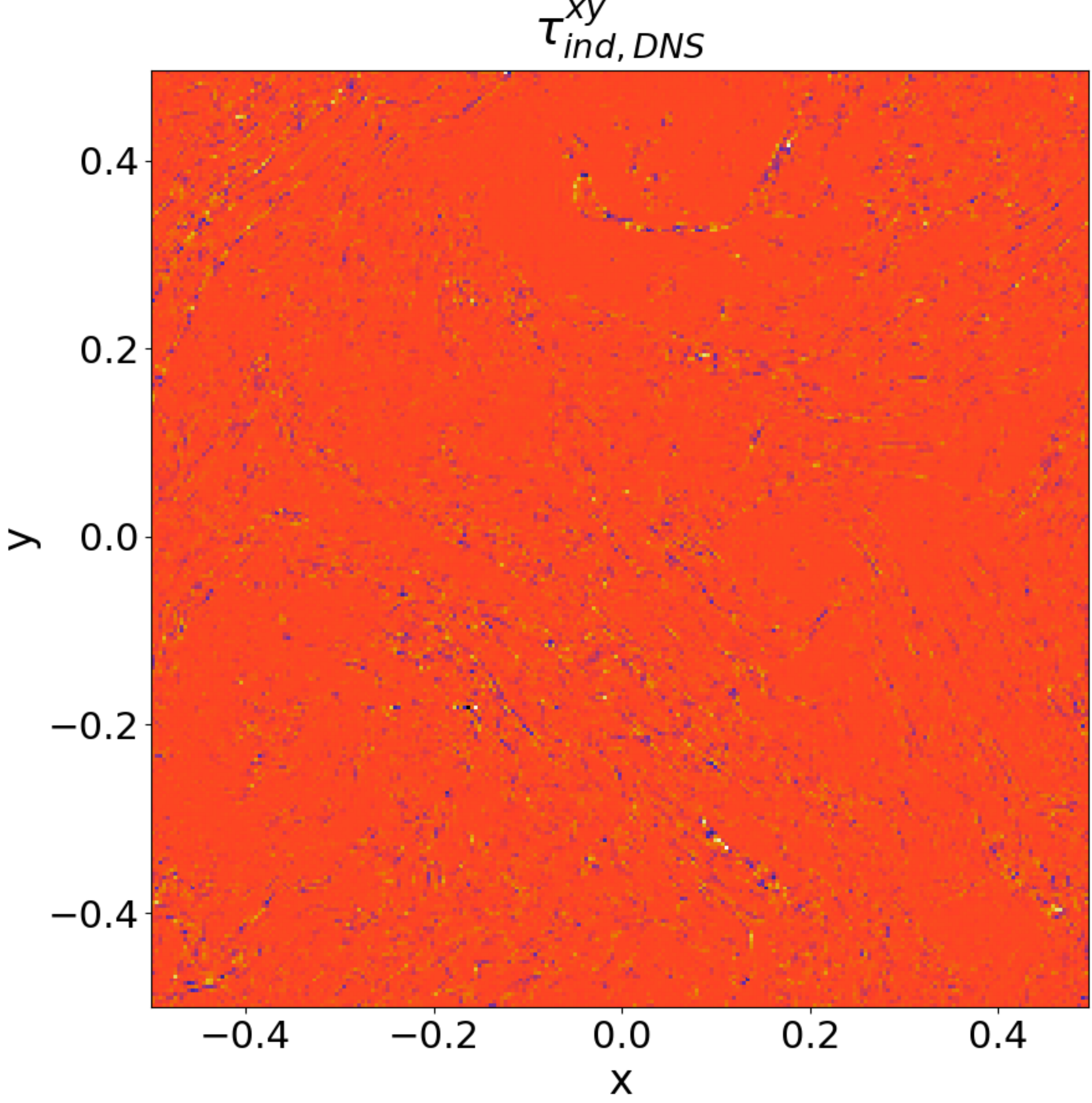}
\includegraphics[height=0.23\textheight]{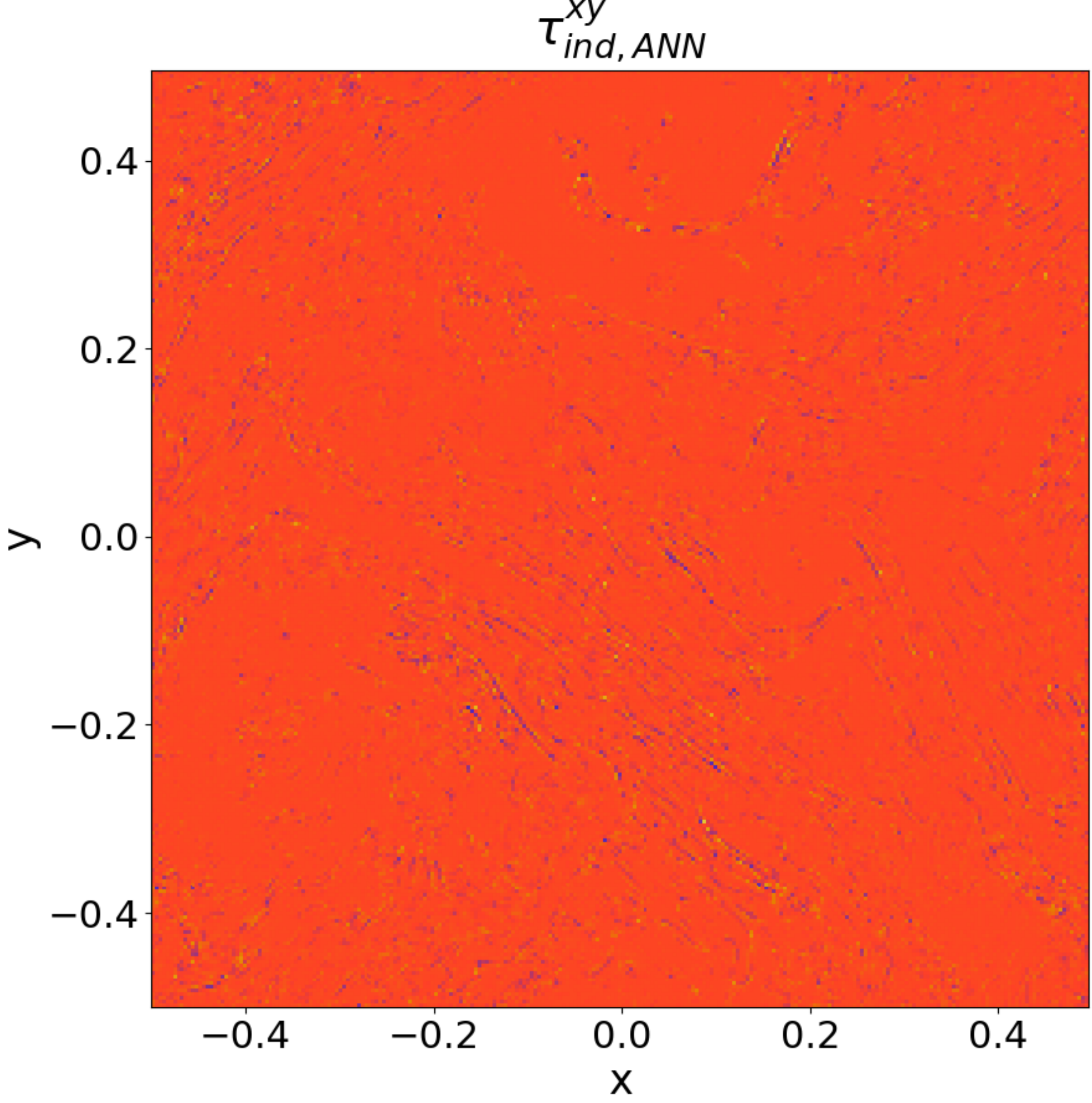}
\includegraphics[height=0.23\textheight]{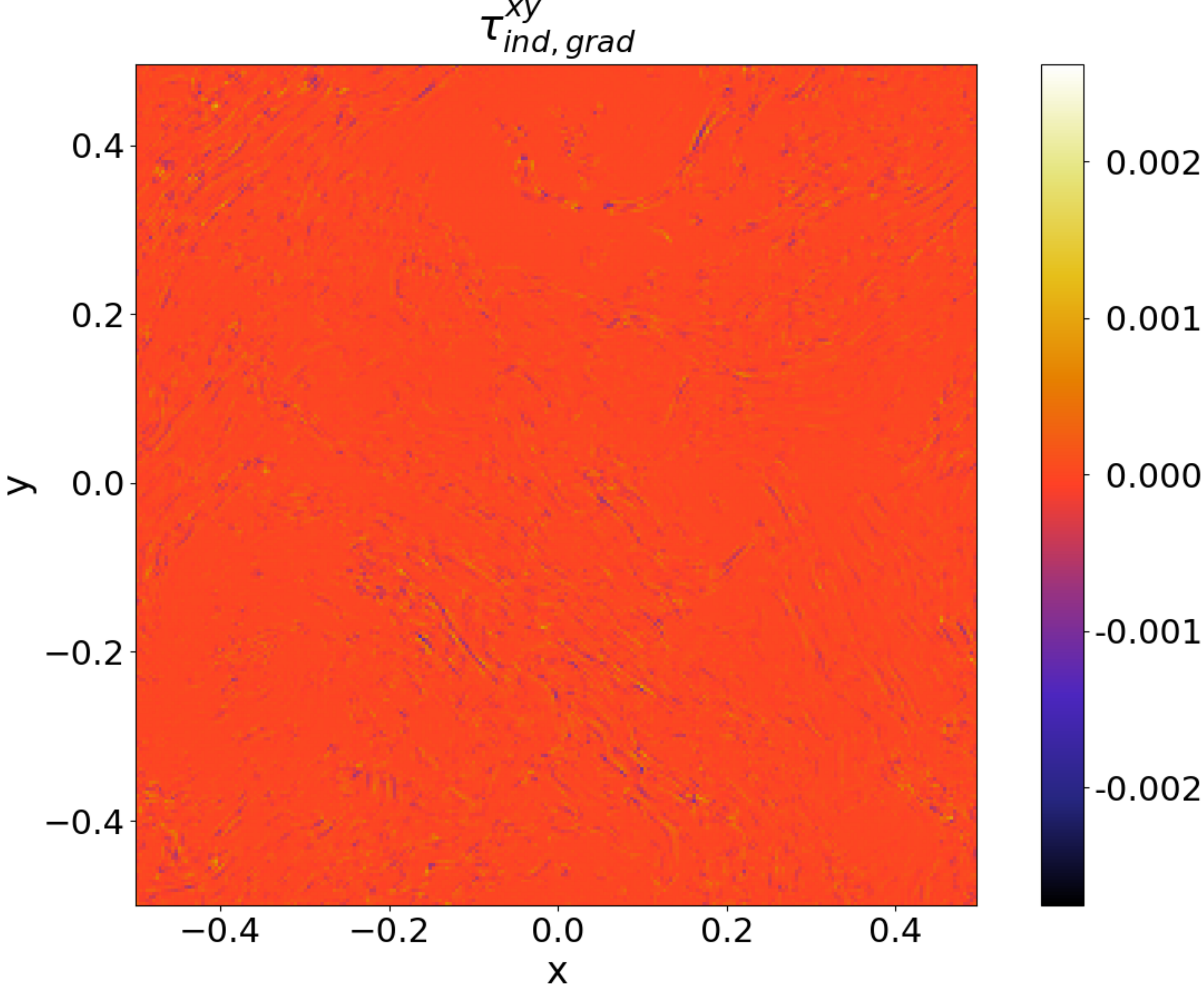}
\caption{Same as \Cref{fig:tau_mag} for the SGS tensor components of $\tau_{ind}$.}
\label{fig:tau_ind}
\end{figure}

\begin{figure}[h]
\centering
\includegraphics[height=0.23\textheight]{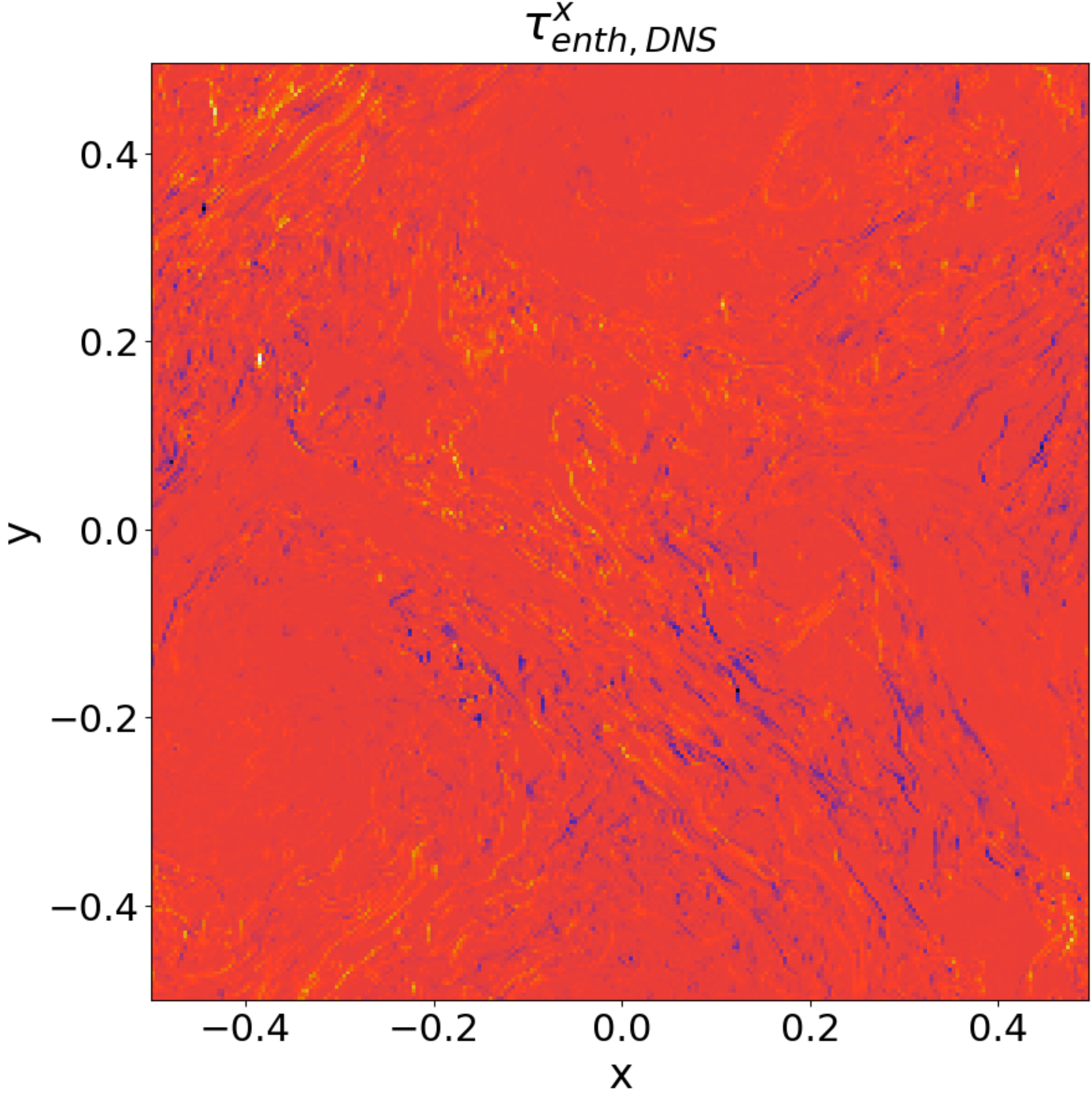}
\includegraphics[height=0.23\textheight]{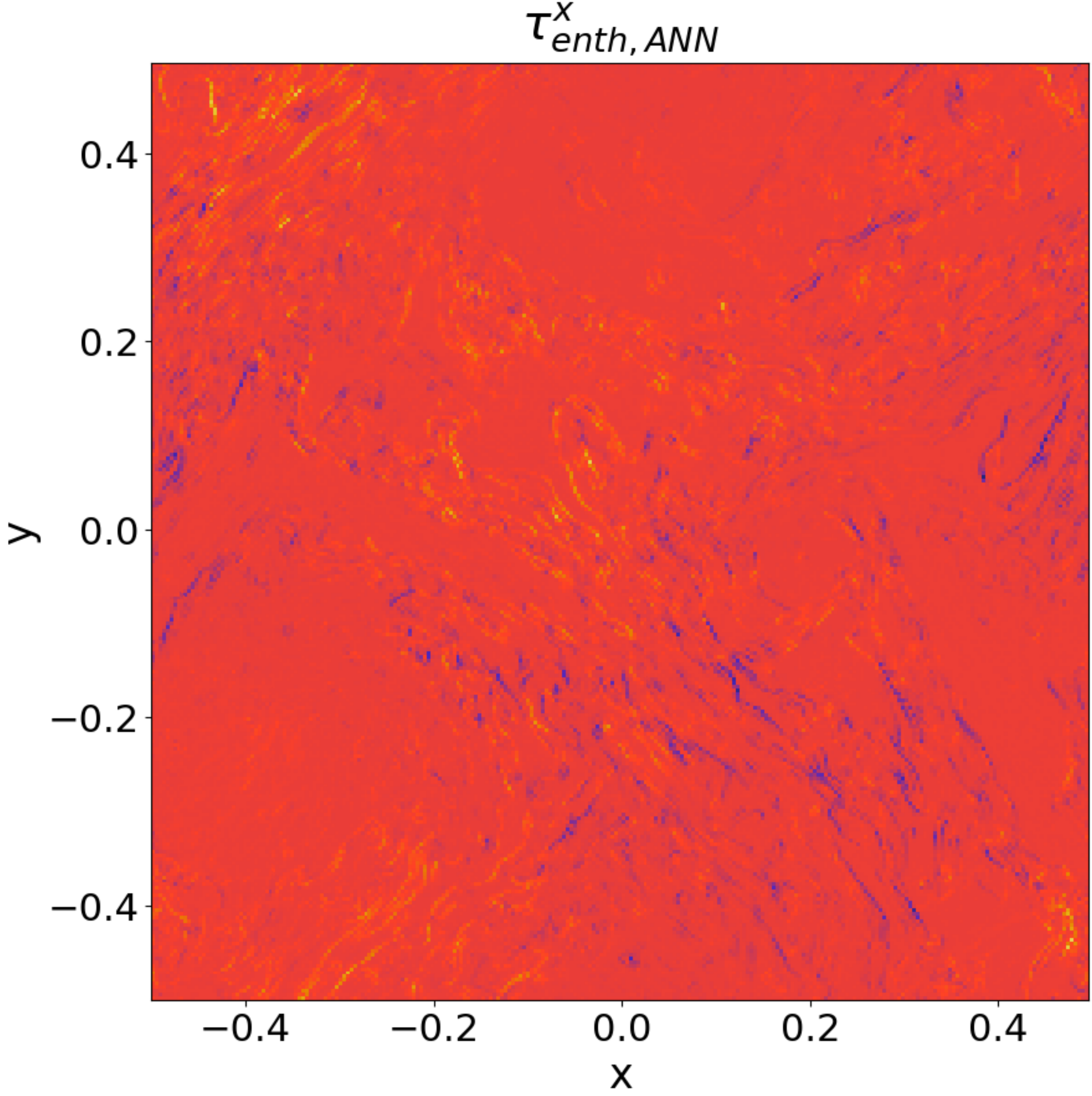}
\includegraphics[height=0.23\textheight]{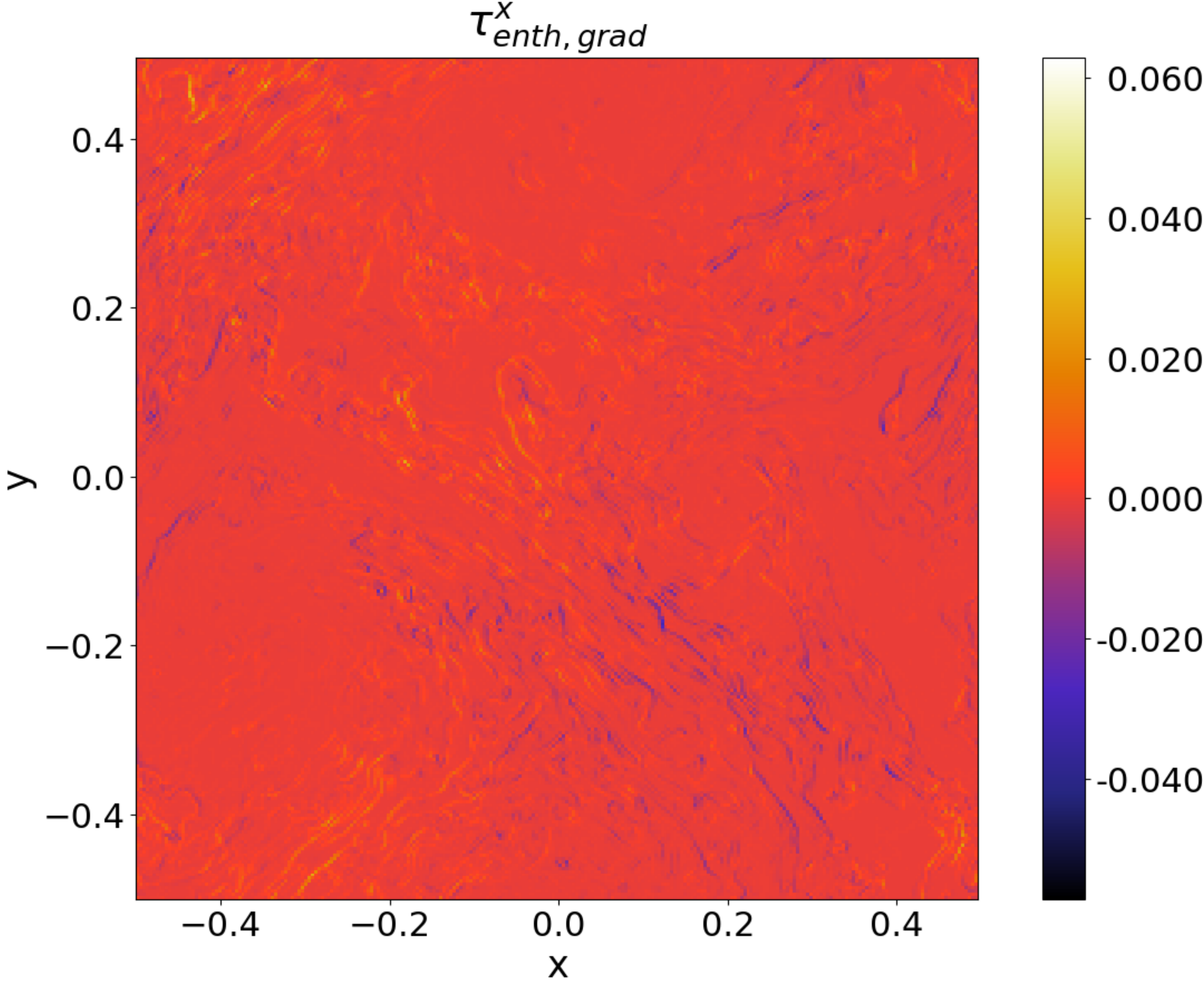}

\includegraphics[height=0.23\textheight]{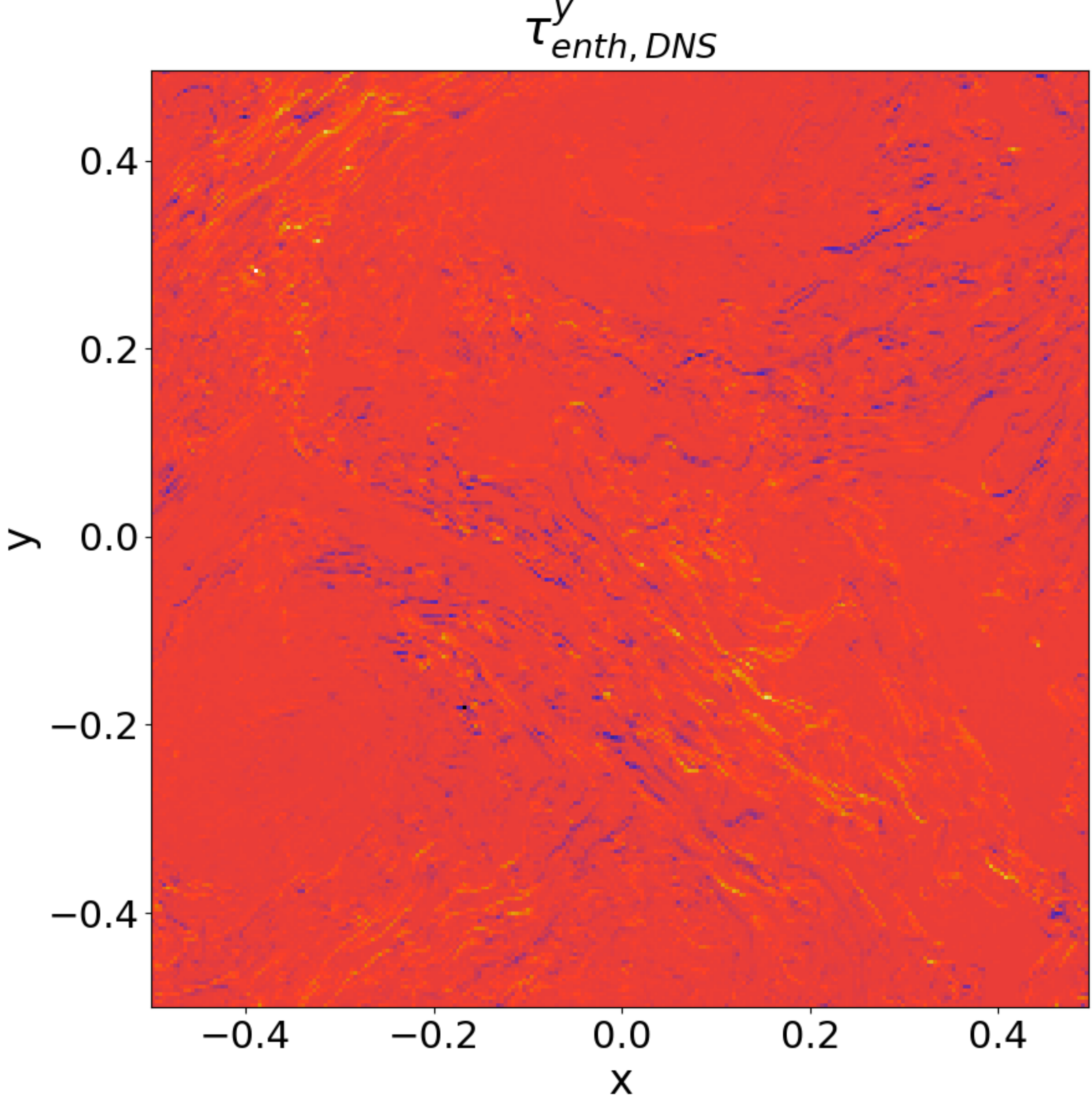}
\includegraphics[height=0.23\textheight]{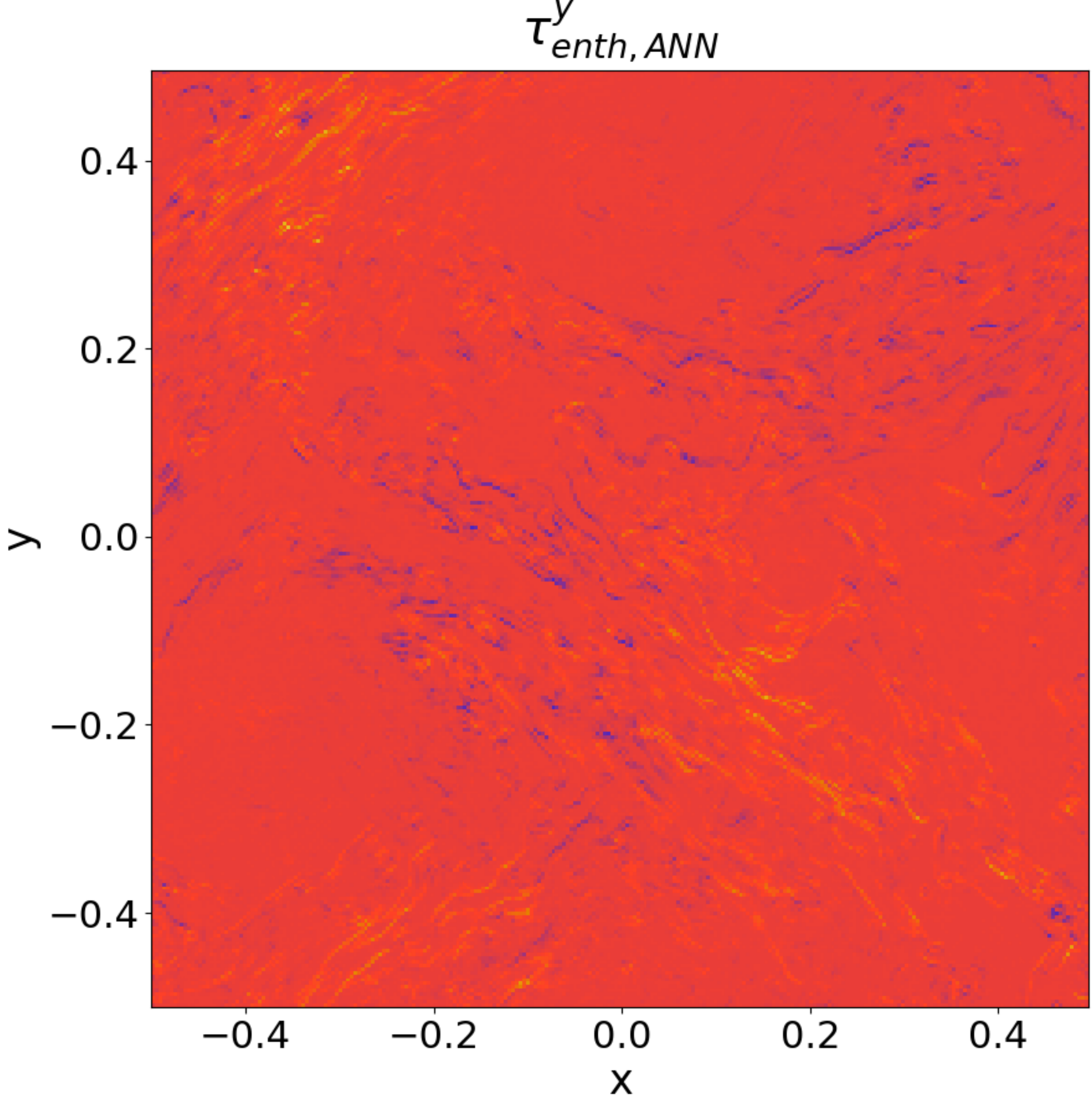}
\includegraphics[height=0.23\textheight]{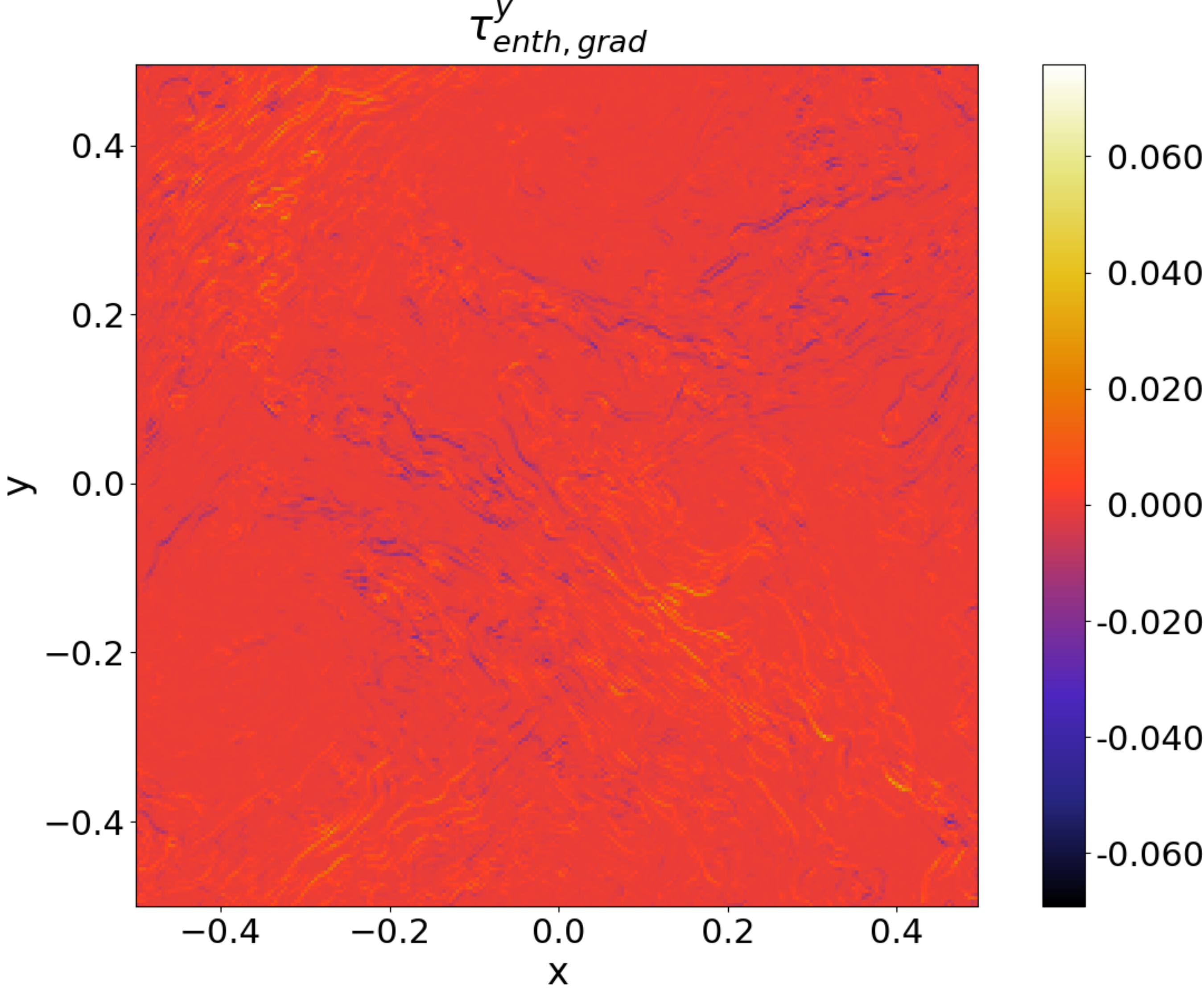}
\caption{Same as \Cref{fig:tau_mag} for the SGS tensor components of $\tau_{enth}$.}
\label{fig:tau_enth}
\end{figure}

Here we present the plots of the SGS tensors $\tau_{kin}$, $\tau_{ind}$, and $\tau_{enth}$ in \Cref{fig:tau_kin,fig:tau_ind,fig:tau_enth} respectively.  Each figure provides the value of $\tau$ computed from the DNS data $\tau_{DNS}$, the ANN model $\tau_{ANN}$, and the gradient model $\tau_{grad}$.

\end{document}